\documentclass[sigconf, nonacm]{acmart}
\usepackage{graphicx}
\usepackage{balance}  % for  \balance command ON LAST PAGE  (only there!)

\newcommand\norm[1]{\left\lVert#1\right\rVert}
\usepackage{subfigure}
\usepackage{times}
\usepackage{multirow}
\usepackage{algorithm}
\usepackage{algorithmic}
\usepackage{color}
\usepackage{listings}
\usepackage{amsmath}
\usepackage{cleveref}
\usepackage{courier}
\usepackage{fontenc}
\usepackage{amsmath}
\usepackage{latexsym}
\usepackage{bm}
\usepackage{subscript}
\usepackage{tgheros}
\usepackage[skip=0pt]{caption}
\usepackage[labelfont=bf]{caption}

\crefformat{section}{\S#2#1#3} % see manual of cleveref, section 8.2.1
\crefformat{subsection}{\S#2#1#3}
\crefformat{subsubsection}{\S#2#1#3}
\newcommand{\eat}[1]{}

\newcommand\vldbdoi{XX.XX/XXX.XX}
\newcommand\vldbpages{XXX-XXX}
\newcommand\vldbvolume{14}
\newcommand\vldbissue{1}
\newcommand\vldbyear{2020}
\newcommand\vldbauthors{\authors}
\newcommand\vldbtitle{\shorttitle} 
\newcommand\vldbavailabilityurl{http://vldb.org/pvldb/format_vol14.html}
\newcommand\vldbpagestyle{plain}

\eat{
\vldbTitle{Lachesis: Automated  Generation of Partitionings for Big Data Applications}
\vldbAuthors{Jia Zou,  Amitabh Das, Pratik Barhate, Arun Iyengar, Binhang Yuan, Dimitrije Jankov, Chris Jermaine }
\vldbDOI{https://doi.org/10.14778/xxxxxxx.xxxxxxx}
\vldbVolume{12}
\vldbNumber{xxx}
\vldbYear{2019}}

\begin{document}

% ****************** TITLE ****************************************

\title{Lachesis: Automatic Partitioning for UDF-Centric Analytics}

% possible, but not really needed or used for PVLDB:
%\subtitle{[Extended Abstract]
%\titlenote{A full version of this paper is available as\textit{Author's Guide to Preparing ACM SIG Proceedings Using \LaTeX$2_\epsilon$\ and BibTeX} at \texttt{www.acm.org/eaddress.htm}}}

% ****************** AUTHORS **************************************


% 1st. author
\author{Jia Zou Amitabh Das\\
        Pratik Barhate}
\affiliation{%
  \institution{Arizona State University}
}
\email{(jia.zou, adas59, pbarhate)@asu.edu}

\author{Arun Iyengar}
\affiliation{
    \institution{IBM T.J.Watson Research Center}
}
\email{aruni@us.ibm.com}

% 2nd. author
\author{Binhang Yuan  Dimitrije Jankov
Chris Jermaine}
\affiliation{
    \institution{Rice University}
}
\email{(by8, dj16, cmj4)@rice.edu}

\begin{abstract}
Partitioning is effective in avoiding expensive shuffling operations. However, it remains a significant challenge to automate this process for Big Data analytics workloads that extensively use user defined functions (UDFs), where sub-computations are hard to be reused for partitionings compared to relational applications. In addition, functional dependency that is widely utilized for partitioning selection is often unavailable in the unstructured data that is ubiquitous in UDF-centric analytics.
We propose the \textit{Lachesis} system, which represents UDF-centric workloads as workflows of analyzable and reusable sub-computations. \textit{Lachesis} further adopts a deep reinforcement learning model to infer which sub-computations should be used to partition the underlying data. This analysis is then applied to automatically optimize the storage of the data across applications to improve the performance and users' productivity.
\end{abstract}

\maketitle
\pagestyle{plain}

\eat{
\pagestyle{\vldbpagestyle}
\begingroup\small\noindent\raggedright\textbf{PVLDB Reference Format:}\\
\vldbauthors. \vldbtitle. PVLDB, \vldbvolume(\vldbissue): \vldbpages, \vldbyear.\\
\href{https://doi.org/\vldbdoi}{doi:\vldbdoi}
\endgroup
\begingroup
\renewcommand\thefootnote{}\footnote{\noindent
This work is licensed under the Creative Commons BY-NC-ND 4.0 International License. Visit \url{https://creativecommons.org/licenses/by-nc-nd/4.0/} to view a copy of this license. For any use beyond those covered by this license, obtain permission by emailing \href{mailto:info@vldb.org}{info@vldb.org}. Copyright is held by the owner/author(s). Publication rights licensed to the VLDB Endowment. \\
\raggedright Proceedings of the VLDB Endowment, Vol. \vldbvolume, No. \vldbissue\ %
ISSN 2150-8097. \\
\href{https://doi.org/\vldbdoi}{doi:\vldbdoi} \\
}\addtocounter{footnote}{-1}\endgroup

\ifdefempty{\vldbavailabilityurl}{}{
\vspace{.3cm}
\begingroup\small\noindent\raggedright\textbf{PVLDB Artifact Availability:}\\
The source code, data, and/or other artifacts have been made available at \url{\vldbavailabilityurl}.
\endgroup
}
}

\lstnewenvironment{DSL}
  {\lstset{
        aboveskip=5pt,
        belowskip=5pt,
        mathescape=true,
        basicstyle=\ttfamily\small,
        morekeywords={Set, Vector, Map, HashMap, bool, select,
  multiselect, aggregate, join, partition, member, method, construct, true, false, if, else, CREATE TABLE, PARTITION BY,
  self, literal, vector, return, for push_back, function, enum, sort, string, double},
        literate={~} {$\sim$}{1},
        showstringspaces=false}\vspace{0pt}%
   \noindent\minipage{0.47\textwidth}}
   {\endminipage\vspace{0pt}}
%}%

\lstnewenvironment{SQL}
  {\lstset{
        aboveskip=5pt,
        belowskip=5pt,
        escapechar=!,
        mathescape=true,
%        language=SQL,
        basicstyle=\linespread{0.94}\ttfamily\small,
        morekeywords={JOIN, FROM, WHERE, SELECT},
        deletekeywords={VALUE, PRIOR},
        showstringspaces=false}
        \vspace{0pt}%
        \noindent\minipage{0.47\textwidth}}
  {\endminipage\vspace{0pt}}

\newcommand{\littlesection}[1]{\vspace{5pt}\noindent\textbf{#1}}

\vspace{-10pt}
\section {Introduction}
\label{sec:introduction}

%What is the problem? %Why is it important?
Big Data analytics systems such as Spark~\cite{zaharia2010spark}, Hadoop~\cite{white2012hadoop},  Flink~\cite{alexandrov2014stratosphere}, and TupleWare~\cite{crotty2015tupleware} have been designed and developed to address analytics on unstructured data which cannot be efficiently represented in relational schemas. Users can easily represent unstructured data as nested objects. By supplying user-defined functions (UDFs) written in the host language, such as Python, Java, Scala, or C++, users can use control structures
such as conditional statements and loops to express complex computations. Such systems provide high flexibility and make it easy to develop complex analytics on top of unstructured data, which accounts for most of the world's data (above $80\%$ by many estimates~\cite{turner2014digital}). 

{\color{red}{
Most Big Data analytics frameworks are deployed on distributed clusters and require to partition a large dataset horizontally across multiple machines~\cite{borthakur2008hdfs}. Because a large dataset can be involved in multiple join-based analytics workloads, finding the optimal partitioning is a non-trivial task~\cite{rao2002automating, agrawal2004integrating, nehme2011automated, zhou2012advanced, klonatos2014building, eltabakh2011cohadoop, boehm2016systemml, dittrich2010hadoop++}. Therefore, it is urgent to automate this partitioning process. Existing works in physical database design~\cite{rao2002automating, agrawal2004integrating, nehme2011automated, zhou2012advanced, klonatos2014building, hilprecht2020learning} can well automate the partitioning for relational datasets. As illustrated in Fig.~\ref{fig:3steps}, they enumerate partitioner candidates based on foreign keys and select the optimal candidate using a cost-based approach. 

However, it remains a significant challenge to automate this process for UDF-centric analytics workloads, where sub-computations are opaque to the system and hard to be reused and matched for partitionings compared to relational applications. First, functional dependency that is widely utilized for partitioning selection is often unavailable in the unstructured data. Second, while the cost model based on relational algebra is widely used for selecting optimal partitioner candidates for relational applications, there is no widely acceptable cost model for UDF-centric applications due to the opaqueness of UDFs and objects, as well as the complexity of the underlying systems~\cite{shi2014mrtuner}. \eat{In this paper, we identify and address three problems associated with automatic horizontal partitioning for UDF-centric analytics workloads:

}

\eat{
Therefore, we argue for designing a new toolkit to automatically partition data of arbitrary types for UDF-centric analytics.}

\begin{figure}[H]
\centering
\includegraphics[width=3.4in]{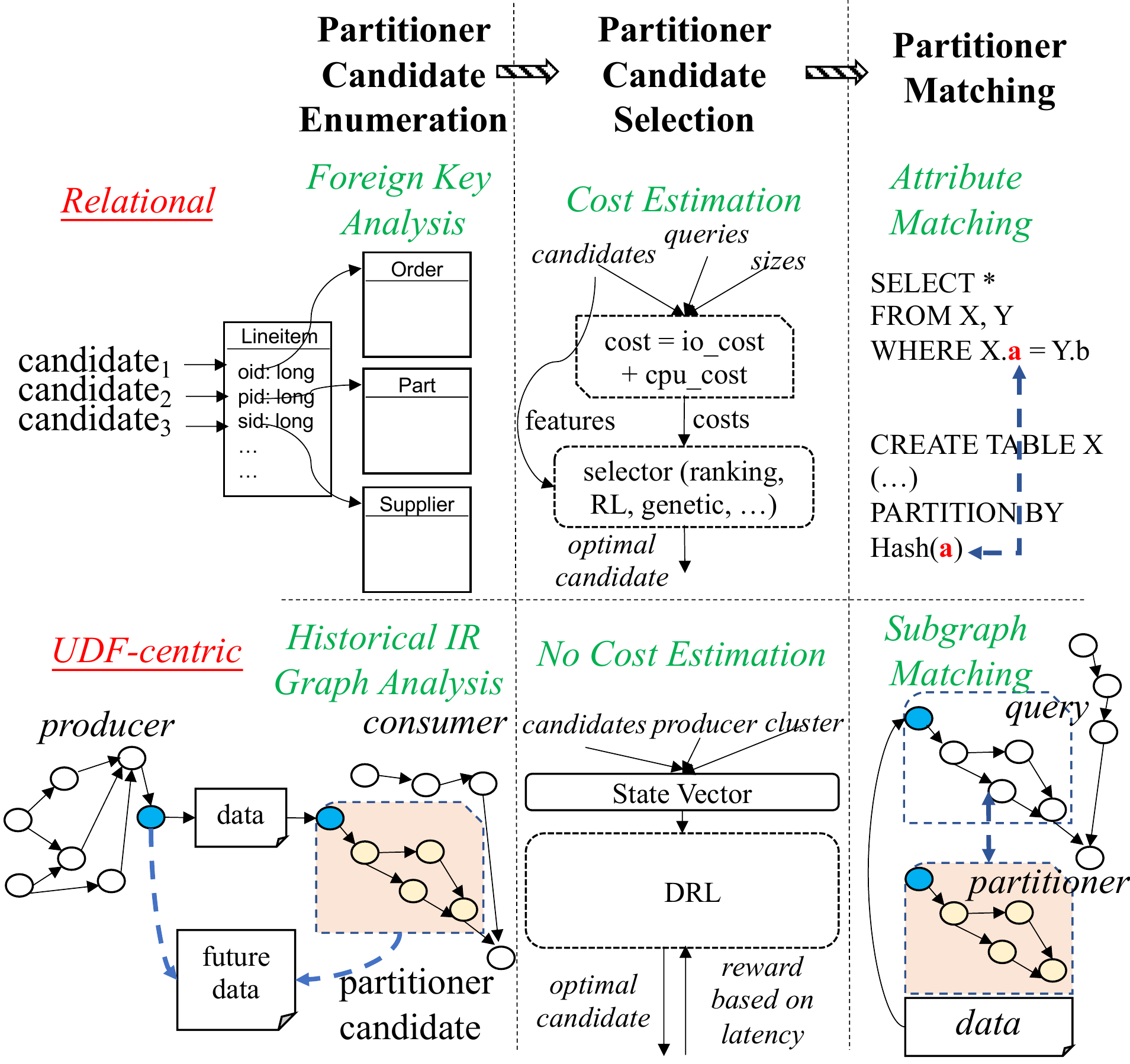}
\vspace{-1.0em}
\caption{ \small  \textit{Lachesis} vs. relational physical database design}
%\vspace{0.5em}
\label{fig:3steps}
\end{figure}

\vspace{-10pt}
\noindent
\textbf{Motivating Example.} We have three datasets: (1) a collection of reddit \textit{comments} objects in JSON ($\{c\}$); (2) a collection of reddit \textit{author} objects ($\{a\}$) in CSV; and (3) a collection of the \textit{subreddit} community objects ($\{sr\}$) also in JSON. We need to rank comments by their impacts. The impact score ($I_{c}$) of a comment ($c$) is determined by some dynamic logic as illustrated in Eq.~\ref{eq:importance}. It is computed based on a classifier that predicts whether the author or the subreddit community is more important in determining the impact score for this comment. If the author is more important, the impact score is computed as the maximum of the comment's impact ($c.ck$), and the author's impact ($a.lk$). Otherwise, the impact score is computed as the number of subscribers ($sr.ns$) of the subreddit channel associated with the comment. The classifier may use arbitrary algorithm, such as a complex deep learning neural network or a simple conditional branch such as \texttt{c.score > x}.

\vspace{-10pt}
\begin{equation}
  I_{c} =
    \begin{cases}
      \max\{a.lk, a.ck\} & if ~classify(c) ~is ~true\\
      sr.ns  & otherwise
    \end{cases}  
\label{eq:importance}
\end{equation}

The comment importance scores can be computed via a UDF-customized three-way join, of which the UDF that defines the join selection predicate is illustrated in Listing.~\ref{code1}. 

Unlike SQL applications where in many cases people can simply follow the foreign keys to perform co-partitioning, UDF-centric applications may involve arbitrary logic such as the  \texttt{classify()} in the above example. Even the application programmer cannot easily figure out the optimal partitioning. To make it worse,  UDF-centric applications running over complex objects are almost opaque to the system, for example, the system does not understand what is happening inside the \texttt{join\_selection()} function as illustrated in Listing.~\ref{code1} and thus automatic enumeration, selection, and matching of partitionings for this problem become difficult. }} 

\vspace{-3pt}
\begin{lstlisting}[language=C++,frame=single,caption={\color{red}{UDF-centric join selection predicate}},label=code1, breaklines=true, morekeywords={Set, Vector, Map, HashMap, bool, select,
  multiselect, aggregate, join, partition, member, method, construct, true, false, if, else, CREATE TABLE, PARTITION BY,
  self, literal, vector, return, for push_back, function, enum, sort, string, double}, basicstyle=\scriptsize, columns=fullflexible]
bool join_selection (string comment_line, string author_line, string subreddit_line) {
    string new_comment_line = schema_resolve(comment_line); //preprocessing
    json c = my_json::parse(new_comment_line); //parsing comment json object
    if (classify(c) == true) { //need to join with authors
        string c_a = c["author"];//derive author name from comment
        vector<string> r = my_csv::parse(author_line);//parsing author CSV file
        string a_name = r[1]; //derive name from author
        return (c_a == a_name);
    } else { //need to join with subreddits
        string c_sr  = c["subreddit"]; //derive subreddit name from comment
        json sr = my_json::parse(subreddit_line);//parsing subreddit json object
        string sr_name = sr["name"]; //derive name from subreddit 
        return (c_sr == sr_name);
   }
}
\end{lstlisting}

\vspace{-10pt}
\begin{figure} [H]
\centering
   \includegraphics[width=3.4in]{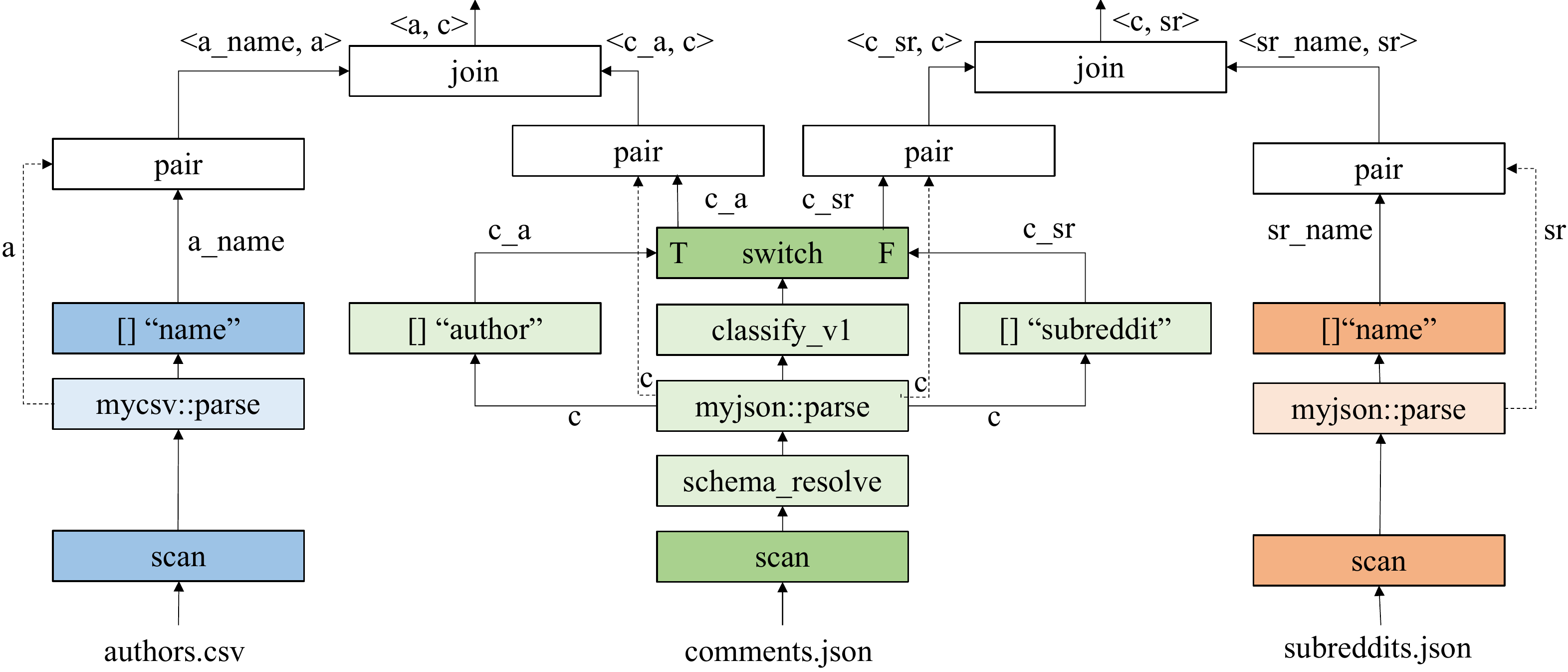}
\caption{\label{fig:graph-ir-example}
{\color{red}{\small IR graph for Listing. 1:  Partitioner candidates for authors, comments, and subreddits, are illustrated in different colors.}}
}
\end{figure}

\vspace{-10pt}

\vspace{5pt}
% our solution
\noindent
\textbf{\textit{Lachesis}: Automatic Partitioning.}
To address the problems, we propose a data partitioning optimizer for UDF-centric workloads, called as \textit{Lachesis}~\footnote{\small Lachesis is the name of a Greek god, who partitions lots and assigns fates to people. (https://en.wikipedia.org/wiki/Lachesis)}. \textit{Lachesis} allows user code to be translated into an Intermediate Representation (IR) that the system can reason with. For example, the code in Listing.~\ref{code1} is translated to a graph IR as shown in Fig.~\ref{fig:graph-ir-example}. We mainly focus on three problems in this work: 

\noindent
\textit{Problem 1. Partitioner Candidate Enumeration.} In UDF-centric processing, a partitioner candidate can be arbitrary logic that is deeply embedded in a UDF, which is hard to identify. To address the problem, we first abstract a partitioner candidate as a two-terminal graph ~\cite{bein1992optimal, riordan1942number, duffin1965topology} that has only one unique root node (i.e., a source node that has no parents) and one unique leaf node (i.e., a target node that has no children). Then we convert the problem into a subgraph searching and merging problem. {\color{red}{(Sec.~\ref{sec:problem2} and Sec.~\ref{sec:matching})}} \eat{For example, in Fig.~\ref{fig:graph-ir-example}, there exists two partitioner candidates available for partitioning the comment dataset, which can be merged to construct a new partitioner candidate, with the code  illustrated in Fig.~\ref{fig:comments-partitioner-candidate}. {\color{red}{(Sec.~\ref{sec:partitionings} and Sec.~\ref{sec:enumeration})}}

}

\noindent
\textit{Problem 2. Partitioner Candidate Selection.} A UDF-based partitioner candidate may involve dynamic control flows and it is hard to predict its runtime behavior. In addition, the cost model for relational partitioner candidate selection~\cite{rao2002automating, agrawal2004integrating, nehme2011automated, zhou2012advanced, klonatos2014building, eltabakh2011cohadoop, boehm2016systemml, dittrich2010hadoop++} cannot describe the overhead of manipulations (i.e, parsing, (de)compression, and (de)serialization) of arbitrary objects~\cite{shi2015clash, armbrust2015scaling}. These issues bring challenges for selecting the optimal partitioner. Therefore, we propose a deep reinforcement learning (DRL)~\cite{sutton1998reinforcement, lillicrap2015continuous, silver2016mastering, mnih2016asynchronous} formulation that is based on a set of unique features extracted from historical workflows for each partitioner candidate, including frequency, recency, selectivity, complexity, key distributions, number and size of co-partitioned datasets, etc.. {\color{red}{(Sec.~\ref{sec:problem1} and Sec.~\ref{sec:rl})}}

\noindent
\textit{Problem 3. Partitioner Matching.} For a runtime query, if a dataset has been partitioned using a UDF-based partitioner, the query optimizer should recognize the partitioning and decide whether a shuffling stage can be avoided. To facilitate such matching, we abstract the UDF matching problem into an IR subgraph isomorphism problem~\cite{cook1971complexity}. \eat{Although the general subgraph isomorphism problem is NP-complete,} We further provide a solution by utilizing the two-terminal characteristics of the partitioner IR graphs. {\color{red}{(Sec.~\ref{sec:problem2} and Sec.~\ref{sec:matching})}}

\noindent
\textbf{Our contributions} can be summarized as:

\noindent
(1) As to our knowledge, we are the first to systematically explore automatic partitioning for UDF-centric applications. We propose {\it Lachesis}, which is an end-to-end cross-layer system that automatically creates partitions to improve workflow performance.

\noindent 
(2) We propose a set of new functionalities for partitioner candidate enumeration, selection, and partitioner matching, based on subgraph searching and merging, DRL with historical workflow analysis, and isomorphic subgraph matching.

\eat{
\noindent
(3) We combine a deep reinforcement learning (DRL) approach based on actor-critic networks~\cite{mnih2015human, silver2016mastering, mnih2016asynchronous} with historical workflow analysis for choosing the optimal partitioning candidate.}

\noindent
(3) We implement the \textit{Lachesis} system and conduct detailed performance evaluation and overhead analysis.

{\color{red}{\section{Background}
\label{sec:background}
\subsection{IR for UDF-centric analytics}
User defined function (UDF) is first proposed as an enrichment of the SQL language, to allow SQL programmers to implement their own functions for processing relational data. Later, the emergence of MapReduce and dataflow platform like Hadoop~\cite{white2012hadoop}, Spark~\cite{zaharia2010spark}, Flink~\cite{alexandrov2014stratosphere}, further integrate the UDFs with high-level languages like Java, Python so that even non-relational data such as texts and images can be easily processed. However, the embedding of opaque functions that lack costing prevents UDF-centric workflows from being automatically optimized. To address the problem, several technologies were proposed in the past.

Froid~\cite{ramachandra2017froid} assumes the underlying data to be \textit{relational}, the query that invokes the UDF must be a SQL query, and the executions of UDFs are based on statement-by-statement interpretation. Then, Froid transforms the imperative statements, conditional blocks, and loops in the UDF into relational algebraic expressions\eat{, so that the UDF is inlined into the calling SQL query as nested subqueries}. 

In contrast, other existing IRs, including Emma/Lara~\cite{alexandrov2015implicit, kunft2019intermediate}, Weld~\cite{palkar2017weld}, PlinyCompute~\cite{zou2018plinycompute}, are designed for automatic optimization of UDF-centric workloads running on \textit{unstructured data}. 

Weld provides a cross-library IR based on a parallel loop operator and several declarative builders and mergers for vectors (i.e., \texttt{vec[T]}), dictionaries (i.e., \texttt{dict[K, V]}), and groups (i.e., \texttt{dict[K, vec[V]]}), to facilitate loop fusion across libraries. In Weld's implementation~\footnote{https://github.com/weld-project/weld}, a (hash) join is represented at low-level by building and probing the dictionary using the parallel loop operator. \eat{Therefore, the desired partitioning semantic for an input dataset can be identified in one or more UDFs that transform an input instance into the join key \eat{(i.e., \texttt{K})}.} A UDF can be further represented as an abstract syntax tree (AST). Part or all of the AST tree can be replaced by the invocations of opaque C/C++ functions, depending on how much details the programmer wants to expose to the system.

Emma/Lara and PlinyCompute provide an even more declarative IR representation for a $k$-way join operation, which can be abstracted into following expression that is similar to relational calculus:

\begin{equation}
\{(x_1, ..., x_k)   |   p(x_1, ..., x_k), x_i \in X_i, 1\leq i \leq k\}
\end{equation}

\vspace{-5pt}
\noindent
except that $p$ can be represented as a UDF that processes arbitrary objects and returns a boolean value, such as the \texttt{join\_selection} as illustrated in Listing.~\ref{code1}. Similar to Weld, the UDFs are represented as ASTs . Our proposed Lachesis approach is designed based on top of this $k$-way join representation.

\vspace{-5pt}
\subsection{Storage Requirements}
This work mainly considers two types of data partitioning: (1) persistent partitioning, which is to persist the partitionings of the data in the underlying storage, so that it can be reused across applications; and (2) intra-application partitioning, where the partitionings are enforced at runtime, can only live within the lifetime of an application, and is only visible to the application. Many distributed UDF-centric frameworks, such as Spark~\cite{zaharia2010spark}, support intra-application partitioning by allowing users to supply a partitioner in the application. In addition, SystemML~\cite{boehm2016systemml}, which is a linear algebra library built on top of Spark, automates the intra-application partitioning for various matrix manipulations. Intra-application partitioning is helpful for iterative joins like in PageRank, where the online repartitioning overhead can be amortized over multiple iterations in the same application. But such partitioning is inadequate for a broad class of workloads, such as data pre-processing, where a dataset is joined only once for each application. Making it worse, a non-partition-preserving operator such as {\texttt{map}} may easily remove an intra-application partitioning~\cite{boehm2016systemml}. 

However, to our surprise, most popular Big Data frameworks such as Spark do not support persistent partitioning. That's because their simple storage APIs~\cite{borthakur2008hdfs, HBase} cannot convey the partitioning information between the storage layer and the computation layer~\cite{zou2019pangea, zou2020architecture}. 
We have attempted to manually create persistent partitionings for Spark applications, but that only seems possible for  Hive~\cite{thusoo2009hive} tables via the bucketBy operator~\cite{luu2018spark}. However, most UDF-centric analytics tasks cannot represent their arbitrary data in Hive tables. 
CoHadoop~\cite{eltabakh2011cohadoop} allows programmers to manually enforce co-partitioning of HDFS files using MapReduce jobs and then specify this co-location relationship by changing the HDFS interface and namenode implementation, but the function used by the partitioning is hidden from the system. Therefore, the partitioning cannot be automatically reused or matched without the programmer's knowledge. \eat{HadoopDB~\cite{abouzeid2009hadoopdb} replaces HDFS using relational
databases. Then it allows the user to specify the partition key for
each table to partition data at loading time.}

\noindent
\textbf{Implementation, Deployment, and Elasticity.} We target at supporting both persistent and intra-application partitionings, thus we choose to implement \textit{Lachesis} on PlinyCompute~\cite{zou2018plinycompute}, which is a distributed UDF-centric analytics system implemented in C++. We use our previous work, Pangea  ~\cite{zou2019pangea}, as the storage, which allows to pass the partitionings of datasets to the computation layer so that the latter can utilize such information to avoid the shuffling.  \textit{Lachesis} can be easily extended to support other distributed frameworks that compile UDF-centric workloads into analyzable IRs, and use a storage that maintains partitioning information and communicates such information to the computations. 

The partitionings can be even pushed to a cloud storage like S3 that is disaggregated with the computation cluster, as long as the UDF specifying how a dataset is partitioned into multiple S3 objects is stored somewhere and queryable to the computations. 

To achieve elasticity, when cluster nodes are dynamically added and removed, we can leverage existing live data migration mechanisms~\cite{borthakur2008hdfs, lim2010automated} to minimize the service interruption for adding/removing nodes. In addition,  we can map partitions to cluster nodes using an elastic strategy such as lazy consistent hashing, following the SnowFlake's elastic architecture design~\cite{vuppalapati2020building}. \eat{Then minimal shuffling is required for adding/removing computational nodes.} These extensions are all orthogonal to this work.}}

\section{Problem Definition}
In this section, we analyze and formalize the problems. Then we summarize the challenges and main ideas. 

\vspace{-10pt}
\subsection{Assumptions and Targeting Workloads}
\label{sec:assumption}
The \textit{Lachesis} approach is based on following assumptions: 

\noindent
(1) The \textit{write-once read-many assumption} that \textit{once a dataset is written, it will be read many times}. It indicates that creating persistent partitionings while storing the data can benefit multiple workloads that take the data as input. Such a pattern is observed in a number of real-world traces~\cite{chen2012interactive}. For example, according to the publicly available Yahoo! cloud trace~\cite{yahootrace}, $83\%$ of total stored bytes have been accessed for more than once; 
$28\%$ of the bytes were accessed for even more than $100$ times.

\noindent
(2) The \textit{recurrent workflow assumption} that \textit{a majority of workflows are re-executions } on different or incremental datasets, as widely observed in recent Microsoft and other production traces ~\cite{jyothi2016morpheus, jindal2018computation, chen2012interactive}. Therefore, we can extract partitioner candidates from historical executions of workflows and reuse these for future datasets. For example, if the \textit{Comment-Loader} application loads a comment dataset collected in $2019$ to storage; and then the \textit{Reddit-Feature-Extractor} application joins the comments dataset with the authors dataset to create feature vectors for topic recommendation, the system will think that the workflow \textit{Comment-Loader} $\rightarrow$ \textit{Reddit-Feature-Extractor} may recur. Then in the future, if  \textit{Comment-Loader} loads a new comment dataset collected in $2020$ to storage, the input partitioning desired by the \textit{Reddit-Feature-Extractor} may be a good candidate for pre-partitioning the new data loaded by  \textit{Comment-Loader}.  

\eat{\noindent
(3) The \textit{intermediate representation (IR) assumption} that an analytics application written in an object-oriented language like Python, Scala, Java, C++, can be mapped to a graph-based IR, where each node represents an atomic computation and each edge represents either a data flow that is output from the source node and consumed by the destination node, or a control flow so that the destination node gets executed only if the source node finishes and triggers it. There are many existing IRs that can serve this role such as Weld IR~\cite{palkar2017weld}, Lara IR~\cite{kunft2019intermediate}, Spark's query graph IR~\cite{zaharia2012resilient, armbrust2015spark}, TensorFlow's computational graph IR~\cite{abaditensorflow}, etc., and these IRs are usually derived by embedding a domain specific language (DSL) or a library of APIs into the object-oriented languages. In this work, we choose to use PlinyCompute's lambda calculus IR~\cite{zou2020architecture} because in its design, after compilation, each atomic computation is executable separately, which means any subgraph of computations can be executed from the IR. This greatly simplifies the implementation of \textit{Lachesis}.} %A formalization of the DSL/IR that we use in this work can be found in a separate work~\cite{zou2018plinycompute}.

\eat{
\noindent
(3) The \textit{shared-nothing architecture assumption} that we focus on the automated persistent partitioning and co-location problem in a shared-nothing distributed architecture~\cite{stonebraker1986case}, which is followed by most of the high-performance and scalable DBMSs, including Teradata~\cite{xu2008handling}, Netezza~\cite{francisco2011netezza}, Greenplum~\cite{stonebraker2010sql}, and also used by most of the high-end internet platforms, such as Amazon, Akamai, Yahoo, Google, and Facebook~\cite{shared-nothing}.}

{\color{red}{
\textit{Lachesis} is focused on \textit{identifying the optimal horizontal partitioner candidates for datasets of arbitrary types}, in UDF-centric analytics workloads that involve shuffle-operations such as equi-join, group-by, and aggregations. 
We do not consider the vertical partitioning of arbitrary objects in \textit{Lachesis}, because it is more complicated and requires to reason with the object layout, and thus less popular in UDF-centric systems~\cite{zaharia2010spark, alexandrov2014stratosphere, crotty2015tupleware}. 

Once a partitioner candidate is selected, it will be used to extract partition key(s) from each object in the dataset, and the objects that have the same keys will be dispatched to the same node. Therefore, all types of joins that can be converted into equi-join, such as array join~\cite{duggan2015skew} based on the equality of dimensions, attributes, or both; and similarity/fuzzy join based on the equality of locality sensitive hashing~\cite{chen2019customizable}, can benefit from our work. 

In addition, how to map keys to nodes is not a focus of this work and we use a simple hashing mechanism for that in our implementation. More advanced mapping techniques, e.g., skew-aware mapping and elastic mapping as used in array join~\cite{duggan2014incremental, duggan2015skew}, and sparsity-aware recursive mapping as used in deduplication for the partitioning of band-join~\cite{li2020near}, are all orthogonal to our work, and can be incorporated to our proposed framework.
}}

\subsection{Problem Formulation}
\label{sec:formulation}

We first give an overview of the \textit{Lachesis}' workflow. When a data is going to be written to storage, \textit{Lachesis} will recommend a candidate set of IR fragments based on historical producer-consumer patterns. There are three situations: {\color{red}{(1). If a dataset is created by a producer job that has no historical consumers, no partitioner candidates can be identified, and thus the dataset will be partitioned using a default policy such as the round-robin policy~\cite{gupta2011gpfs}. (2). If the producer has one or more historical consumers, then, one or more partitioner candidates may be identified. Then, \textit{Lachesis} evaluates all candidates and selects one using a deep reinforcement learning approach~\cite{sutton1998reinforcement}. (3). If an existing dataset is identified to have bad organizations using certain external algorithms~\cite{idreos2007database}, \textit{Lachesis} can be applied to identify the optimal partitioner candidate for reorganizing the dataset.}}  \eat{The selected partitioner is applied to data when it is being materialized. Then, when a join is about to be performed on the data, the optimizer first matches the application IR with input datasets' partitioner, and then decides whether such partitioning can be utilized for physical optimization, e.g. to perform a local join without shuffling data. }

\textit{Lachesis} focuses on two processes: (1) when a dataset is going to be stored or reorganized, the system attempts to automatically enumerate, select, and create the optimal partitioning; (2) for running applications, the system attempts to match, recognize, and utilize existing partitionings to avoid unnecessary shuffling of data. In this section, we formalize the representation of IR and partitionings in UDF-centric workflows, as well as the two processes.

\subsubsection{IR and Partitioner Candidates}
\label{sec:partitionings}
\eat{
A UDF-centric analytics application is usually coded in an object-oriented language such as Python, Scala, C++, which has embedded a DSL (e.g. PlinyCompute's lambda calculus) or a library of APIs (e.g. Spark RDD/DataFrame/DataSet APIs). Then after compilation or interpretation, it returns an IR that is a directed acyclic graph (DAG).  The IR later goes through multiple optimization passes to generate high performance code for executing the workload.}

In this work, we define that for any workload $w$, there exists a mapping $h$ that transforms $w$ into an IR graph $a=h(w)=(V, E, S, O)$. Each node ($v \in V$) represents an atomic computation. This set of atomic computations varies with IR designs, but usually contains three categories of operators:

\noindent
(1) \textit{Lambda abstraction functions} such as a function that returns a literal (a constant numerical value or string), a member attribute or a member function from an object; unary functions such as \texttt{exp}, \texttt{log}, \texttt{sqrt}, \texttt{sin}, \texttt{cos}, \texttt{tan}, etc; or opaque unary functions if the programmer prefers not to expose the logic, such as \texttt{classify()}, \texttt{parse()}, \texttt{schema\_resolve()}, etc. 

\noindent
(2) \textit{Higher-order lambda composition functions} such as binary operators: \texttt{\&\&}, \texttt{||}, \texttt{\&}, \texttt{|}, \texttt{<},\texttt{>}, \texttt{==}, \texttt{+}, \texttt{-}, \texttt{*}, \texttt{/}, \texttt{pair}, conditional operator like \texttt{switch? on\_true:on\_false}; etc.

\noindent
(3) \textit{Collection-based operators} such as \texttt{scan} and \texttt{write} that reads/writes a collection of objects; \texttt{apply} that applies a lambda calculus expression (i.e., composed of lambda abstractions and higher order composition functions) to a collection of objects (like \textit{map}); and \texttt{join}, \texttt{aggregate/fold}, \texttt{flatten}, \texttt{filter}, etc. 

\noindent
Each edge ($e \in E$) represents a data flow or a control flow from the source node to the destination node, as mentioned. $S \subset V$ is the set of all \texttt{scan} nodes. $O \subset V$ is a set of \texttt{write} nodes. For example, the IR derived from Listing.~\ref{code1} is illustrated in Fig.~\ref{fig:graph-ir-example}.

As illustrated in the partitioner matching part of Fig.~\ref{fig:3steps}, in relational partitioning problems, a partitioner is simply a set of attributes of a relation that can be easily matched to the WHERE clause of a join query. But in UDF-centric workflows, a partitioner candidate is implicitly specified in UDFs. For example, the partitioner candidate defined in Fig.~\ref{fig:comments-partitioner-candidate} is implicitly specified in the \texttt{join\_selection()} function of Listing.~\ref{code1}.

A partitioner candidate is represented as a two-terminal directed acyclic graph (DAG) that has one root source node that has no parents (e.g., the \texttt{scan} node associated with the dataset to be partitioned) and one leaf target node that has no children (e.g., the \texttt{switch} node in Fig.~\ref{fig:comments-partitioner-candidate}). The two-terminal graph that represents the partitioner candidate must be a subgraph in a historical consumer workload's IR graph. The leaf target node in  the subgraph must connect to a \texttt{pair} node for \texttt{join} in the parent graph. Given a \texttt{scan} node $s_D \in S$ that reads from the dataset $\mathcal{D}$, we can enumerate the partitioner candidates of $\mathcal{D}$ as a set of subgraphs of IR graph $a$, denoted as $\mathcal{F}_{D}$. Each $f_k = (V_k, E_k, S_k,O_k) \in \mathcal{F}_{D}$  satisfies $V_k \subset V, E_k \subset E, S_k = \{s_D\}$, and $\norm{O_k}=1$. 

\vspace{-10pt}
\begin{figure} [H]
\centering
   \includegraphics[width=3.5in]{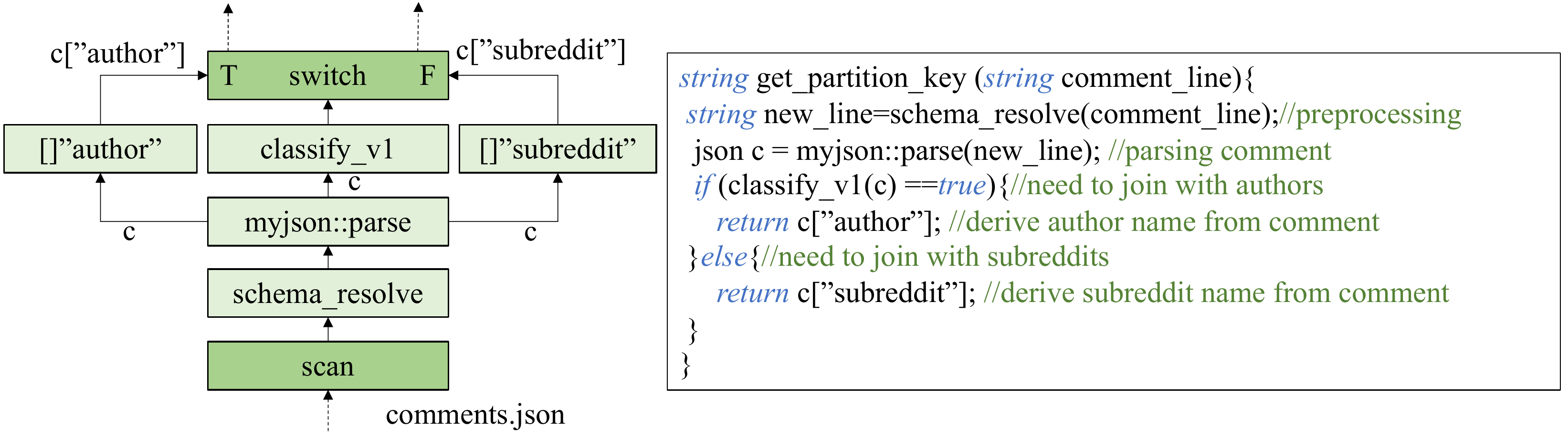}
\caption{\label{fig:comments-partitioner-candidate}
{\color{red}{\small The 2-terminal graph that represents the partitioner candidate of the comments dataset extracted from Fig.~\ref{fig:graph-ir-example}}}
}
\end{figure}

In the next two sections, we will formalize the partitioning creation and matching processes respectively.

\subsubsection{Process 1. Creation of Partitionings}
\label{sec:problem1}
We first present a high-level definition of the problem, as follows. A producing workload $p$ is going to write $\mathcal{D}$, which is a collection of $n$ objects $\mathcal{D}=\{d_i\}, (0\leq i < n)$, to a distributed storage $\mathcal {C}$ that consists of $m$ nodes, $\mathcal{C}=\{c_j\}, (0\leq j < m)$. The problem is first to find a horizontal partitioning $g: {\mathcal{D} \rightarrow \mathcal{C}}$,  so that the overall latency of the producer $p$ and consuming workloads of $\mathcal{D}$ is minimized, as denoted in Eq.~\ref{eq:cost}. The set of $l$ consuming workloads are represented as $\mathcal{W}=\{w_k\}, (0\leq k < l)$, $lat_p$ represents the latency of the producer,  and $freq_k$ and $lat_k$ denote the execution frequency and latency of $w_k$ respectively. Then the selected partitioning $g_{opt}$ needs to be automatically applied while storing $\mathcal{D}$ to the cluster $\mathcal{C}$.

\begin{equation}
\label{eq:cost}
g_{opt}=\arg \min_{g: {\mathcal{D} \rightarrow \mathcal{C}}}(lat_p +\sum_{\forall w_k \in \mathcal{W}}{({freq_k}\times lat_k)})
\end{equation}

We further formulate a more detailed model by lowering down the partitioning functions ($g$). There exist $m^n$ different partitioning functions, to prune which, we only consider well-known partition strategies such as hash partitioning, range partitioning, round robin partitioning, and random partitioning~\cite{zhou2010incorporating}. A hash partitioner is defined by a function $f_{keyProj}$ that extracts the partition key from a data item, where the key must have a hash function defined. For this type of partitioner, given $f_{keyProj}$, the corresponding $g$ is defined as $g_{hh}^{f_{keyProj}}$ $(d_i) = hash$ $(f_{keyProj}(d_i))\%m,$ $\forall d_i \in \mathcal{D}$. Range partitioners are similar, except that the partition key must have a comparator defined for sorting; and $g$ is accordingly defined as $g_{rn}^{f_{keyProj}}(d_i) = range(f_{keyProj}(d_i))\%m$. Round robin and random partitionings do not require any functions. The former is defined as $g_{rr}(d_i) = next\_int()\%m, \forall d_i \in \mathcal{D}$, and the latter is denoted as $g_{rm}(d_i) = random()\%m$. Therefore, given a set of $q$ different $f_{keyProj}$ for partitioning the dataset $D$, denoted as $\mathcal{F}=\{f_i\}, 0\leq i < q$, the search space includes all $2q$ combinations of the two partition strategies (i.e. hash or range) and the $q$ functions, plus the round robin and random strategies, represented as  $\mathcal{G}^\mathcal{F}=\{g_{hh}^{f_0}, ..., g_{hh}^{f_{q-1}}\} \cup \{g_{rn}^{f_0}, ..., g_{rn}^{f_{q-1}}\} \cup \{g_{rr}, g_{rm}\}$. Thus Eq.~\ref{eq:cost} can be lowered into Eq.~\ref{eq:cost1}:
\begin{equation}
\label{eq:cost1}
g_{opt} = \arg \min_{g \in \mathcal{G}^\mathcal{F}}(lat_p +\sum_{\forall w_k \in \mathcal{W}}{({freq_k}\times{lat_k})})
\end{equation}

{\color{red}{
Latency is influenced by numerous factors such as CPU costs, I/O costs, hardware parallelism, memory size, network bandwidth. Because of the lack of a widely accepted cost model for UDF-centric analytics, instead of detailing all of these factors, we choose to use a DRL approach to optimize the objective purely based on the observed latency (used to compute reward) and features regarding each partitioner candidate as well as the data and the environment, which we will describe in detail in Sec.~\ref{sec:rl}.
}}

\subsubsection{Process 2. Match of Partitionings}
\label{sec:problem2}

{\color{red}{
Supposing the comments dataset is partitioned by the candidate as illustrated in Fig.~\ref{fig:comments-partitioner-candidate}, when the \texttt{Reddit-Feature-Extractor} (Fig.~\ref{fig:graph-ir-example}) runs to process it, the system should recognize that the partitioner assoicated with the data is a desired partitioning of this consuming workload. Furthermore, if  the subreddits and authors datasets are also co-partitioned for the workload, the query optimizer will schedule local joins to avoid the shuffling stage.}}

A partitioner candidate, except for the random or the round robin partitioners, is a pair of $f_{keyProj}$ and its partition strategy (hash or range). Supposing a partitioner candidate, with its IR represented as $f_D=(V_D, E_D, S_D, O_D)$, has been applied to a dataset $\mathcal{D}$.\eat{ ($f_D$ not only specifies the IR graph of $f_{keyProj}$, but also specifies the partition strategy through the label of the leaf \texttt{partition} node.)} Then if an application $w \in \mathcal{W}$ reads from $\mathcal{D}$ and we have $a=h(w)=(V, E, S, O)$ as the IR graph of $w$, there must exist a \texttt{scan} node $s_D \in S$ that reads from $\mathcal{D}$, denoted as $s_D = a.find\_scanner(\mathcal{D})$. In addition, if there exists a subgraph of $a$, which is equivalent to $f_D$, the system's query scheduler can simply avoid the execution of this subgraph, because the partitioning represented by this subgraph has already been applied to $\mathcal{D}$. The identification of such subgraphs can be abstracted into a subgraph isomorphism problem~\cite{cook1971complexity} : \textit{Given two graphs $f_D=(V_D, E_D, S_D, O_D)$ and $a=(V, E, S, O)$, a subgraph isomorphism from $f_D$ to $a$ is to find a function $f: V_D \rightarrow V$ such that if $(u, v) \in E_D$, then $(f(u), f(v)) \in E$ and if $s \in S_D$, then $f(s) \in S$. }

\vspace{-10pt}
\subsection{Summary of Challenges}
 \eat{we are not given the set of workloads $\mathcal{W}$. In addition,} We are focusing on UDF-centric applications, which is very different with relational applications where SQL queries are easy to reason about, and a partitioning predicate is easily to be extracted (i.e. searching in WHERE clause) and reused (i.e. appending a PARTITION BY predicate). The specific challenges include:

\noindent
\textbf{(1) Workload Enumeration.} Given an incoming/existing dataset $\mathcal{D}$, how to obtain the set of consuming workloads $\mathcal{W}$?

\noindent
\textbf{(2) Enumeration of Partitioner Candidates.} How to obtain the set of partitioner candidates $\mathcal{F}$ for storing/reorganizing a dataset $\mathcal{D}$? 

\noindent
\textbf{(3) Optimization.} How to solve the optimization problem illustrated in Eq.~\ref{eq:cost1}, with the lack of a widely-accepted cost model?

\noindent
\textbf{(4) Match of Partitionings.} How to efficiently solve the subgraph isomorphism problem, which is NP-complete~\cite{cook1971complexity}?

\noindent
These challenges are addressed in \textit{Lachesis} based on following ideas: 

\noindent
(1) We utilize historical workflow execution information to predict future workloads based on the workload recurrence patterns. 

\noindent
(2) Partitioner candidates or existing/desired partitionings are just a special type of subgraphs. Recognizing such subgraphs in an IR may be simpler than the general subgraph isomorphism problem. 

\noindent
(3) A DRL-based approach that models the dynamic factors purely through rewards of past decisions may solve the optimization problem with good adaptivity and also avoid the costs of profiling the hardware environments as required in a cost model approach.

\section{Our Solutions}
\subsection{Creation of Partitionings}
\subsubsection{Workload Enumeration}
\label{sec:workflow}
Given a producer $p$ that is going to write a dataset $\mathcal{D}$ to the storage, and a set of $nw$ historical workloads $\mathcal{W'}=\{w'_i\}, (0\leq i<nw)$, how to enumerate the set of workloads $\mathcal{W}$ that may process $\mathcal{D}$ in the future?

Based on the \textit{recurrent workflow assumption}, if there exists $w'_i \in  \mathcal{W}'$ with $h(w'_i)=(V'_i, E'_i, S'_i, O'_i)$ that is isomorphic to $p$ with $h(p)=(V,E,S,O)$, which means an isomorphism bijection $f: h(p) \rightarrow h(w'_i)$ exists, then for $o_D \in V$ that is the node outputting $\mathcal{D}$ in $h(p)$, there must exist $o_D' \in O'_i$ so that $f(o_D) = o_D'$. Further more, if $\exists$ $w'_j \in \mathcal{W'}$ with $h(w'_j)=(V'_j, E'_j, S'_j, O'_j)$, $s_D \in S'_j$, satisfying that the dataset read by $s_D$ is created by $o_D'$, we can conclude that $w'_j$ once consumed the output of $w'_i$, so it may consume the output of $p$ in the future (because of the isomorphism between $h(p)$ and $h(w'_i)$), and thus we have $w'_j \in \mathcal{W}$. 

We encapsulate the above process into a historical workflow analysis component. It first reconstructs low-level workflow information from execution logs, which is illustrated in Fig.~\ref{fig:lowlevelgraph}, where each node represents an execution of a workload, identified by (app\_id, timestamp) and each edge represents a historical dataset created by its source node, and consumed by its destination node. It then further condenses the low-level graph into a skeleton graph~\cite{rodrigues2006supergraph, wang2015schema} by merging nodes that have the same IRs and thus expect exactly the same partitionings, as illustrated in Fig.~\ref{fig:supergraph}. In the skeleton graph, each edge represents a list of historical execution runs in the form of (app\_id, timestamp, input\_data\_id, output\_data\_id). Given a currently running application belonging to group$1$ that is going to write a dataset to the storage, based on the skeleton graph in Fig.~\ref{fig:supergraph}, \textit{Lachesis} will predict that applications from group$2$ and group$4$ may process the dataset in the future. The matching of $h(w'_i)$ to $h(p)$ is achieved by offline computing a hash signature for each workload's IR DAG graph ($h(w'_i)$) through enumerating, sorting, and concatenating all distinct paths that connect a \texttt{scan} node to a \texttt{write} node~\cite{valiant1979complexity}. These signatures are stored into a hash table and then the lightweight online process matches the signature of the producer's IR graph ($h(p)$) against the hash table.

\subsubsection{Partitioner Candidate Enumeration}
\label{sec:enumeration}
Given a set of consuming workloads $\mathcal{W}$ enumerated for  $\mathcal{D}$, $\forall w_i \in \mathcal{W} (0\leq i <nw)$, we can further enumerate a set of partitioner candidates, with each being a subgraph of $a_i=h(w_i)=(V, E, S, O)$. 

\begin{figure*}
    \centering
    \begin{minipage}{0.2\textwidth}
      \centering
      \subfigure[Low-level Graph ]{%
      \label{fig:lowlevelgraph}
      \includegraphics[width=1.2in]{low-level-graph.pdf}  
      }%
      \vspace{0pt}
      \subfigure[Skeleton Graph]{%
      \label{fig:supergraph}
      \includegraphics[width=1.2in]{supergraph.pdf}
      }
      \caption{\label{fig:workflow-graphs} \small
          Workflow representation (each node is a workload IR graph) 
      }
    \end{minipage}
    %\hspace{10pt}
    \begin{minipage}{0.75\textwidth}
        \centering
        \includegraphics[width=5in]{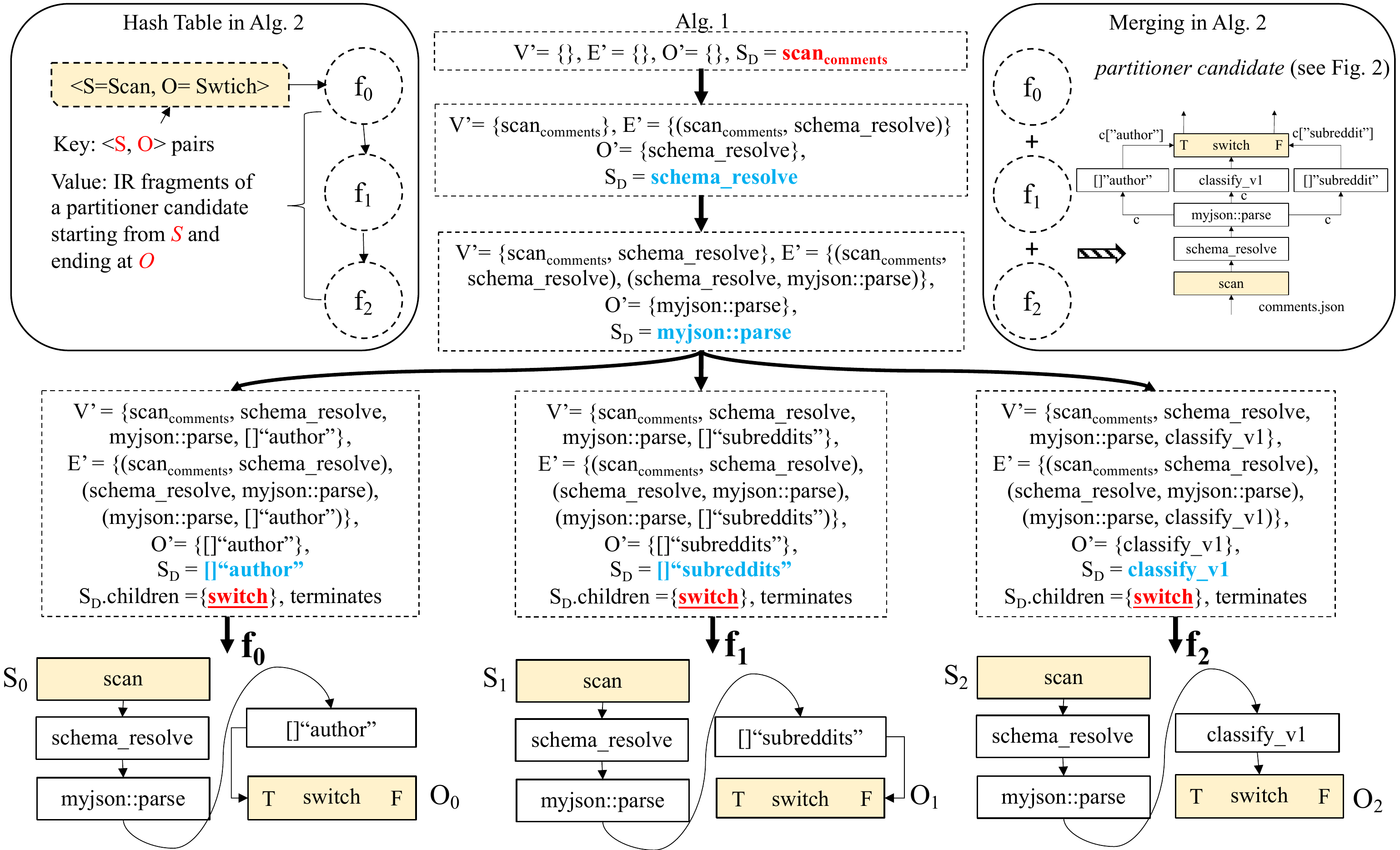}
        \caption  {\small {\color{red}{Running example for Alg. 1 and Alg. 2}}}
        \label{fig:alg1}
    \end{minipage}%
\end{figure*}

As mentioned in  Sec.~\ref{sec:partitionings}, the subgraph representing a partitioner candidate must satisfy that the sole root node is a \texttt{scan} node ($s_D \in S$) that reads from $\mathcal{D}$, and the unique leaf node is a node that connects to a  \texttt{pair} node  followed by a \texttt{join} node. \eat{Here, a source node is a special type of node that has one or more children nodes but no parent nodes and a target node is another special type of node that has one or more parent nodes but no children node.}  Obviously satisfying these conditions makes the subgraph sufficient to serve as a partitioner candidate. To efficiently identify such subgraphs, we propose a two step approach, as illustrated in Fig.~\ref{fig:alg1}. The first step is to recursively traverse $a_i$ and enumerate all distinct paths that start at the \texttt{scan} node $s_D$ and end at any of the \texttt{pair}$\rightarrow$\texttt{join} paths. We formalize this process in Alg.~\ref{alg:search-partitioner-candidate}. The second step is to merge all paths that connect the same \texttt{scan} node and the same \texttt{leaf} node into one graph to serve as one partitioner candidate, as illustrated in Alg.~\ref{alg:merge-partitioner-candidate}. Thus we can formalize the process of enumerating all partitioner candidates from $\mathcal{W}$: $\hat{h}_{\mathcal{W} \rightarrow \mathcal{F}} = {\bigcup_{w_i \in W}} \{merge(search(h(w_i)), \mathcal{D}, \emptyset)\}$.

\eat{
\begin{figure*} 
\centering
   \includegraphics[width=6in]{Alg1&Alg2.pdf}
\caption{\label{fig:alg1&2}
{\color{red}{Running example for Alg. 1 and Alg.2.}}
}
\end{figure*}

\begin{figure}[H]
\centering\subfigure[Low-level Graph ]{%
   \label{fig:lowlevelgraph}
   \includegraphics[width=1.5in]{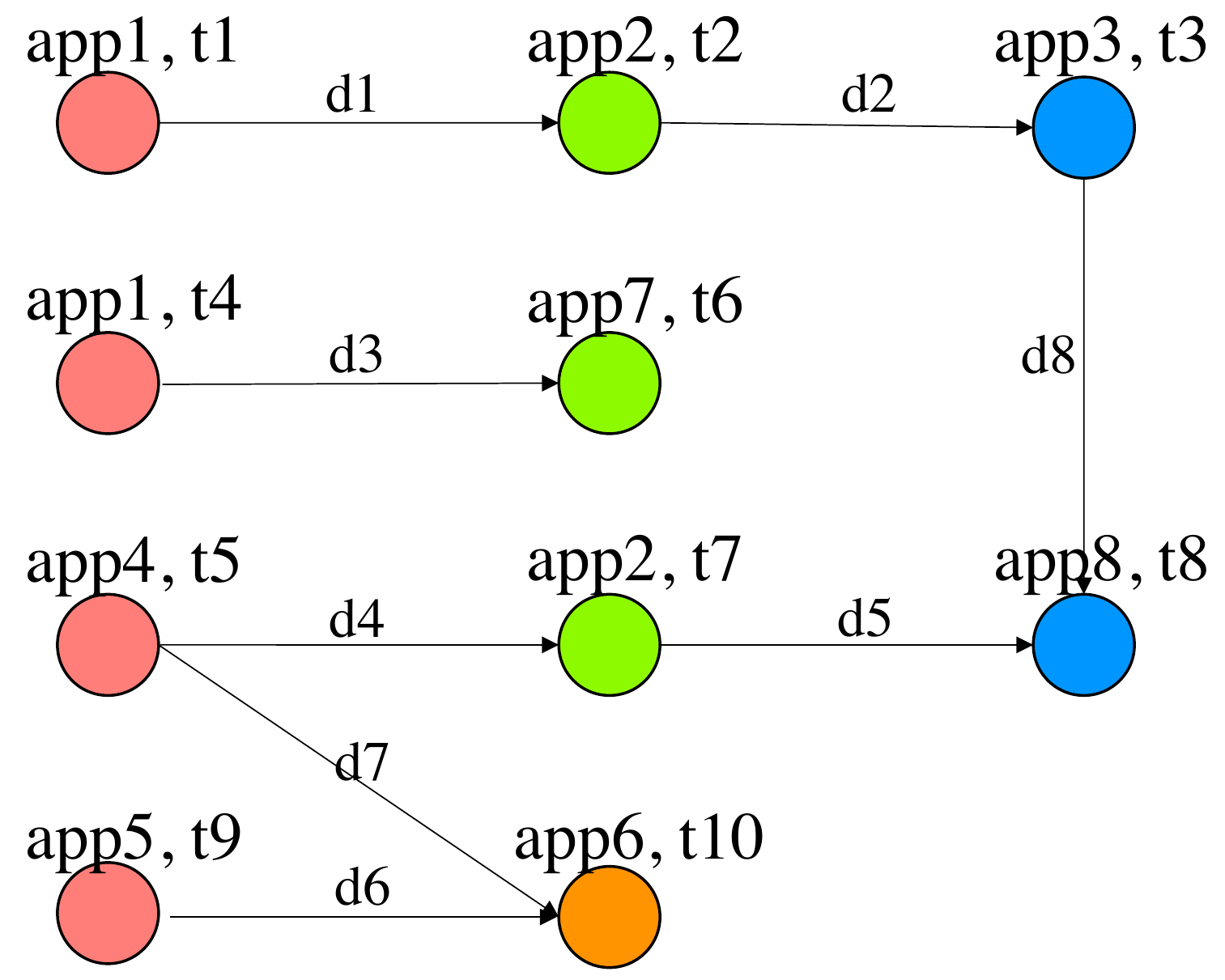}  
}%
\subfigure[Skeleton Graph]{%
  \label{fig:supergraph}
  \includegraphics[width=1.5in]{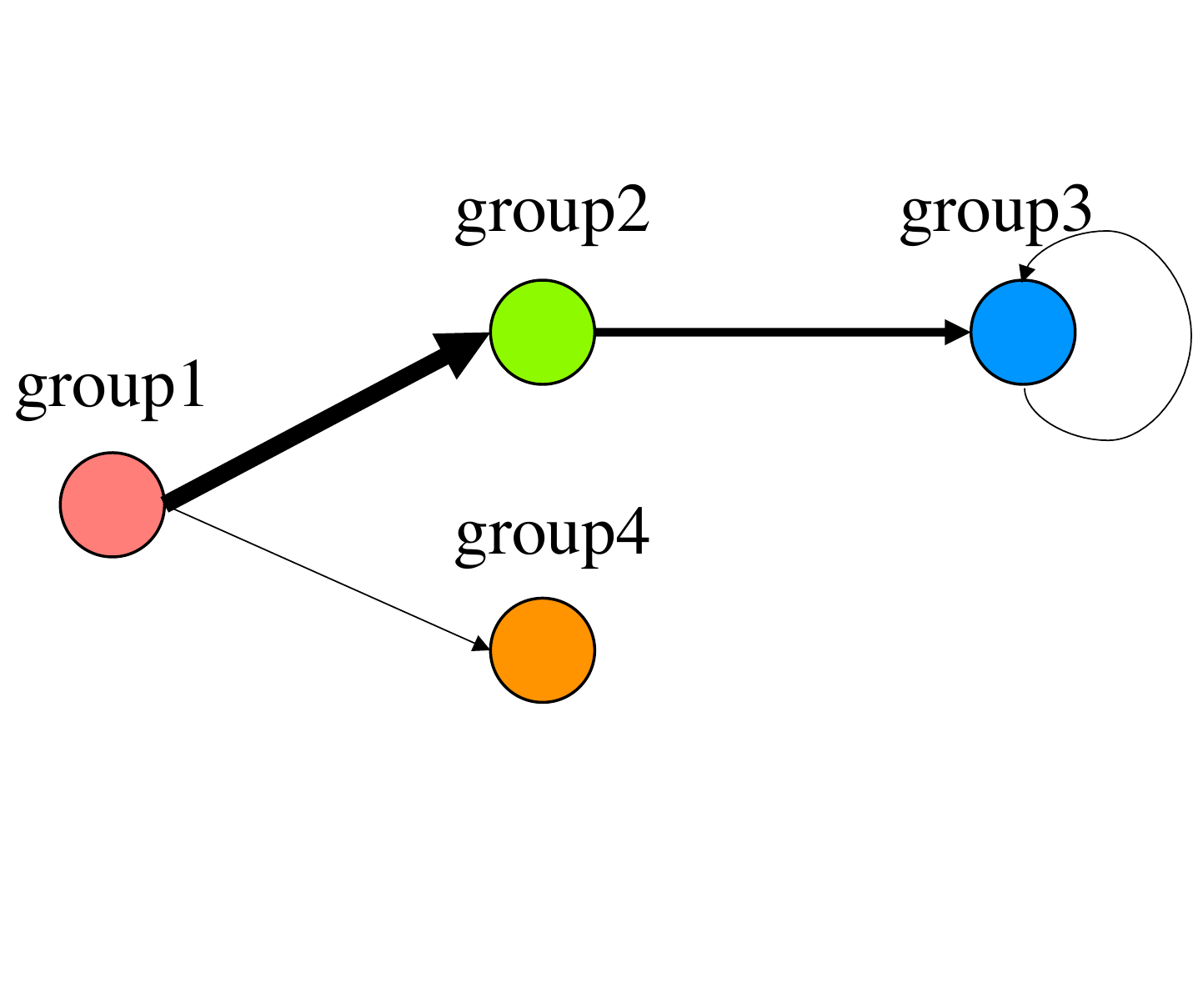}
}

\caption{\label{fig:workflow-graphs}
Workflow representation (each node in both graphs can be further represented as IR graph). 
}
\end{figure}}

\eat{
\begin{figure}
\centering\subfigure[Hash Table]{%
   \label{fig:hashtable}
   \includegraphics[width=3.5in]{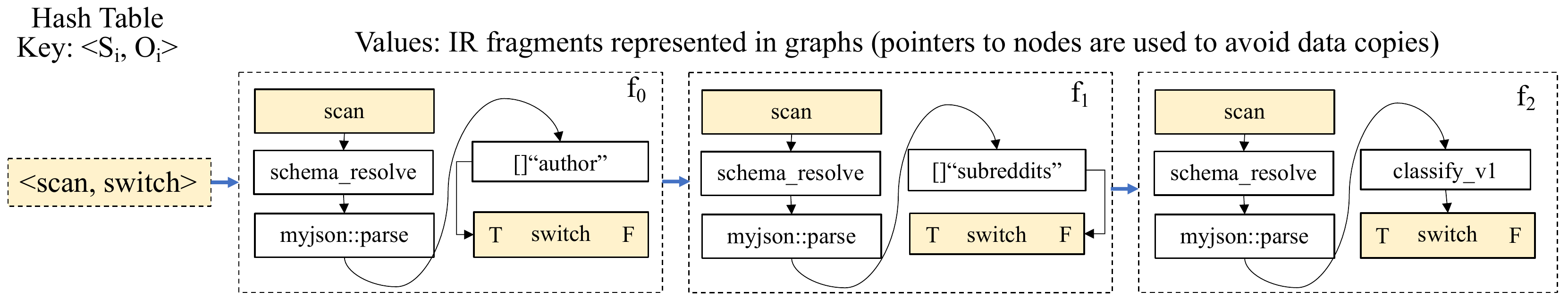}  
}%
\vspace{1pt}
\subfigure[Merged Partitioner Candidate]{%
  \label{fig:comments-partitioner-candidate}
  \includegraphics[width=3.5in]{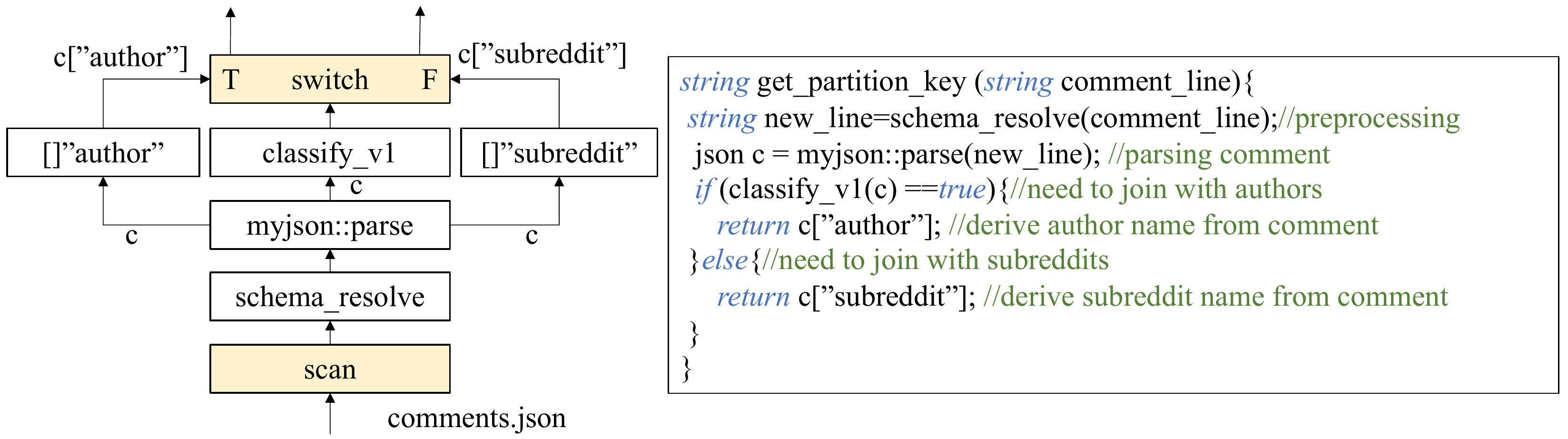}
}

\caption{\label{fig:alg2}
Running example for Alg 2: IR fragments that have the same key will be merged into a partitioner candidate. 
}
\end{figure}
}

\begin{algorithm}[!ht]\small
\caption{\bf $search(a_i, s_D, F_i)$}
\label{alg:search-partitioner-candidate}
\begin{algorithmic}[1]
\STATE INPUT1: $a_i = (V, E, S, O)$ (the IR graph of one of $\mathcal{D}$'s consuming workloads $w_i \in \mathcal{W}$)  \STATE INPUT2: $s_D$ (the \texttt{scan} node in $a_i$ that reads from $\mathcal{D}$)
\STATE INPUT3 and OUTPUT: $F_i$ (a list of partial partitioner candidates for $\mathcal{D}$ extracted from $a$)
\STATE $F_i \leftarrow \phi$
\STATE $V' \leftarrow \{s_D\}$;  $E' \leftarrow \emptyset$; $S' \leftarrow \{s_D\}$; $O' \leftarrow \emptyset$
\FOR{$v_k$ in $s_D.children$}
  \STATE $V'\leftarrow V' \cup \{v_k\}$
  \STATE $E'\leftarrow E' \cup \{edge(s_D, v_k)\}$
  \STATE $O' \leftarrow \{v_k\}$
  \IF {$\nexists path (v_k, pair, join)$}
    \STATE $F^t\leftarrow \emptyset$
    \STATE $search((V-V')\cup\{v_k\},E-E',(S-S')\cup\{v_k\},O'), v_k, F^t)$
    \FOR{$f^t=(V^t, E^t, S^t, O^t) \in F^t$}
      \STATE $F_i \leftarrow F_i \cup \{(V' \cup V^t, E' \cup E^t ), S', O^t\}$
    \ENDFOR
  \ELSE
    \IF {$E' \neq \emptyset$}
      \STATE $F_i \leftarrow F_i \cup \{(V', E', S', O')\}$
    \ENDIF
  \ENDIF
\ENDFOR
\RETURN $F_i$
%\EndProcedure
\end{algorithmic}
\end{algorithm}

\begin{algorithm}[!ht]\small
\caption{\bf $merge(F_i)$}
\label{alg:merge-partitioner-candidate}
\begin{algorithmic}[1]
\STATE INPUT: $F_i$ (a list of partial partitioner candidates output from Alg.~\ref{alg:search-partitioner-candidate})
\STATE OUTPUT: $F'_i$ (a list of partitioner candidates)
\STATE $hashmap \leftarrow \emptyset$
\STATE $F'_i \leftarrow \emptyset$ 
\FOR{$f_k=(V_k, E_k, S_k, O_k) \in F_i$}
   \IF{$hashmap.count((S_k, O_k)) \neq 0$}
      \STATE $(V^t, E^t, S^t, O^t) \leftarrow hashmap[(S_k, O_k)]$
      \STATE $hashmap[(S_k, O_k)] \leftarrow (V^t \cup V_k, E^t \cup E_k, S_k, O_k)$
      
   \ELSE
      \STATE $hashmap[(S_k, O_k)] \leftarrow f_k$
   \ENDIF
\ENDFOR
\FOR {$((S_k, O_k), f_k) \in hashmap$}
    \STATE $F'_i = F'_i \cup \{f_k\}$
\ENDFOR
\RETURN $F'_i$
\end{algorithmic}
\end{algorithm}

\subsubsection{DRL-based Optimization} 
\label{sec:rl}
Once a set of partitioner candidates are enumerated, the next step is to select the optimal one to apply. There are existing works targeting similar data partitioning optimization problems in OLAP relational databases~\cite{rao2002automating, agrawal2004integrating, nehme2011automated, shaikhha2018building, hilprecht2020learning, hilprecht2019learning, eadon2008supporting,zamanian2015locality,lu2017adaptdb}. These works, including recent RL-based partitioning advisors~\cite{hilprecht2020learning, hilprecht2019learning}, are largely depending on the functional dependency~\cite{ramakrishnan2000database} and cost models of relational databases, which may not exist in   UDF-centric analytics. \eat{In addition, the optimization process depends on a number of hard-to-predict factors, such as  workload recurrence patterns, CPU/IO speed in different environments, etc. We solve the optimization problem  using a DRL approach.}
{\color{red}In this work, we choose a DRL approach based on the actor-critic network~\cite{silver2016mastering}, which integrates the best of both worlds of value-based (e.g. Q-Learning~\cite{gu2016continuous,watkins1992q})}
{\color{red}  and policy-based (e.g. proximal policy optimization~\cite{schulman2017proximal})RL.
We use A3C algorithm~\cite{mnih2016asynchronous}, which is a state-of-the-art
algorithm for learning the actor-critic network, and widely adopted in various domains~\cite{mao2017neural}. It also allows to use multiple learning agents to accelerate the training process.}

The actor-critic network is based on policy gradient. It takes a \textit{state} vector, which describes the environment, as
input, and outputs \textit{policy}, which is represented as the probability distribution in the \textit{action}
space. Then, the critic network also takes \textit{state} as input,
and outputs the expectation of value function that will be used
together with \textit{reward} to compute the policy gradient to improve
the learning for both of the actor and critic networks.

The optimization goal of the model is to minimize the cumulative
processing latency of current and future applications. We formulate the DRL problem in detail as follows.

\eat{
\begin{figure}[H]
\centering
\includegraphics[width=3.4in]{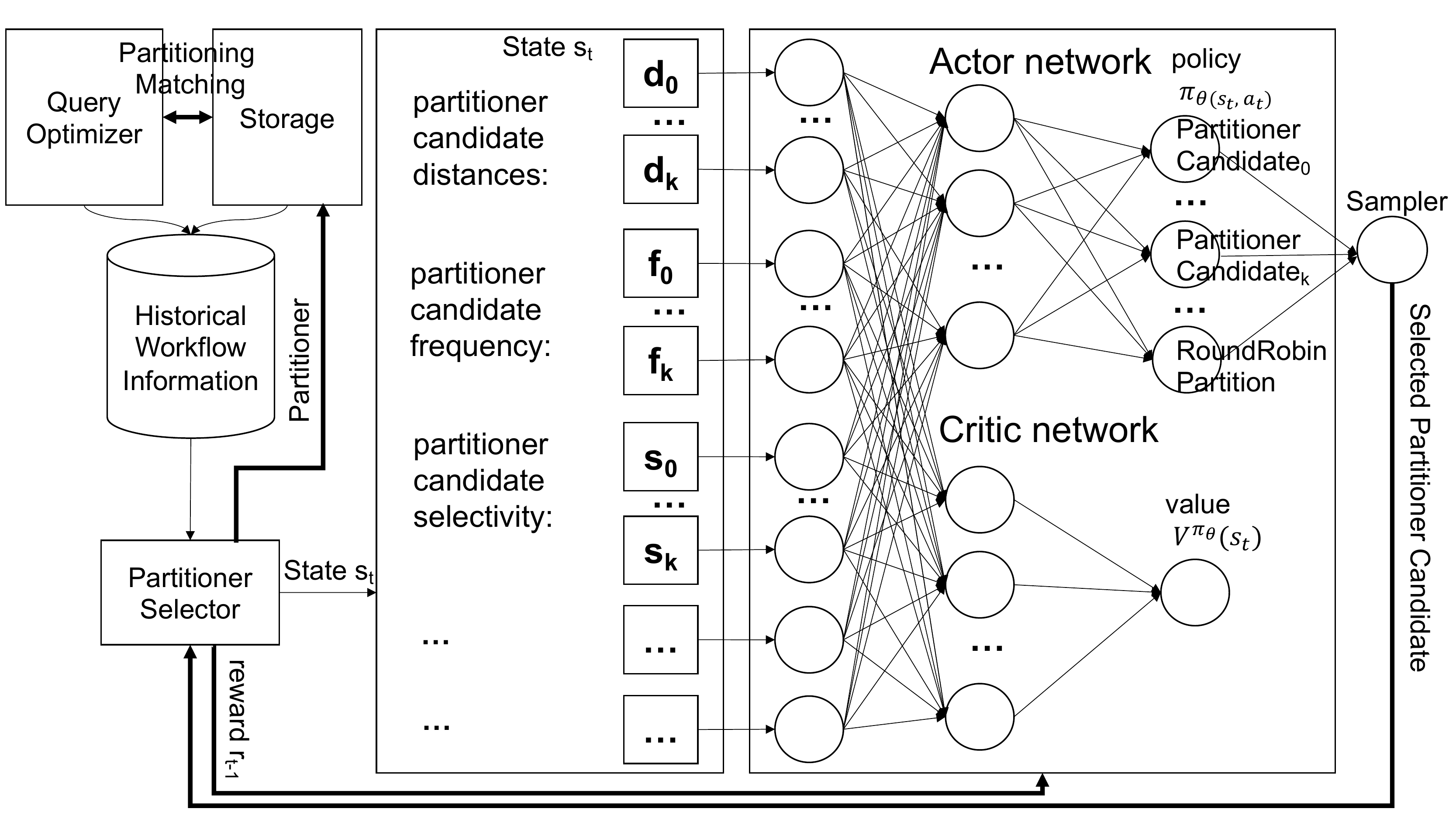}
\vspace{-1.0em}
\caption{ \small Overview of \textit{Lachesis}'s RL approach.}
\vspace{0.5em}
\label{fig:input-features}
\end{figure}}

\noindent
\textbf{State.} To formualte the \textit{state} feature vector, we first consider following features for each of the $k$-most recent partitioner candidates:

\noindent
1. \texttt{frequency} indicates the total number of historical executions of the IR where the partitioner candidate is extracted from. 

\noindent
2. \texttt{distance} indicates the average time interval between the most recent two runs in the candidate's IR group mentioned in Fig.~\ref{fig:supergraph}. 

\noindent
3. \texttt{recency} indicates the timestamp of the most recent run of applications in the candidate's IR group. 

\noindent
4. \texttt{complexity} computes the number of nodes at the longest path from the root node  to the leaf node  in the subgraph that represents the partitioner candidate. \eat{Each weight estimates the time complexity of the partitioner candidate based on scaling historical measurements with corresponding input data sizes~\cite{shi2014mrtuner}.}

\noindent
5. \texttt{selectivity} indicates the ratio of the average size of the partition keys extracted to the average size of source objects. This metric measures the amount of data that should be shuffled at runtime if this partitioner candidate is desired but not selected.

\noindent
6. \texttt{key\_distribution} indicates the average number of unique values
  generated by hashing the output of the partitioner candidates in historical runs. The key distribution affects the system load balance. If the output keys are skewed, most of the objects may be stored on the same worker instance, while only a small portion of objects are distributed in other workers.
  
\noindent
7. \texttt{num\_copartitioned} indicates the number of existing datasets that will be co-partitioned with the data if this partitioner candidate is selected. These datasets are identified by searching their partitionings in the IR where this partitioner candidate is extracted from. 

\noindent
8. \texttt{size\_copartitioned} indicates the total sizes of the  co-partitioned datasets mentioned above.

{\color{red}{
We compute the pearson correlation coefficient (PCC) ~\cite{lee1988thirteen}, which is a measure of the linear correlation coefficient between two random variables, for the reward (which we will describe later) and each of aforementioned features. The results show that \eat{when queries repeat for just a few times and without temporal locality,} \texttt{frequency}, \texttt{num\_copartitioned}, \texttt{size\_copartitioned} are the top three features that are mostly correlated with the reward. In addition, the PCC of \texttt{recency} and \texttt{distance} to the reward will increase with the temporal locality of the workload patterns. While the rest of the features have significantly less PCC with reward, they are also useful in avoiding some bad partitioner candidates, e.g., an aggregation/join key extraction function that maps all elements into a few keys with skewed distribution by using \texttt{key\_distribution}.
}}

\eat{If round robin partitionings are selected as candidates, their \texttt{complexity} is defined as $0$, \texttt{selectivity} is defined as $1$, and \texttt{key\_distribution} is the average number of data elements in historical runs.}

\vspace{-10pt}
Besides the features that describe each of top $k$ partitioner candidates, the other features we use include the estimated size of the dataset to be dispatched, the number of workers, the number of cores and sizes of available memory and disk space on each worker. All features will be normalized before being used. 

\eat{
{\color{red}{
The first three features (\texttt{frequency}, \texttt{distance}, and \texttt{rece- ncy}) measure the probability of the recurrence of the applications that expect the partitioner candidate. The other features (\texttt{complexity}, \texttt{selectivity}, and \texttt{key\_distribution}) measure the performance overhead of applying the partitioner candidate. 

One problem we meet is that the number of partitioner candidates for each dataset could vary. A similar problem is that when we try to include the sizes and partitioning scheme of each existing dataset as features, we find the number of datasets could vary significantly along the time. It is hard to encode such dynamic information into a fixed-length feature vector. To solve the problem, instead of considering all partitioner candidates, we select the top $k$ partitioner candidates using a simple score function that considers the partitionings of existing datasets as illustrated in Eq.~\ref{}.
}}
}
\eat{It is obvious that the modeling of the overall partitioning costs and benefits are not only non-linear with these factors, but also depending on a lot of dynamic environmental factors, e.g. relative CPU speed and I/O speed, number of nodes in the cluster, evolution in workloads and workflows, and so on. While it is difficult to profile and model all of these dynamics, we choose to use a deep reinforcement learning approach for selecting the optimal partitioner candidate.}

\noindent
\textbf{Action Space and Policy.} Upon receiving the \textit{state} vector $s_t$, the RL agent needs
to send back an action $a_t$ that corresponds to the selected partitioner candidate. The RL agent selects actions based
on a policy, defined as a probability distribution over candidate
lambdas: $\pi = \pi_{\theta}\{s_t, a_t\} \rightarrow [0, 1]$. Here
$\theta$ is the hidden parameter that controls the policy, which is
represented by the actor neural network~\cite{mnih2016asynchronous}. {\color{red}{The action space can be extended to sample and select more than one actions, for creating multiple replicas, with each organized using a different partitioning~\cite{zou2020architecture, zou2019pangea}. In addition,  it can be extended to enable replicating a dataset to partial or all cluster nodes by adding a new dimension that represents the replication factor.}}

\noindent
\textbf{Reward Function.} \textit{Lachesis} also
needs to compute reward $r_{t-1}$ for last action $a_{t-1}$. Because latency will increase with data size, we define the
reward function to be the performance speedup of the
total throughput of applications that consume the dataset for which action $a_{t-1}$ is
applied, compared to a baseline throughput. The baseline is the average throughput of the historical executions of these applications. The reward function is formalized as below.  $W_{t-1}^{t}$ represents all workloads that have processed the dataset partitioned at time $t-1$, during the period from time $t-1$ to $t$. $W'$ represents all historical workloads used for workload enumeration.

$r_{t-1} =\frac{\sum_{w \in W_{t-1}^{t}}{\sum_{\mathcal{D} \in w.input}{size(\mathcal{D})}}/\sum_{w \in W_{t-1}^{t}}{latency(w)}}{\sum_{w' \in W'}{\sum_{\mathcal{D} \in w'.input}{size(\mathcal{D})}}/\sum_{w' \in W'}{latency(w')}}$

\noindent
\textbf{Policy Gradient.} 
Policy gradient methods estimate
the gradient of the expected total reward by computing the gradient of
cumulative
discounted reward with respect to
the policy, which can be represented as ~\cite{mnih2016asynchronous}: 

$\nabla_{\theta} E[\sum_{t\geq0}\gamma^t r_t | \pi_{\theta}] =
E[\nabla_{\theta}log\pi_{\theta}(s, a) A^{\theta}(s, a)| \pi_{\theta}] $

$A^{\theta}(s, a)$ is called advantage function that indicates how
much better an action is compared to the expected. 
Each update of the \textit {actor network} follows the policy gradient to
reinforce actions that lead to better rewards:

\eat{θ←θ+α ∇θlogπθ(st,at)A(st,at)+β∇θH(πθ(·|st)),}

$\theta \leftarrow \theta + \alpha\nabla_{\theta}log\pi_{\theta}(s_t,
a_t)A(s_t, a_t) + \beta\nabla_{\theta}H(\cdot|s_t)$.

Here,  $\alpha$ is the learning rate; $H(\cdot)$ is the entropy of the
policy, which is to encourage exploration in the action space; and
$\beta$ is used to control the emphasis in exploration over exploitation.

To compute the advantage function $A(s_t, a_t)$, we need estimate
the value function $V^{\pi_{\theta}}(s)$
as $Q(s, a)$. The \textit{critic network} is responsible to
learn the estimate of the value function from observed rewards. All the details of derivation can be found in reference
\cite{mnih2016asynchronous}.

\noindent
\textbf{An End-to-End Algorithm.}
Based on the proposed techniques for workflow enumeration, partitioner candidates enumeration and optimization, we give Alg.~\ref{alg:end-to-end} to describe the end-to-end process for creating a partitioning.

\begin{algorithm}[!ht]\small
\caption{\bf $partitioning\_creation(p, \mathcal{D}, \mathcal{W'})$}
\label{alg:end-to-end}
\begin{algorithmic}[1]
\STATE INPUT1: $p$ (the producer workload) 
\STATE INPUT2: $\mathcal{D}$ (the dataset to store and partition) 
\STATE INPUT3: $\mathcal{W}'$ (the set of historical workloads)
\STATE $\mathcal{W} \leftarrow match(p, \mathcal{W}')$ \COMMENT{Sec.~\ref{sec:workflow}}
\STATE $\mathcal{F} \leftarrow \emptyset$
\FOR{$w_i \in \mathcal{W}$}
    \STATE $a_i \leftarrow h(w_i)$ \COMMENT{via DSL/IR: Sec.~\ref{sec:partitionings}}
    \STATE $s_D \leftarrow a_i.find\_scanner(\mathcal{D})$\COMMENT{Sec.~\ref{sec:problem2}}
    \STATE $F_i \leftarrow merge(search (a_i, s_D, \emptyset))$ \COMMENT{Sec.~\ref{sec:enumeration}: Alg.~\ref{alg:search-partitioner-candidate} and Alg.~\ref{alg:merge-partitioner-candidate}}
    \STATE $\mathcal{F} \leftarrow \mathcal{F} \cup F_i$
\ENDFOR
\STATE $g_{opt} \leftarrow \min_{g \in \mathcal{G}^\mathcal{F}}(lat_p+\sum_{\forall w_k \in \mathcal{W}}{(freq_k \times lat_k)})$\COMMENT{Sec.~\ref{sec:rl}}
\FOR{$\forall d_i \in \mathcal{D}$}
   \STATE store $d_i$ to the node $g_{opt}(d_i)$
\ENDFOR
\end{algorithmic}
\end{algorithm}

\vspace{-10pt}
\subsection{Matching of Partitionings}
\label{sec:matching}
In this section, we discuss how to  match the partitioning of an input dataset to the running application (Process 2). While subgraph isomorphism problem is NP-complete~\cite{cook1971complexity}, we can utilize the two-terminal DAG characteristics of the subgraph associated with a partitioner candidate \eat{(i.e., each of such subgraphs has only one root source node that can be a \textit{scanner} node and one leaf target node that must connect to a join \textit{pair} node)} to provide an efficient solution.

Given a dataset $\mathcal{D}$, we can obtain the IR graph of its partitioning through the storage interface, denoted as $f_D=(V_D, E_D, S_D=\{s\}, O_D=\{o\})$. Also given a running application $w$ that takes $\mathcal{D}$ as one of its inputs, we can obtain $w$'s IR graph $a = h(w)=(V, E, S, O)$. We first locate the scanner node $s_D \in S$ that connects to $\mathcal{D}$, and create an isomorphic mapping from $s$ to $s_D$, because these two are root nodes and they must match with each other due to the uniqueness of root node. Then we recursively visit each descendant (denoted as $v$) of $s_D$ in IR DAG $a$ (we use depth-first search for this step). Each time meeting a $v$ that is a \texttt{pair->join} path, we create a candidate isomorphic subgraph $IG^{(s_D,v)}$ that connects the root node $s_D$ and the leaf node $v$. This step can be accelerated by indexing all \texttt{join} nodes when constructing the IR graph. Then for each candidate isomorphic subgraph, we  create a signature for each distinct path from $s_D$ to $v$ by concatenating the node labels along the path. We can thus derive a unique signature to identify each subgraph by further concatenating all path signatures sorted in lexicographical order. By matching the signatures of the $f_D$ and each candidate subgraph, we can find all isomorphic subgraphs. The algorithm is illustrated in Alg.~\ref{alg:matching}.
It can be further optimized by using a hashmap to store the signatures of candidate subgraphs.  

\begin{algorithm}[!ht]\small
\caption{\bf $partitioning\_match(f_D, ssset_D, a, P)$}
\label{alg:matching}
\begin{algorithmic}[1]
\STATE INPUT1: $f_D=(V_D, E_D, S_D=\{s\}, O_D=\{o\})$ (IR of the dataset's partitioning) 
\STATE INPUT2: $ssset_D$ (sorted set of signatures for all paths in $f_D$)
\STATE INPUT3: $a = (V, E, S, O)$ (IR of the running consumer workload) 
\STATE INPUT4: $P=\{p_i\} \subset V$ (the set of \textit{pair->join} paths)
\STATE OUTPUT: $I$ (the set of subgraphs in $a$ that is isomorphic to $f_D$)
\STATE $I \leftarrow \phi$
\STATE $s_D = a.find\_scanner(\mathcal{D})$
\FOR{$p_i \in P$}
    \STATE $path\_set = a.find\_all\_paths(s_D, p_i)$
    \IF{$path\_set \neq \phi$}
        \STATE $sig\_set \leftarrow \phi$
        \FOR{$path \in path\_set$}
            \STATE $sig \leftarrow create\_signature(path)$
            \STATE $sig\_set \leftarrow sig\_set \cup \{sig\}$
        \ENDFOR
    \ENDIF
    \IF{$ssset_D$ equals to $sorted(sig\_set)$}
        \STATE $I \leftarrow I \cup \{IG^{(s_D, p_i)}\}$
    \ENDIF
\ENDFOR
\RETURN $I$
\end{algorithmic}
\end{algorithm}

\section{System Implementation}
\label{sec:training-design}

We implement \textit{Lachesis} on top of a baseline system,  PlinyCompute~\cite{zou2018plinycompute}, which is a UDF-centric analytics framework, using the Pangea storage~\cite{zou2019pangea}\footnote{\small The \textit{Lachesis} code is available: https://github.com/asu-cactus/lachesis}.
\textit{Lachesis} stores information regarding historical application executions, including the paths, sizes, partitionings of datasets, IRs, runtime statistics such as execution latency, output data sizes, for each job stage, in a SQLite database. The producer and consumer relationships among applications are reconstructed, and  re-executions of workloads are detected and labeled, which provides a full picture of historical workflow executions. Given a producer that materializes a dataset, the historical workflow analyzer can efficiently supply a set of applications that once have processed the datasets output from the same type of producers. Based on the \textit{recurrent workflow assumption}, each application in the set may rerun and process the dataset in the future. Therefore, any relevant partitioning computations extracted from these applications may be considered as a partitioner candidate for this dataset. The historical workflow analyzer is also responsible for selecting the top $k$ partitioner candidates and extracting features for DRL inferences. 

\noindent
\textbf{DRL Model Training.} 
\eat{
\textit{Lachesis} trains a deep reinforcement learning model (DRL) using the actor-critic network~\cite{sutton1998reinforcement} to serve as the partitioner selector. Each time a dataset is to be stored, the historical workflow analyzer recommends a set of partitioner candidates, and sends a state vector that describes various features for each of the candidates, to a TensorFlow-based DRL server. The DRL server predicts and samples a probability distribution over an action space that consists of all candidates to select a partitioner candidate, which is then applied to partition the dataset at storage time. The overall throughput variations for recent workloads are computed to serve as the reward. The DRL model is able to automatically adapt to changes in hardware environments, workload characteristics and data schemas by learning from the historical rewards. More details are described in Sec.~\ref{sec:rl}.}
The RL model is deployed using
TensorFlow. \eat{The Python server accepts the partitioner candidate features and
measured the performance benefits resulting from last decisions as \textit{state} and \textit{reward} respectively to feed into the neural network 
and sends back the optimal partitioner candidate as \textit{action}.} 
{\color{red}{Ideally, the training would occur with actual data loading
and workload execution. However, this will be slow because
to compute the \textit{reward}, the RL client needs to wait for all of the related datasets are loaded using the
partition scheme specified in recent \textit{action}s and related queries are
all executed. It is impractical to run each workflow many times. To avoid this overhead, existing DRL approach for relational data partitioning chooses to bootstrap the model using traces generated by a cost model~\cite{hilprecht2019learning}. While this is a reasonable solution for relational database, there is no widely-accepted cost models for UDF-centric analytics~\cite{shi2014mrtuner}. 

To significantly alleviate the training overhead, we  propose to  transform the running statistics of a limited set of actual query executions into the estimated statistics of a large set of diversified workloads generated by randomly combining queries with varying frequencies. We first run a few selected queries. (One requirement is that some of these queries should be latency-sensitive to the partitionings of their input data, e.g., queries involving join and aggregation.)
Then, we
enumerate all possible partition schemes for the inputs of these queries. Furthermore,  we run
the queries and measure each query's latency for each possible partition scheme.}}
\eat{
Data augmentation, which is to increase the size and diversity of training data by applying random (but realistic) transformations such as image rotation, is commonly used in deep learning problems~\cite{}.
} 

Thus, the training component works like a simulator. It first
samples a workload, identifies the partitioner candidates, and forms
the \textit{state} based on the historical statistics of these partitioner candidates. Then it sends the
\textit{state} to the RL server and obtains the action for partitioning. Instead of actually partitioning the data and running the queries from the sampled workload, it
simply look-up the latency related to this action and computes reward from historical latency statistics of these queries.
{\color{red}{In this way, we can generate an unlimited number of workloads for training, and we do not need to actually run workloads with the specified frequency. The increased number of workloads
 result in more robust representations of important features such as frequency, recency, number of co-partitioned datasets, etc., as mentioned in Sec.~\ref{sec:rl}.}}

For each partitioner candidate $L_i$ that is extracted from query $Q_j$ , we can obtain
statistics such as reference distance ($d_{j}$), frequency ($f_{j}$), recency ($r_{j}$),complexity ($c_i$), selectivity ($s_{ij}$), number of distinct keys ($k_{ij}$). (Other features such as number ($n_i$) and size ($s_i$) of co-partitioned datasets are computed for each synthetic workload at training time.)
\eat{At training time, we randomly
generate workloads by sampling 
combinations of the selected queries. For example, \{$\langle Q_a, 0.5\rangle$,
  $\langle Q_b, 0.5\rangle \}$ represents a workload that consists of two
queries $Q_a$ and $Q_b$, with the same frequency.}  For example,
if \{$L_1$, $L_2$\} are partitioner candidates extracted from $Q_a$, and
\{$L_2$, $L_3$\} are extracted from $Q_b$, then
we can enumerate a set of partitioner candidates for the datasets associated with this workload as
\{$L_1$, $L_2$, $L_3$\}. For $L_1$ and $L_3$, their features are ($d_{a}$,
$f_{a}$, $r_{a}$, $c_{1}$, $s_{1a}$, $k_{1a}$, $n_{1}$, $s_{1}$) and ($d_{b}$,
$f_{b}$, $r_{3}$, $c_{3}$, $s_{3b}$, $k_{3b}$, $n_{3}$, $s_{3}$); For $L_2$, because it is
applied to both  $Q_1$ and  $Q_2$, its feature vector is
(avg($d_{a}$, $d_{b}$), avg($f_{a}$+$f_{b}$), avg($r_{a}$+$r_{b}$), $c_2$, max($s_{2a}$,
$s_{2b}$), min($k_{2a}$, $k_{2b}$), $n_{2}$, $s_{2})$). Here, we
use the maximum value for selectivity and the minimum
value for number of distinct keys, mainly because we want to
encourage partitioning using partitioner candidates for job stages that have large
selectivity and avoid partitioning using lambda terms that lead to a small
number of distinct hash keys.

\eat{
There are other components in the \textit{Lachesis} system, including a data rebalancer, which adjusts existing data partitionings in case of node removal, node addition,  changes in workload characteristics, and evolution of data schema,  by moving only a portion of data across the cluster nodes; and a data replicator, which creates heterogeneous replications~\cite{zou2019pangea} of data by selecting and applying multiple persistent partitionings. Because these components are not in the critical path of the automatic partitioning workflow, we omit the details in this work.}

\section{Evaluation}
\label{sec:new-exp}
In this section, we mainly want to answer following questions:

\noindent
(1) What are the performance gains that can be achieved by \textit{Lachesis}'s automatic persistent partitioning for different types of Big Data analytics applications? (Sec.~\ref{sec:reddit}, ~\ref{sec:tpch}, ~\ref{sec:pagerank})

\noindent
(2) How much online and offline overhead is incurred during the automatic partitioning process? (Sec.~\ref{sec:overhead})

\noindent
(3) How effective is the DRL training process? (Sec.~\ref{sec:training})

\eat{
\noindent
(4) How will the amount of history affect the effectiveness? (Sec.~\ref{sec:history})}

\vspace{-10pt}
\subsection{Workloads, Baselines, and Environment Setup}
\label{sec:environment}

\subsubsection{Workloads}
\eat{\textit{Lachesis} is built on top of a PlinyCompute~\cite{zou2018plinycompute}. When \textit{Lachesis} is enabled, all job execution information will be logged, and each data loading or materialization computation will trigger a request to \textit{Lachesis}. }To answer the first question, we implement a set of representative workloads including:

\noindent
{\color{red}{(1) {\textbf{Reddit data integration workflows.}} We implement two dynamic workflows: one is the motivating example illustrated in Fig.~\ref{fig:graph-ir-example}; the other is a deep learning model inference workflow.

}}

\eat{
\noindent
(3) {\textbf{Linear Algebra Operations.}} We mainly select four representative linear algebra workloads: dense matrix multiplication,  sparse matrix multiplication, gram matrix (given a matrix $\textbf{X}$, compute
$\textbf{X}^T \textbf{X}$)~\cite{zou2018plinycompute}, and the least squares linear regression (given a matrix of features $\textbf{X}$ and
responses $\textbf{y}$, compute 
$\hat{\pmb{\beta}} = (\textbf{X}^{T} \textbf{X})^{-1} \textbf{X}^{T} \textbf{y}$)~\cite{zou2018plinycompute}. The producer workload loads matrix data into the storage, which will be processed by above mentioned linear algebra operations.}

\noindent
(2) {\textbf{TPC-H Queries.}} {\color{red}We implement ten TPC-H queries (Q1, Q2, Q3, Q4, Q6, Q12, Q13, Q14, Q17, Q22). Each table is represented as a collection of C++ objects and each predicate is wrapped as a UDF. The producer workloads load eight TPC-H datasets ( lineitems, orders, customers, parts, suppliers, partsupp, regions, and nations), to the storage, and then the queries run to process the loaded data.}

\noindent
(3) {\textbf{PageRank Analytics workflow.}} We implement a web analytics workflow that consists of two workloads: pre-processing the web pages, and running PageRank iterations on the processed  pages~\cite{page1999pagerank}. 
\vspace{-15pt}
{\color{red}{
\subsubsection{Baselines}
\label{performance-gain}

For TPC-H benchmarks, we compare Lachesis to the partition schemes suggested by a commercial distributed database. For other workloads, because there are no existing automatic partitioners that can work with UDF-centric workloads to process arbitrary data types, we measure the performance speedup by comparing the consuming workload's latency of applying  \textit{Lachesis} {\color{red}{to different baselines listed as follows. 

\noindent
(1) Heuristics that are typically used
by a database administrator~\cite{zamanian2015locality, hilprecht2020learning}: one is to co-partition all datasets with the most frequent joined dataset (i.e. Heuristics(a)) and the other is to co-partition all datasets with the largest dimension table (i.e. Heuristics(b)). 

\eat{
\noindent
(2) Cost-based optimizer, which we implement for UDF-centric workflows. The system first enumerates the partitioner candidates as described in Sec.~\ref{sec:enumeration}. Then given a dataset, for each of its partitioner candidate, we use a cost-based query optimizer to estimate the total I/O costs and CPU costs for all queries that may run on the dataset. Then we the best partitioner candidate with the lowest costs. While the idea is similar to the physical database design of commercial relational databases such as IBM DB2~\cite{rao2002automating}, Microsoft SQLServer~\cite{agrawal2004integrating, nehme2011automated}, etc.. Our optimizer is based on the UDF-centric IR and all our proposed functionalities for extraction, enumeration, matching, and historical information analysis of partitioner candidates.}

\noindent
(2) The round robin dispatching strategy, which is to dispatch each page of data to a cluster node in order. This is an effective way to guarantee load balance for large dataset, and is adopted by many  storage systems such as IBM GPFS~\cite{gupta2011gpfs}.}}

\noindent
(3) For Reddit Workflow 1 and the PageRank workflow, we also compare to a reactive approach~\cite{idreos2007database}.}}

\subsubsection{Environment Setup}
 We mainly use  three AWS clusters; (1) Environment 1, which is a cluster that has three r4.2xlarge instances. Each r4.2xlarge instance has $8$ CPU cores, $61$GB memory, up to $10$ GB network connection. (2) Environment 2, which has eleven r4.2xlarge instances. (3) Environment 3, which  has six m2.4xlarge instances. Each m2.4xlarge instance has $8$ CPU cores, $68$ GB memory, up to $1$ GB network connection. In each cluster, one instance serves as the master and the rest of the instances serve as workers. Each instance has $200$GB EBS SSD.

\subsection{Reddit Data Integration Workflows}
\label{sec:reddit}
We compare  the performance of two workflows by using different partitioning strategies. In each workflow, three workloads are responsible for loading the Reddit comments in JSON format {\color{red}{(up to $100$ millions of comments objects, $128$ gigabytes in total)}}, Reddit authors data in CSV format {\color{red}{($78$ millions of authors objects, $10$ gigabytes in total)}}, and subreddits data in CSV format {\color{red}{($3.7$ gigabytes in total)}}.\footnote{Reddit datasets are download from http://files.pushshift.io/reddit/.} 

{\color{red} We store the raw files in a S3 bucket, and the time to copy all the files from S3 to the master node of the cluster is $76$ seconds. Such loading happens only once and the loaded data will be repeatedly processed and the following measurements do not include this time.}

\begin{figure}
\centering\subfigure[speedup]{%
   \label{fig:latency-workflow1}
   \includegraphics[width=3.2in]{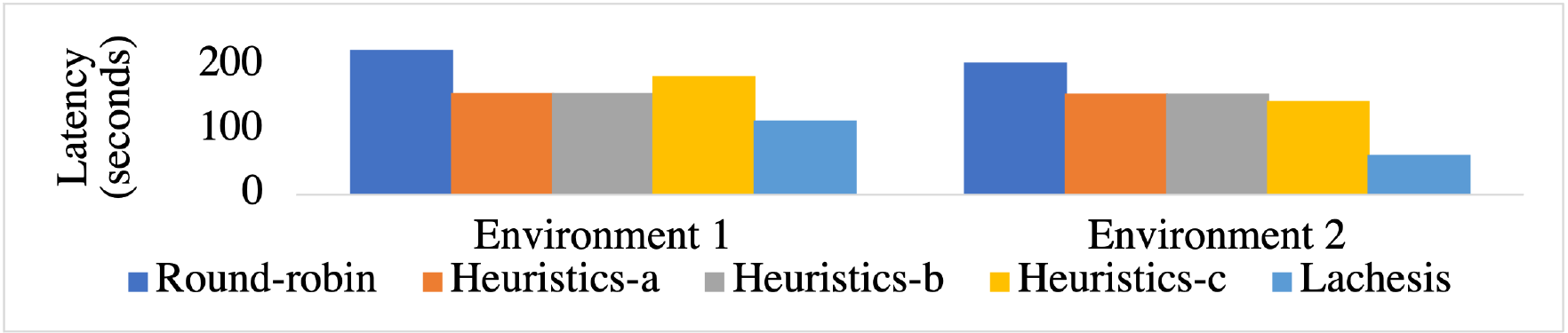}  
}%
\vspace{-5pt}
\subfigure[{\color{red}shuffled bytes}]{%
  \label{fig:shuffle-workflow1-environment2}
  \includegraphics[width=3.2in]{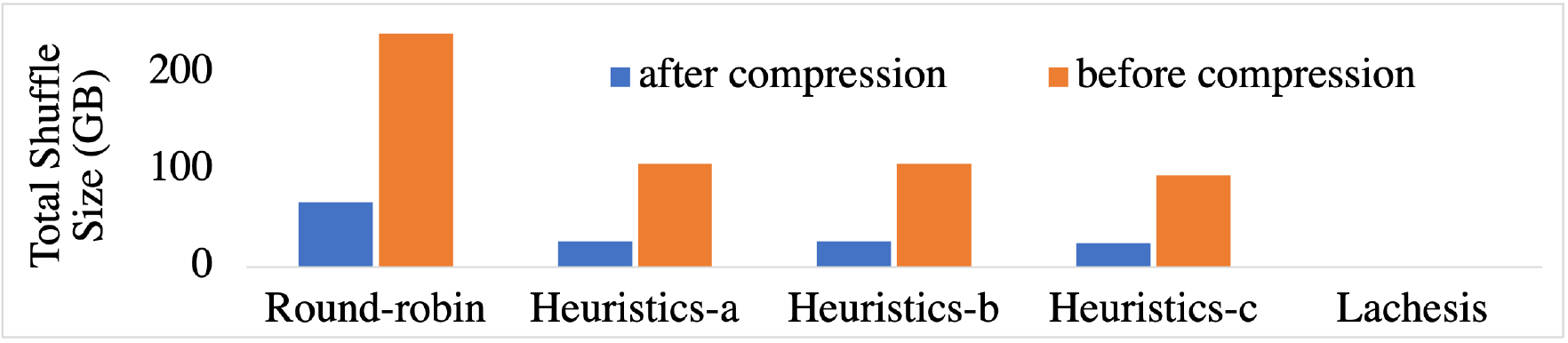}
}
\caption{\label{fig:workflow1}\small
Reddit Workflow-1
}
\end{figure}

{\color{red}{
\subsubsection{Workflow-1: three-way dynamic join}
\label{sec:workflow1}
For this experiment, we use $10\%$ of data from the three Reddit datasets in Environment 1; and use all data in Environment 2. After loading the data to storage, the workflow performs a three-way join similar to Listing.~\ref{code1}, depending on a classifier that checks the value of a comment's score. In this workflow, the classifier results in around $50\%$ of comments to join with authors, and $50\%$ of comments to join with subreddits. 

Because the workflow contains only one three-way join, both heuristics-a and b choose to partition the comment objects based on author name. Existing physical database design advisors enumerate partition keys based on foreign keys. These tools will either choose to partition the Reddit comments dataset along the author name, or along the subreddit channel. So we added heuristics-c to partition along the subreddit channel for completeness. Only our Lachesis approach can identify and exploit the UDF-based partitioner candidate as illustrated in Fig.~\ref{fig:comments-partitioner-candidate}. As a result, Lachesis achieves $1.4\times$ speedup in the two-worker cluster and $2.4\times$ speedup in the ten-worker cluster, as shown in Fig.\ref{fig:latency-workflow1}.
The total amount of data shuffled in the Environment 2 is illustrated in Fig.~\ref{fig:shuffle-workflow1-environment2}. Furthermore, we observe that with the round-robin partitioning, in Environment 2, in average, it needs to shuffle $32$ gigabytes to co-locate with the authors data; and $29$ gigabytes to shuffle for co-locating with the subreddits data, after compression using Snappy v1.5. \eat{But once we pre-partition using heuristic-c, not only the shuffle overhead for subreddits is eliminated, but also the shuffle overhead for authors is reduced from $32$ to $25$ gigabytes as a side effect. This indicates that once the comments are co-located with subreddits, a significant portion of authors are also co-located with part of the comments automatically, which do not need to be shuffled again. This is also observed for using Heuristics-a and b, which reduce the required amount of shuffle data for co-partitioning with subreddits from $29$ to $27$ gigabytes.}

}}

\begin{figure}
\centering
   \includegraphics[width=3.5in]{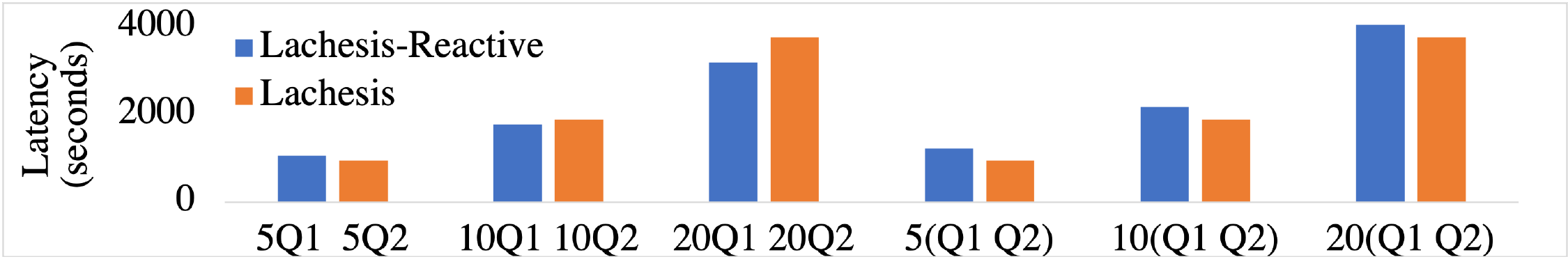}
\caption{\label{fig:reddit-reactive} \small
{\color{red}Comparison of Lachesis-reactive and Lachesis in Environment 1. $k$Q1 $k$Q2 represents executing Q1 for $k$ times and then Q2 for $k$ times. $k$(Q1Q2) represents consecutively executing Q1 and Q2, and repeat it for $k$ times.}
}
\end{figure}

{\color{red}{As mentioned in the beginning of Sec.~\ref{sec:formulation}, \textit{Lachesis}' functionalities for enumerating and matching partitioner candidates can be used to repartition existing datasets based on new incoming queries (i.e. reactive approach). We thus design several scenarios that are composed of repeated executions of two queries: Q1, a two-way join of reddit authors and comments; and Q2, the three-way dynamic join that we just discussed. The results are illustrated in Fig.~\ref{fig:reddit-reactive}. When Q1 is repeatedly executed, the reactive approach chooses to co-partition the comments and authors datasets by the UDF that parses the author name; Then when Q2 is repeatedly executed, the reactive approach chooses to further partition a subset of the comments dataset by the UDF that parses the subreddit name. If the queries are executing in a pattern of Q1, Q2, Q1, Q2, ..., the reactive approach only chooses to repartition once based on the author-extraction UDF. \textit{Lachesis} and Heuristics-a/b always choose to pre-partition the comments and authors datasets based on the author-extraction UDF in all of these patterns. We can see that if a query execution pattern  repeatedly occurs, the \textit{Lachesis-Reactive} approach can achieve similar or slightly better performance.}}

{\color{red}{
\subsubsection{Workflow-2: Reddit Comment Classification based on Deep Neural Network (DNN)} 
We implement a DNN model serving workflow\footnote{{\color{red}The model architecture is consistent with the simple FFN example in TensorFlow~\cite{ffn}, but implemented in PlinyCompute~\cite{zou2018plinycompute} using Tensor Relational Algebra~\cite{jankov12declarative, yuan2020tensor}.}} that classifies whether a reddit comment should join with author info or subreddit channel info, as mentioned in the motivating example in Sec.~\ref{sec:introduction}. Considering that features are extracted from all the attributes regarding the comment object, occurrences of words in a large dictionary, and ngrams, this is a high-dimensional machine learning problem. 

This workflow consists of six jobs: (1) blocking, which is to extract a pre-computed feature vector from a comment object through an index, and block the feature vectors of a batch of comments into many $2$-dimensional $1000 \times 1000$ matrix blocks; (2) layer-1, which passes the feature blocks to the first fully-connected layer that has $1000$ neurons and a $1000$-dimensional bias vector with Relu activation, and outputs $y_1$; (3) layer-2, which passes $y_1$ to the second fully-connected layer that has $2000$ neurons and a $2000$-dimensional bias vector with Relu activation, and outputs $y_2$; (4) layer-3, which is the output layer that consists of $2$ neurons for the two labels and a $2$-dimensional bias vector, outputing $y_o$; (5) flattening, which applies softmax activiation to $y_o$ to get the probability distribution over the labels for each sample, and then flattens each result tensor block to a set of label objects indexed by the comment identifier; (6) labeling, which joins the batch of reddit comments with the label objects, so that each comment's label attribute is filled with the prediction.

We run the experiment with $200000$ to $1$ million features in Environment 1, using a batch size of $1000$. All values are using double precision. We observe that in this case, by only partitioning the persistent datasets such as reddit comments, the weight and bias matrices of layer-1, layer-2, and layer-3, it achieves only moderate performance gain, labeled as \textit{Lachesis-persistent}  in Fig.~\ref{fig:reddit-serving}. By additionally co-partitioning intermediate data such as the output of the blocking, layer-1, layer-2, layer-3, and flattening, we can achieve the maximum performance gain, labeled as \textit{Lachesis-full}  in Fig.~\ref{fig:reddit-serving}. 

Although this workflow has more than ten datasets that are inputs to join operations. However, for each dataset, usually only one or two partitioner candidates exist. So the \textit{Lachesis-full} result is similar to Heuristics-a and Heuristics-b. Given the many datasets involved and the complexity of the workflow, it is hard for programmers to manually figure out and manage the partitionings. The productivity brought by automatic partitioning is a great benefit of \textit{Lachesis}. 

A significant portion of overall performance gain ($1.3\times$ to $1.6\times$ speedup) is coming from the blocking stage and the layer-1 stage. The blocking stage transforms and aggregates the batch of comments that is merely $1$ megabytes in total to a set of feature blocks that have $2$GB to $10$GB in total size depending on the number of features. Because \textit{Lachesis} chooses to partition comments by its unique identifier, comments are evenly distributed across worker threads. However, the round-robin approach distributes data by pages, which causes skewed distributions due to the small size of a batch. The layer-1 stage includes a join of a \textit{numNeurons}$\times$\textit{numFeatures} weight matrix and a \textit{numFeatures}$\times$\textit{batchSize} input matrix, which can benefit from \textit{Lachesis} by avoiding shuffling. Without Lachesis, for $1$ million features, we observed  about $8.5$GB gets shuffled ($5.6$GB w/ compression), in Environment 1. 
%; and about $16$GB gets shuffled  ($13.8$GB w/ compression), in Environment 2

The latency of the layer-2, layer-3, flattening, and labeling  will not change with the increase in number of features because the joins in these stages are determined by factors such as the number of neurons and the batch size. Despite of small shuffle sizes in these stages, \textit{Lachesis} can still achieve about $15\%$ performance gain because the removal of the shuffling phases reduces the CPU costs.
}}
\eat{
\begin{figure}
\centering\subfigure[Environment 1]{%
   \label{fig:reddit-serving-1}
   \includegraphics[width=1.6in]{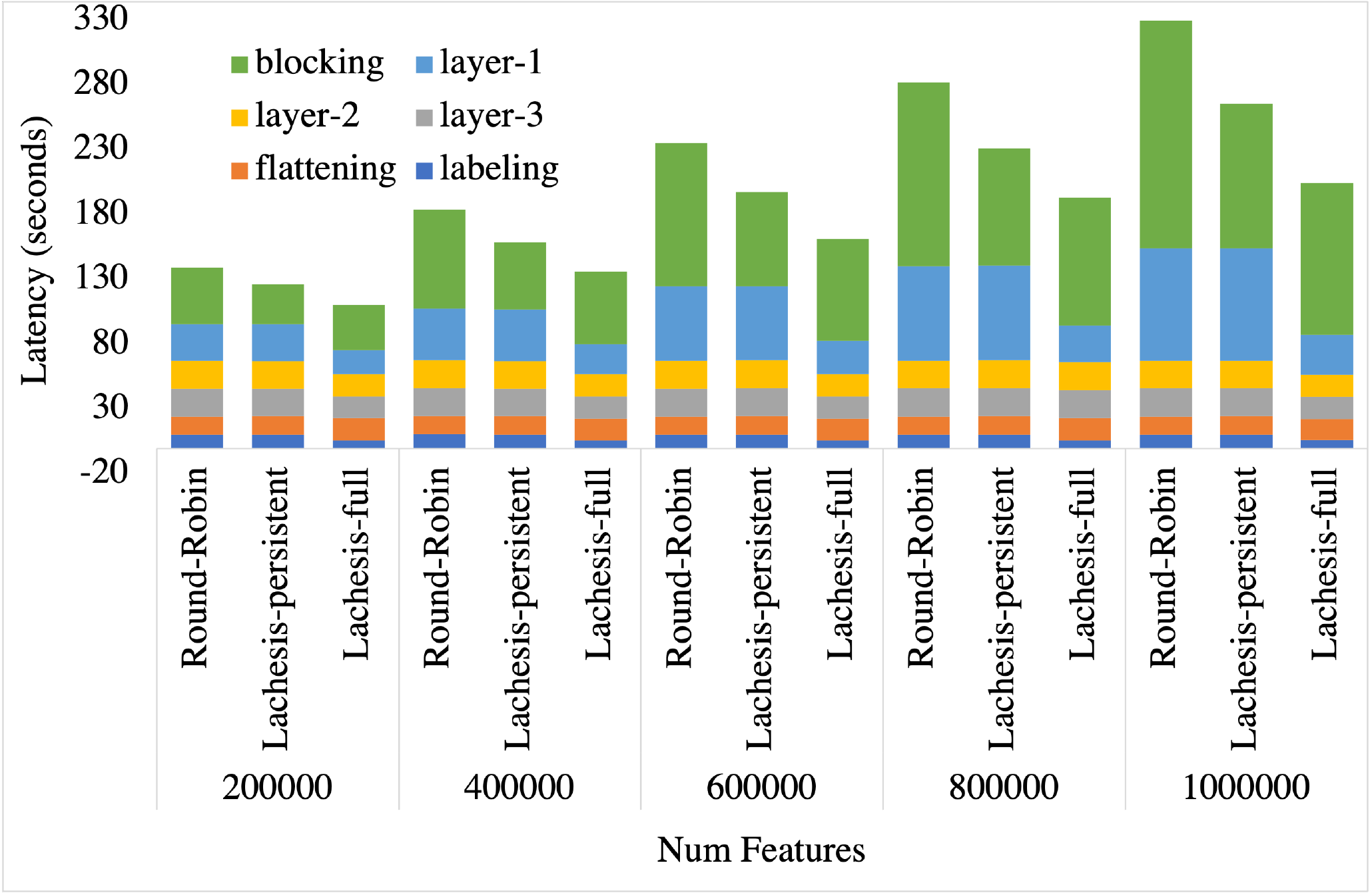}  
}%
\hspace{-5pt}
\subfigure[Environment 2]{%
  \label{fig:reddit-serving-2}
  \includegraphics[width=1.6in]{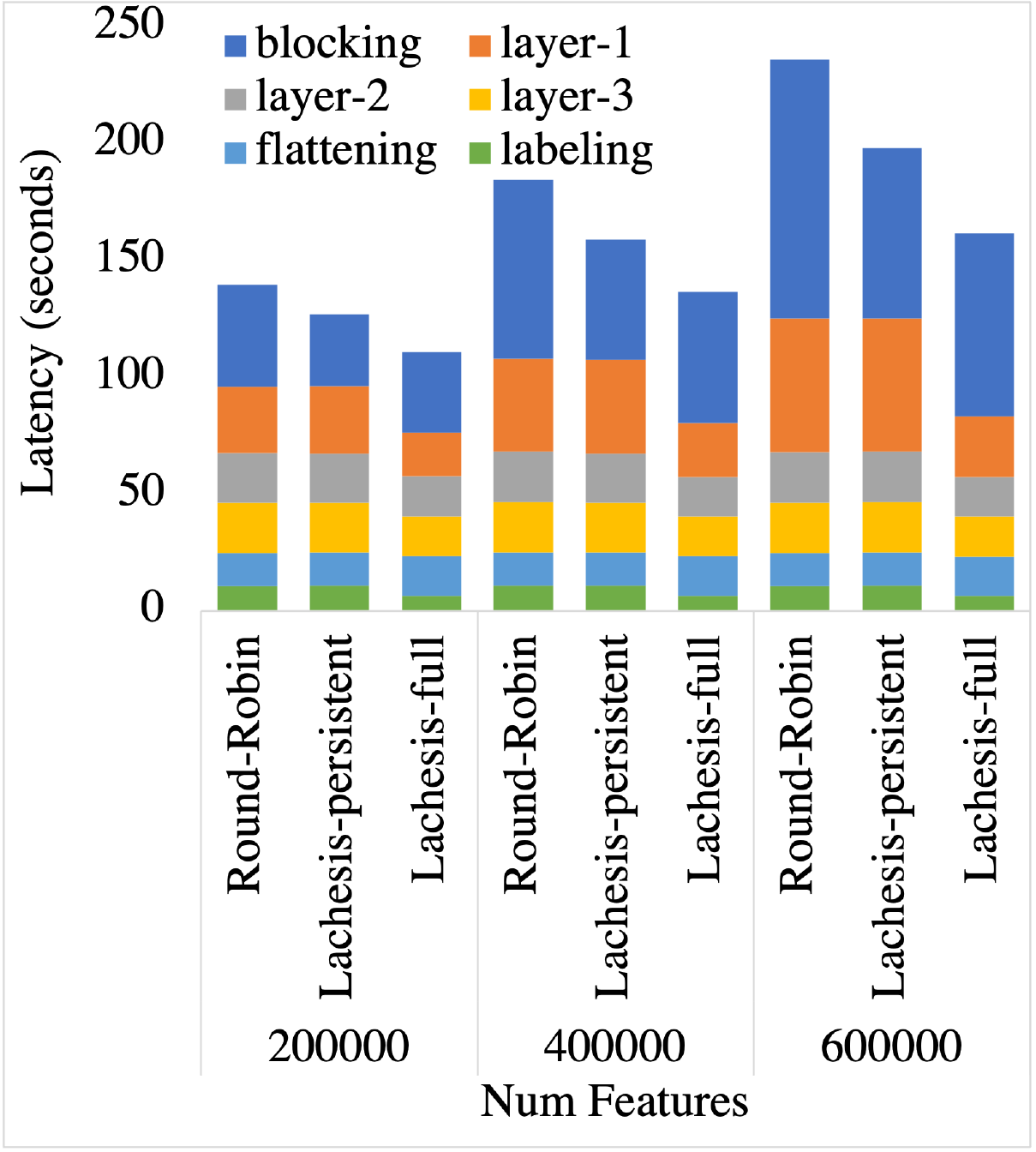}
}
\caption{\label{fig:reddit-serving}\small
{\color{red}{Performance of Workflow 2 for a batch of inferences}}
}

\end{figure}}

\begin{figure}
\centering
   \includegraphics[width=3.2in]{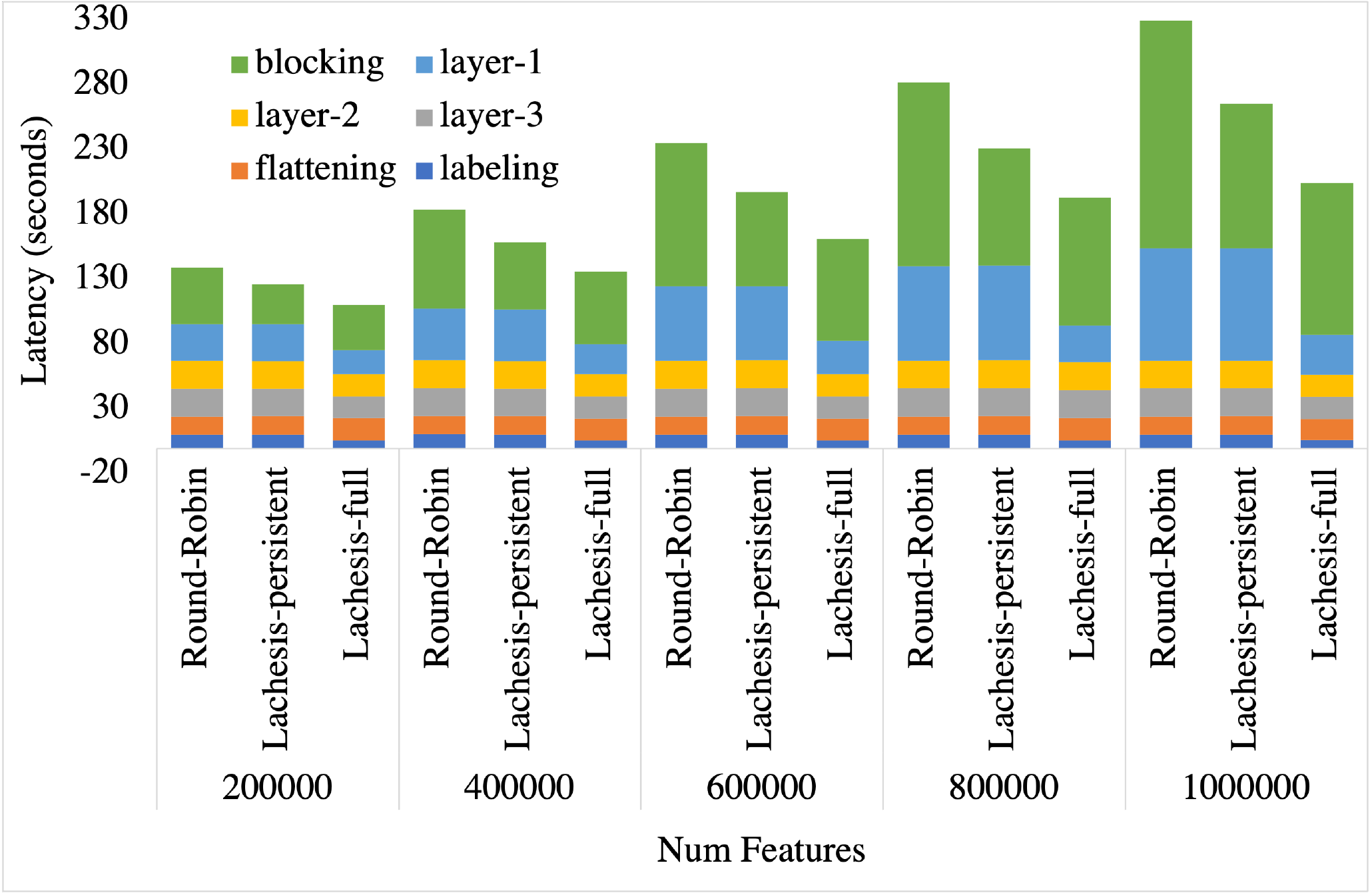}
\caption{\label{fig:reddit-serving} \small
{\color{red}{Performance of Workflow 2 for a batch of inferences}}}
\end{figure}

\begin{figure}
\centering\subfigure[Environment 1: SF-10]{%
   \label{fig:tpch-environment1}
   \includegraphics[width=3.2in]{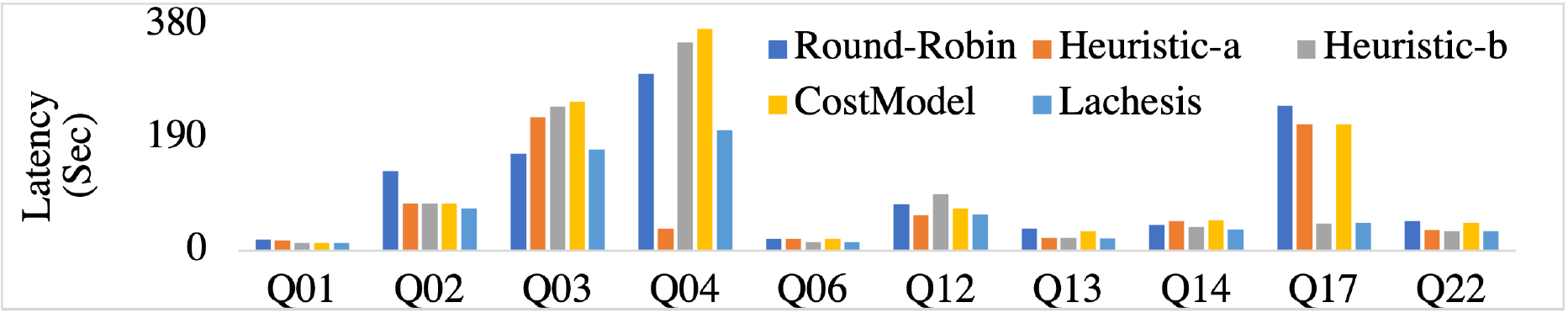}  
}%
\vspace{-5pt}
\subfigure[{\color{red}Environment 2: SF-100}]{%
  \label{fig:tpch-environment2}
  \includegraphics[width=3.2in]{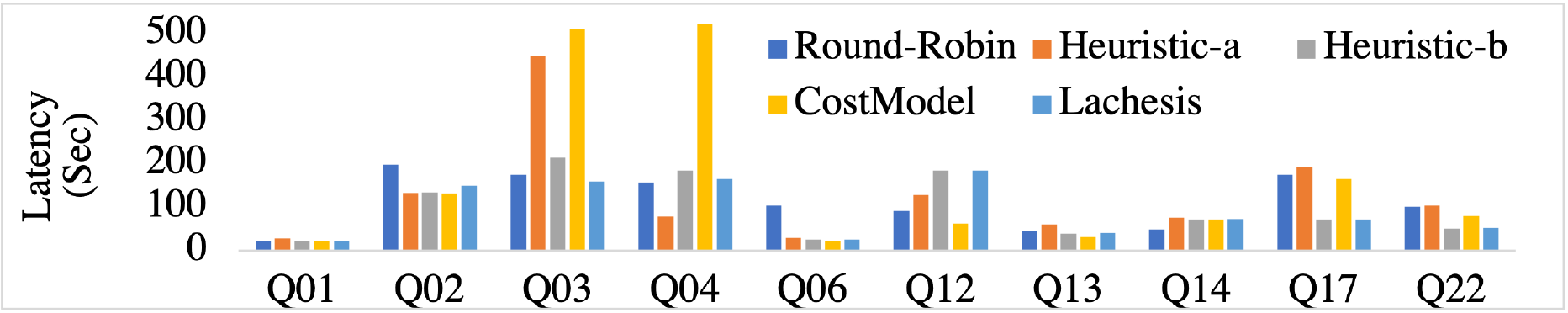}
}
\caption{\label{fig:tpch-detailed} \small
{\color{red}{TPC-H performance (refactored with objects and UDFs)}} 
}
\end{figure}

{\color{red}{
\subsection{TPC-H Refactored with Objects and UDFs}
\label{sec:tpch}
We implement all eight TPC-H tables as eight C++ classes. Then we implement ten TPC-H queries that involve aggregations and/or joins and can be represented using PlinyCompute computations. We load data that is generated using dbgen at scale \textit{SF-10} into a Environment 1; and load data at scale \textit{SF-100} into Environment 2. For this experiment, we compare to round-robin (RR) partitioning, Heuristics-a, Heuristics-b, and the partitionings automatically created by a commercial distributed database using a cost-based physical database design advisor (denoted as CostModel). The selected partitionings and the total execution latency of ten queries for each partitioning strategy are shown in Tab.~\ref{tab:tpch-partitionings} and Tab.~\ref{tab:tpch-total-latency}. The measured latency for each query is illustrated in Fig.~\ref{fig:tpch-detailed}.

In both environments, \textit{Lachesis} achieves the best performance. In Environment 1, it outperforms the second best strategy, which is Heuristic-a, by $12\%$. In Environment 2, \textit{Lachesis} outperforms Heuristic-b, which is the second best, by $6\%$. The CostModel approach shows the worst performance in both enviornments, which indicates that the cost model of a relational database system is not applicable to a UDF-centric analytics system.

Taking Q17 for example, in both environments, \textit{Lachesis} chooses to co-locate the lineitem table and the part table on partkey. As a result, for this query, Lachesis achieves $5\times$ speedup in Environment 1 and $3\times$ speedup in Environment 2, compared to Hueristics-a.
}}

\begin{table}
\centering
\scriptsize
\caption{\label{tab:tpch-partitionings} {\color{red} \small Comparisons of Partitionings in Environment 2 for TPC-H}}
\begin{tabular}{|l|l|l|l|l|l|} \hline
&RR&Heuristics-a&Heuristics-b&CostModel&Lachesis\\\hline \hline
customer&-&c\_custkey&c\_custkey&c\_custkey&c\_custkey\\ \hline
nation&-&n\_nationkey&n\_nationkey&n\_nationkey&n\_regionkey\\ \hline
partsupp&-&ps\_partkey&ps\_partkey&ps\_partkey&ps\_suppkey\\ \hline
region&-&r\_regionkey&r\_regionkey&r\_regionkey&r\_regionkey\\ \hline
lineitem&-&l\_orderkey&l\_partkey&l\_orderkey&l\_partkey\\ \hline
orders&-&o\_orderkey&o\_orderkey&o\_custkey&o\_orderkey\\ \hline
part&-&p\_partkey&p\_partkey&p\_partkey&p\_partkey\\ \hline
supplier&-&s\_suppkey&s\_suppkey&s\_suppkey&s\_nationkey\\ \hline
\end{tabular}
\end{table}

\begin{table}
\centering
\scriptsize
\caption{\label{tab:tpch-total-latency} {\color{red} Comparisons of  Total Latency for ten TPC-H queries.}}
\begin{tabular}{|l|l|l|l|l|l|} \hline
&RR&Heuristics-a&Heuristics-b&CostModel&Lachesis\\\hline \hline
Environment 1&1088 sec&758 sec&939 sec&1153 sec&\textbf{672} sec\\ \hline
Environment 2&1121 sec&1285 sec&1002 sec&1701 sec&\textbf{944} sec\\ \hline

\end{tabular}
\end{table}

\subsection{PageRank}
\label{sec:pagerank}
\begin{figure}
\centering\subfigure[Shuffled Bytes]{%
   \label{fig:pagerank-shuffled-bytes}
   \includegraphics[width=1.6in]{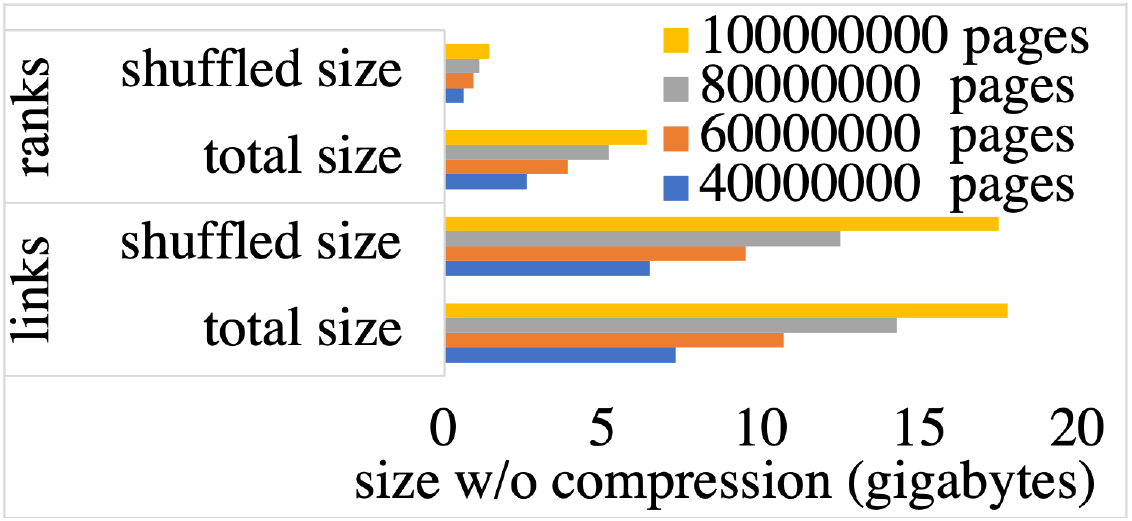}  
}%
\hspace{-5pt}
\subfigure[Shuffling Latency]{%
  \label{fig:pagerank-shuffle-latency}
  \includegraphics[width=1.6in]{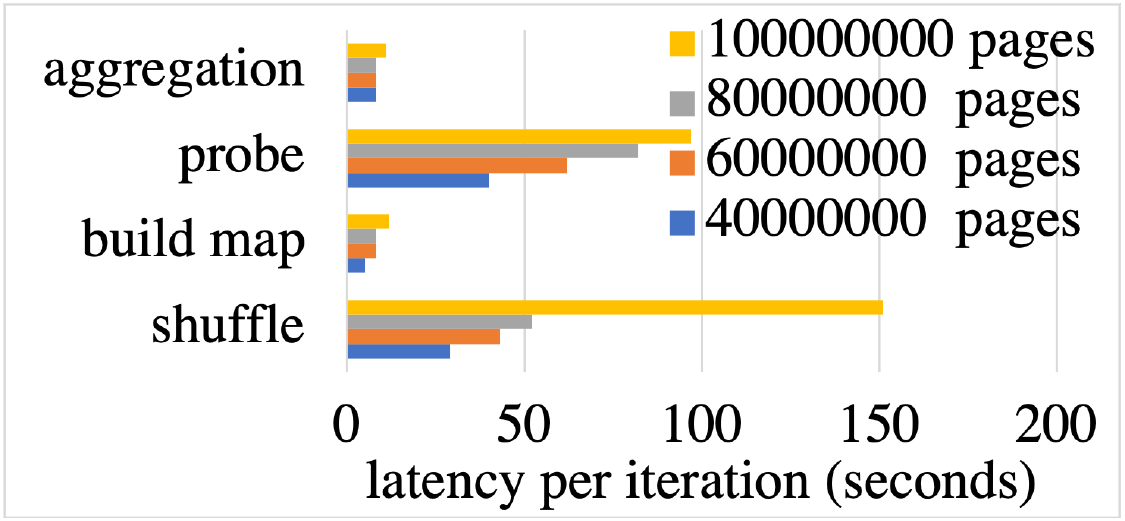}
}
\caption{\label{fig:pagerank-shuffle}\small{
\color{red}Shuffling in PageRank w/o Lachesis (Environment 3)}
\vspace{-5pt}
}
\end{figure}

In the PageRank application, a producer workload extracts a set of Page objects from web pages. Each Page object includes a url member that specifies the page, and a vector of urls this page links to. Then in the consumer workload, each iteration involves a \texttt{join} operation that joins the set of Page objects and the set of Rank objects. Each Rank object includes a url member, and a rank member, which is a double value. we set 
the number of iterations to five by default, and use the default damping factor $0.85$. 
{\color{red}{
Each PageRank iteration requires to join the ranks of links (denoted as \texttt{ranks}) with the link adjacency matrix (denoted at \texttt{links}). Then the output will be used to update \texttt{ranks}. If \texttt{ranks} and \texttt{links} are co-partitioned for the join, all iterations do not require a shuffling. Otherwise, each iteration will require shuffling unless \texttt{ranks} and \texttt{links} are repartitioned.

We benchmark the PageRank application in Environment 3. The producer randomly generates and pre-processes $40$ million to $100$ million Page objects. Each Page object has five neighbors on average. Lachesis chooses to pre-partition the set of Pages and the set of Ranks using the Page object's and Rank object's url member access functions extracted from the IR. We compare Lachesis to round-robin partitioning, and a reactive approach that co-partitions \texttt{ranks} and \texttt{links} after the first iteration. The reactive approach also uses the Lachesis functionality to recognize partitioning candidates.

The results are illustrated in Fig.~\ref{fig:pagerank-perf}. We observe that in Environment 3, \textit{Lachesis} can achieve up to $6.5\times$ speedup by comparing to the round-robin partitioning; and can achieve up to $1.8\times$ speedup by comparing to the reactive approach. In addition, when we increase the number of iterations, the performance gain achieved by the \textit{Lachesis}' pre-partition approach  compared to the \textit{Lachesis}' reactive approach will gradually drop, because the partitioning overhead is amortized to more iterations. 

\begin{figure}
\centering
   \includegraphics[width=3.5in]{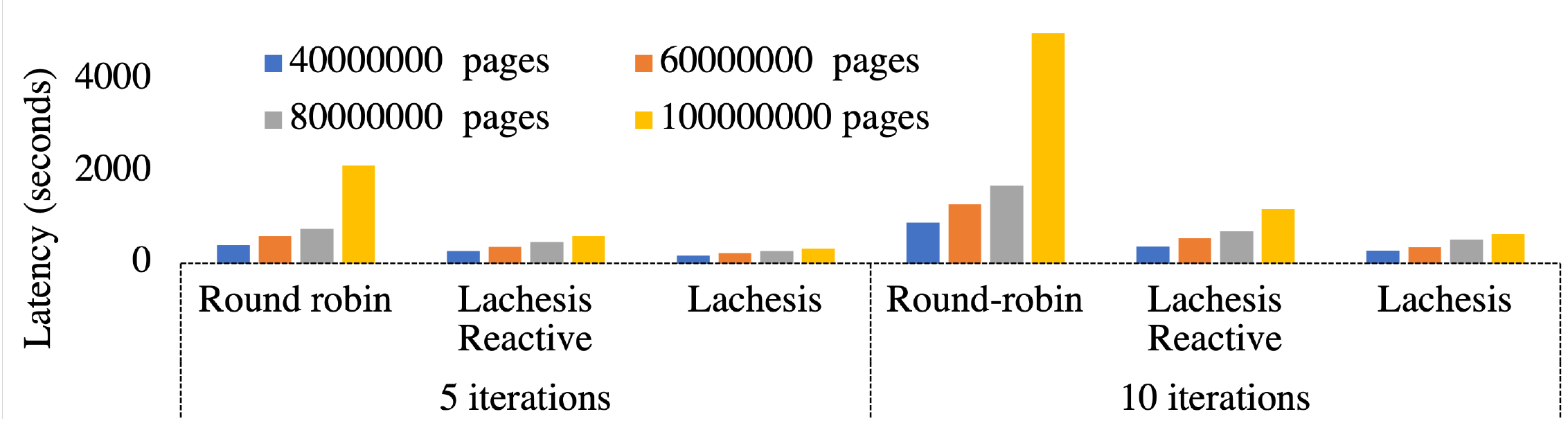}
\caption{\label{fig:pagerank-perf} \small
PageRank performance comparison 
}
\end{figure}

The shuffled bytes information as well as the input data size information is illustrated in Fig.~\ref{fig:pagerank-shuffled-bytes}. Through analysis, as illustrated in Fig.~\ref{fig:pagerank-shuffle-latency}, when round-robin partitioning strategy is utilized for processing $100$ millions of pages, the shuffling of the \texttt{links} and the \texttt{ranks} accounts for $\textbf{75}$\% of the total latency. Through pre-partitioning and repartitioning (as in the reactive approach), the query optimizer can recognize the useful partitioning and remove these two shuffling stages. Thus all the overheads related to shuffling including hashing, data copying, network transferring, synchronization, are eliminated accordingly. We also observe that the shuffling overhead increases significantly faster with the size of inputs than the rest of the overheads, because of the non-determinism in the shuffling process. For example, for shuffling the \texttt{ranks} for $100$ millions Page objects, in one iteration, it takes $190$ seconds on the slowest machine while  less than $100$ seconds on the fastest machine.}}

%\vspace{-5pt}

\subsection{Overhead Analysis}
\label{sec:overhead}
In \textit{Lachesis}, the overheads can be divided into three parts: the offline part that can be amortized to all partitioning requests; the online overhead for the producer that can be amortized to multiple executions of consuming workloads; and the online overhead for the consumer. In this section, we measure and analyze these  overheads.

The offline overheads include creating signatures for historical IR graphs, and creating a skeleton graph from historical workflow graphs. Such overheads are sensitive to the number of workflows, number of workloads in each workflow, and number of operations in each workload.  To better understand the offline overhead for large-scale workflows, we collect above statistics from real world production workflow traces available in the publicly Workflow Trace Archive (WTA) ~\cite{versluis2020workflow}. Then for each trace, we generate workflow graphs following its statistics and apply our algorithms to the synthetic workflow graphs. As illustrated in Tab.~\ref{tab:overhead-analysis}, we find that the measured offline overhead of constructing skeleton graphs and creating signatures in one r4.2xlarge instance for the scale of real-world workflows is merely up to $14$ minutes, which  can be further accelerated by using multiple machines.

\begin{table}
\centering
\scriptsize
\caption{\label{tab:overhead-analysis}\small Offline overhead for real-world traces. We follow the trace source name given by WTA. WF represents the number of workflows; T represents the number of tasks in a workflow; SG-latency denotes the latency for constructing the skeleton-graph; and SN-latency denotes the latency of creating IR graphs signatures. (latency unit: {\color{red}{seconds}})}
\begin{tabular}{|l|r|r|r|r|} \hline

TraceName&WF&T&SG-latency&SN-latency\\\hline \hline
S1. Askalon Old&4,583&167,677&1&1\\ \hline
S2. Askalon New&1,835&91,599&1&1\\ \hline
S3. LANL&1,988,397&475,555,927&26&12\\ \hline
S4. Pegasus&56&10,573&1&1\\ \hline
S5. Shell&3,403&10,208&1&1\\ \hline
S6. SPEC&400&28,506&1&1\\ \hline
S7. Two Sigma&41,607,237&50,518,481&717&3\\ \hline
S8. WorkflowHub&10&14,275&1&1\\ \hline
S9. Alibaba&4,210,365&1,356,691,136&94&39\\ \hline
S10. Google&494,179&17,810,002&8&1\\ \hline
\end{tabular}
\end{table}

\begin{table}
\centering
\scriptsize
\caption{\label{tab:producer-overhead} \small Producer Latency  Comparison (unit: seconds).}
\begin{tabular}{|l|r|r|r|} \hline
data to store&w/ partition&w/o partition&overhead\\\hline \hline
$15$ millions of author objects&{$42$}&{$42$}&$0\%$\\ \hline
$78$ millions of author objects&{$203$}&{$185$}&$10\%$\\ \hline
$20$ millions of comment objects&{$744$}&{$726$}&$2\%$\\ \hline
$112$ millions of comment objects&{$4,505$}&{$4,119$}&$9\%$\\ \hline
\end{tabular}
\end{table}

 \eat{(The overheads of training the DRL model, which is another part of offline overhead, will be described in Sec.~\ref{sec:training})}At runtime, a data storage request will trigger online overheads at the producer's side that cover: (1) communicating with the TensorFlow-based DRL server, which is several milliseconds' overhead as measured; (2) dispatching the data to the storage by using the partitioner automatically selected by the DRL model. This incurs up to $10\%$ overhead as illustrated in Tab.~\ref{tab:producer-overhead}, which is significantly cheaper than shuffle operations at the consumers' side.

The online overhead at the consumer's side for processing a query involves  matching of the query's IR to the partitioners associated with the input datasets to decide whether to avoid the shuffling stage. We measure this overhead by comparing the latency of enabling \textit{Lachesis}, and simply disabling \textit{Lachesis}. The overhead is smaller than one second for most of the workloads.

\subsubsection {Summary}
In this section, we measure various overheads incurred by the \textit{Lachesis} system, including the offline overhead that can be amortized to multiple data storage requests, and the online overhead at the producer's side and at the consumer's side. We see that compared to the significant performance speedup achieved for the consuming workloads, both of the offline and online overheads are relatively small. Particularly the online overhead at the consumer workloads' side is negligible, The net performance gain will be further enlarged according to the \textit{write-once-read-many} assumption that we mentioned in Sec.~\ref{sec:assumption}.

\subsection{Training Overhead and Effectiveness}
\label{sec:training}
{\color{red}{
We choose TPC-H queries (rewritten in UDFs) ~\cite{council2008tpc} to create the statistics for simulating the training process. That's because it involves relatively more partitioner candidates than other workloads we have, and though TPC-H’s UDFs are simple, we find the complexity of UDFs is a relatively less important feature for selecting the optimal partitionings, compared to other features (Sec 4.1.3).  
We first run three
queries in TPC-H workload: Q01, Q02, Q04 using all possible partitioning scheme combinations, to generate statistics for actual runs, {\color{red}{which takes $51$ hours in Environment 1 with SF-10 datasets}}.
There are in total ten partitioner candidates related to
those queries, which can be used to enumerate $432$ partition scheme combinations
across all TPC-H datasets. 
For each partition scheme combination, we run the three
training queries respectively in Environment 1, so that we can obtain statistics for
$1296$ actual runs. 
We also generate statistics for three different queries: Q04, Q12, Q17 in Environment 3, also using the SF-10 datasets. Because these three queries involve only a few tables, it only enumerates $17$ different partitioning scheme combinations for the orders, lineitems, and parts datasets. It takes about three hours to create $51$ actual runs. 
Using the training data augmentation technique proposed in Sec.~\ref{sec:training-design}}, unlimited workloads can be created from these actual runs for training.}

\begin{figure} 
\centering{%
  \includegraphics[width=3.2in]{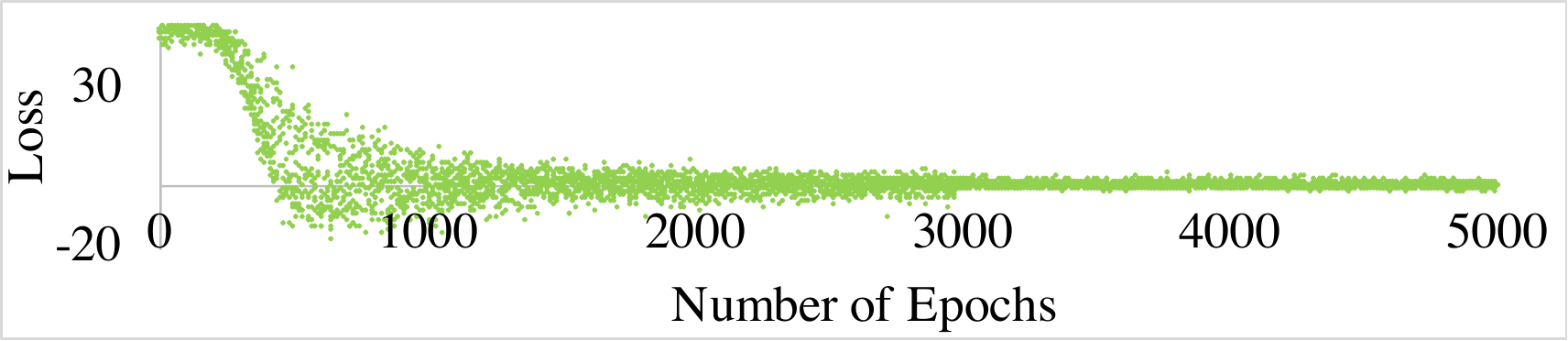}
}%
\caption{\label{fig:training}\small
Training loss in Lachesis
}
\end{figure}

Both the actor and critic neural networks have three  fully
connected layers. The first hidden layer has $128$ neurons and the second hidden layer has $64$ neurons. In both networks, the first two layers use leaky
relu as activation function. For the output layer, the actor network
uses softmax, and the critic network uses linear activation. 
We carefully tune the learning rate ($\alpha$), entropy
weight ($\beta$) and the number of neurons at the hidden layer.
Fig.~\ref{fig:training} illustrates how the training loss
changes with epochs. We use a batch size of $16$, and an epoch has $96$ iterations. The RL-based approach takes about ten hours to run $5000$ epochs with an RL server located on an r4.2xlarge instance. We find that the RL-based approach can be effective in different environments, with different data sizes, and it also requires significant manpower in training, and tuning hyper parameters like entropy value, batch size, model architecture, learning rate, etc..

For each data creation with at least two extracted partitioner candidates, we select the top three partitioner candidates (including round-robin) to formulate the state vector and action space. We find that the RL-based approach can be effective in different environments, with different data sizes, and it also requires significant manpower in training and tuning hyper parameters. 

\eat{
\subsection{Impact of History Collection}
\label{sec:history}
For the PageRank workflow, we test how executing it with different sizes of input data to form varying amounts of history will affect the performance. Fig.~\ref{fig:pagerank-history} shows that if there is no historical execution of PageRank, the performance is the worst. However, if there is at least one execution of the workload, even if the historical execution is for a different input data size, the performance will get optimized similarly. This proves that our feature vector formulation for partitioner candidate is effective and independent with data sizes.

\begin{figure} [H]
\centering{%
  \includegraphics[width=3.4in]{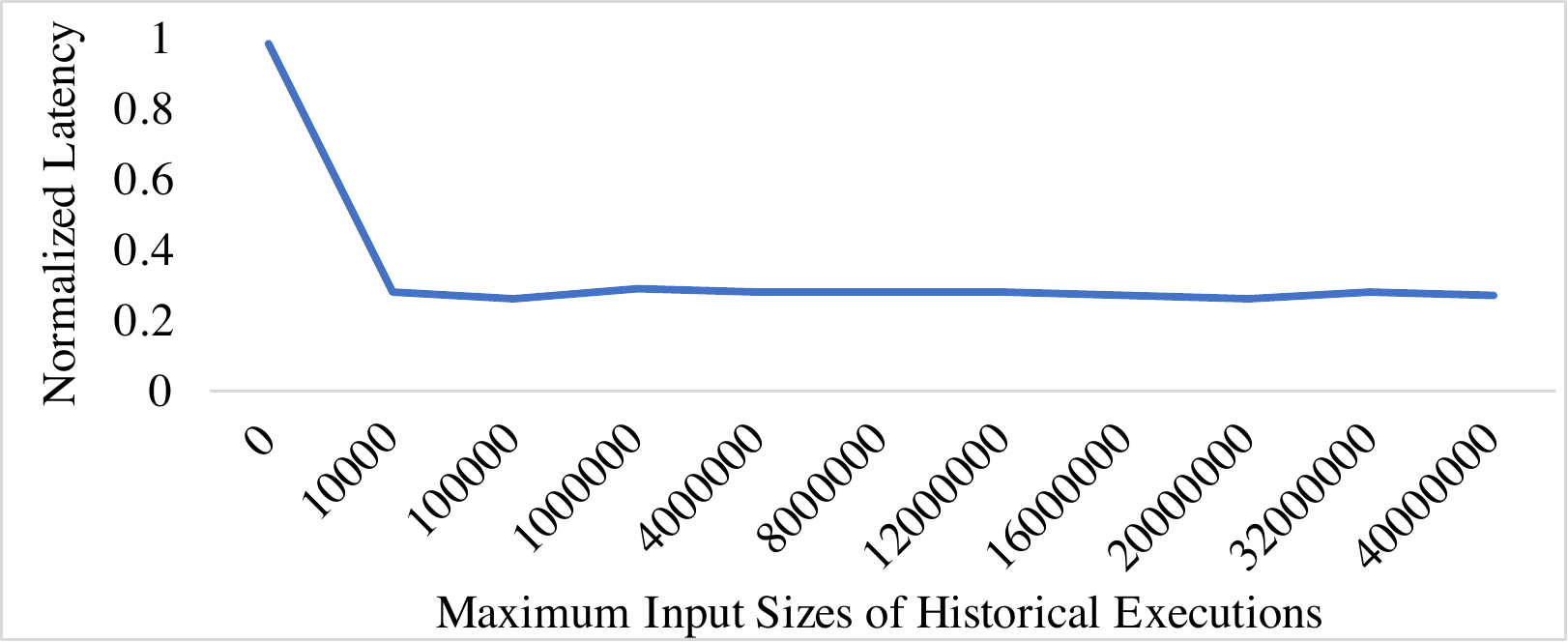}
}%
\caption{\label{fig:pagerank-history}
Normalized performance of a PageRank test case ($8,000,000$ pages and five links per page in average), with varying amount of historical executions, on a small-scale cluster that has one master and two workers (all are AWS r4.2xlarge instances). 
}
\end{figure}

In addition, for the linear algebra workflows, we find that many different workloads (e.g. dense matrix multiplication and gram matrix) share the same desired partitioner. In this case, even if one workload has no historical execution, the system is still able to recommend the right partitioner candidate if any of other partitioning-alike workloads get executed.
}

\eat{
\subsection{Issues with Data Partitioning on Spark}
\label{sec:spark}

In Spark~\cite{zaharia2012resilient, zaharia2010spark}, a partitioning can only live as long as the lifespan of an application run. Such intra-application partitionings cannot be persisted to storage (e.g. HDFS) and reused across different runs of applications due to the gaps between the storage layer and the computation layer~\cite{zou2019pangea, zou2020architecture}. Making it more fragile, a non-partitioning-preserving operator such as {\texttt{map}} may remove a  partitioning~\cite{boehm2016systemml}. Intra-application partitioning in Spark is helpful for iterative joins like in PageRank, where the online repartitioning overhead can be amortized over multiple iterations in the same application. But such partitioning is inadequate for a broad class of workloads, such as data integration and pre-processing, where a dataset is joined only once in each application run. 

We have attempted to manually create persistent partitionings for Spark applications, but that only seems possible for applications that process Hive~\cite{thusoo2009hive} tables by using the bucketBy operators~\cite{luu2018spark}. However it is often difficult for UDF-centric analytics tasks to represent data as Hive tables. \eat{We implement TPC-H Q17 in SparkSQL and store relational data generated at scale-$10$ by dbgen~\cite{council2008tpc} as Hive tables in the above mentioned small scale cluster with three r4.2xlarge instances.  Although we observe up to $90\%$ reduction in the amount of shuffling data (from $2.4$GB to $219$MB, we didn't observe obvious reduction in the query latency, mainly because the Java-based platform is CPU intensive and network shuffling is not the bottleneck.}

We also attempt to compare the persistent partitioning in \textit{Lachesis} with the intra-application partitioning in Spark for the iterative PageRank workloads in the same small-scale cluster. We run PageRank with different numbers of pages. Each run is configured with five PageRank iterations, and we only compare the speedup brought by partitioning mainly because it is hard to directly compare the latency of the two systems due to their complexity. As illustrated in Fig.~\ref{fig:spark-speedup}, Spark partitioning can only achieve up to $1.25\times$ speedup. The relatively lower speedup of Spark is because of two reasons: first, the persistent partitioning doesn't require any shuffling at application runtime, but intra-application partitioning requires one-time online partitioning overhead that is amortized over the five iterations; second, the CPU-intensive nature of Java-based systems like Spark indicates that the shuffling overhead may not be the only performance bottleneck.

Spark v2.4.5 is used for all of the Spark experiments.\footnote{\small The source code for the Spark implementation can be found in https://github.com/asu-cactus/Spark-Partitioning}

\begin{figure} [H]
\centering{%
  \includegraphics[width=3.4in]{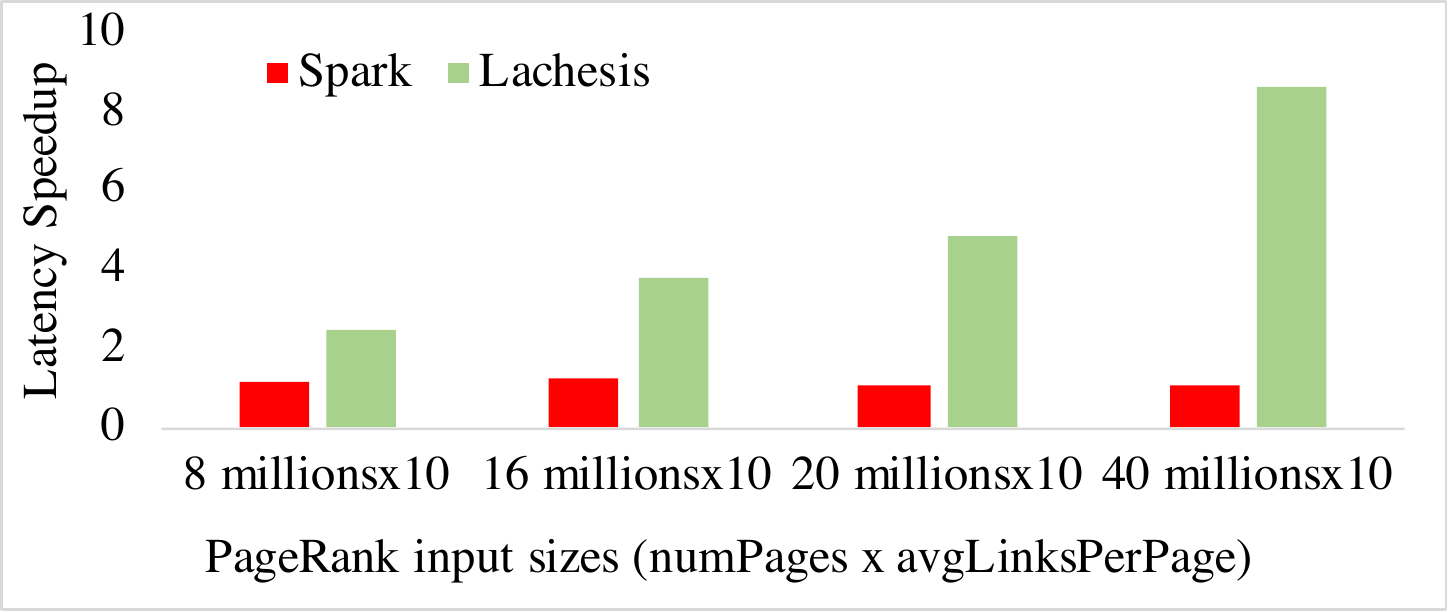}
}%
\caption{\label{fig:spark-speedup}
Speedup comparison of PageRank test cases on a small-scale cluster that has one master and two workers (all are AWS r4.2xlarge instances).}
\end{figure}
}

\eat{\section{Related Works}
\label{sec:survey}
Hilprecht et al.~\cite{hilprecht2019learning, hilprecht2020learning} propose an automatic partitioning advisor using DRL for relational database. They utilize the functional dependency among relational attributes to formulate a network-based state vector. In addition, they use a cost model to bootstrap the DRL model at an offline phase. DRL also has been applied to parameter tuning in relational databases, such as CDBTune~\cite{zhang2019end} and QTune~\cite{li2019qtune}. 
\eat{There are also a number of automatic data partitioning algorithms for
OLTP workloads, like Schism~\cite{curino2010schism}, Sword~\cite{kumar2014sword}, Horticulture~\cite{pavlo2012skew}.
These works are not designed for the UDF-centric analytics workloads and non-relational data, so some of the key ideas such as utilizing functional dependency of relational tables are not applicable to this work, while some other ideas such as replication~\cite{zamanian2015locality} are orthogonal with our work.}}

\eat{
\eat{
IBM DB2 Partition
Advisor~\cite{rao2002automating} recommends candidate partitioning
schemes for physical database design in a shared-nothing
parallel database. Microsoft's AutoAdmin for SQLServer~\cite{agrawal2004integrating}
takes an integrated approach to the problem of choosing indexes,
materialized views, and partitioning, based on optimizer estimated
costs and what-if extensions. But they focus on a single-machine scenario. Nahem and et al~\cite{nehme2011automated} propose an automatic partition
algorithm for MPP database called MESA that deeply integrates with query optimizers and
relies on historical statistics. Legobase~\cite{shaikhha2018building} partitions a relation
differently based on primary key and foreign keys when loading data
and chooses a replica with optimal partition scheme in physical optimization.
AdaptDB~\cite{lu2017adaptdb} partitions data by refining data
partitioning along with relational query executions. 
Eadon et al.~\cite{eadon2008supporting} propose to co-partition relational tables linked via foreign key relationships. Zamanian et al.~\cite{zamanian2015locality} extend this approach to further utilize replication to improve locality.}}

\eat{
\eat{
\noindent
\textbf{Data Partitioning for Big Data Analytics.} As to our knowledge, no Big Data analytics system can automatically create partitionings for general UDF-centric applications. SystemML~\cite{boehm2016systemml} based on Spark can inject intra-application partitionings automatically at runtime. But SystemML is focused on matrix computations, of which the optimal partitionings are easier to search than general data engineering problems coded up with complex UDFs. CoHadoop~\cite{eltabakh2011cohadoop} and Hadoop++~\cite{dittrich2010hadoop++} allow users to specify co-partitionings of HDFS files to benefit join processing in Hadoop by changing the HDFS interface and namenode implementation, but they don't discuss automatic searching of optimal partitionings. HadoopDB~\cite{abouzeid2009hadoopdb} replaces HDFS using relational
databases. Then it allows the user to specify the partition key for
each table to partition data at loading time. \eat{Relational physical database design~\cite{rao2002automating, agrawal2004integrating, nehme2011automated, zhou2012advanced, klonatos2014building} can automatically choose the proper partitionings for storing tables, given a set of known SQL queries. However it is painful to represent and process a bunch of unstructured datasets like the nested product review objects in relational database.}  \eat{Hadoop++~\cite{dittrich2010hadoop++} proposes a co-partitioned join
operator called Trojan join and assumes application programmer understand the data
schema and workloads and trust them to use Trojan join properly.
CoHadoop~\cite{eltabakh2011cohadoop}  provides a lightweight
``locator'' extension
to Hadoop and programmer needs to assign the
same locator to the same partition (same key ranges) in different
files to specify that those partitions need to be co-located. \emph{Different with \textit{Lachesis}, none of above approaches are focused on automatic and persistent partitioning.}  }

\vspace{3pt}
\noindent
\textbf{Automated Partitioning and Physical Database Design for Relational Databases.}} 
IBM DB2 Partition
Advisor~\cite{rao2002automating} recommends candidate partitioning
schemes for physical database design in a shared-nothing
parallel database. Microsoft's AutoAdmin for SQLServer~\cite{agrawal2004integrating}
takes an integrated approach to the problem of choosing indexes,
materialized views, and partitioning, based on optimizer estimated
costs and what-if extensions. But they focus on a single-machine scenario. Nahem and et al~\cite{nehme2011automated} propose an automatic partition
algorithm for MPP database called MESA that deeply integrates with query optimizers and
relies on historical statistics. Legobase~\cite{shaikhha2018building} partitions a relation
differently based on primary key and foreign keys when loading data
and chooses a replica with optimal partition scheme in physical optimization.
AdaptDB~\cite{lu2017adaptdb} partitions data by refining data
partitioning along with relational query executions. 
Eadon et al.~\cite{eadon2008supporting} propose to co-partition relational tables linked via foreign key relationships. Zamanian et al.~\cite{zamanian2015locality} extend this approach to further utilize replication to improve locality. 
Hilprecht et al.~\cite{hilprecht2019learning, hilprecht2020learning} propose an automatic partitioning advisor using DRL for relational database. They utilize the functional dependency among relational attributes to formulate a network-based state vector. In addition, they use a cost model to bootstrap the DRL model at an offline phase, which can be leveraged to improve our training process. DRL also has been applied to parameter tuning in relational databases, such as CDBTune~\cite{zhang2019end} and QTune~\cite{li2019qtune}. 
There are also a number of automatic data partitioning algorithms for
OLTP workloads, like Schism~\cite{curino2010schism}, Sword~\cite{kumar2014sword}, Horticulture~\cite{pavlo2012skew}.
These works are not designed for the UDF-centric analytics workloads and non-relational data, so some of the key ideas such as utilizing functional dependency of relational tables are not applicable to this work, while some other ideas such as replication~\cite{zamanian2015locality} are orthogonal with our work.

\eat{
\vspace{3pt}
\noindent
\textbf{Intermediate Representation for UDF-centric Analytics.}   
Some systems such as TupleWare~\cite{crotty2015tupleware} compile user application code into IR at AST level. However, it is difficult or impossible to reason about the behavior of UDFs that encompass hundreds or thousands of lines of domain-specific code for operating over arbitrary objects. An alternative is to use a DSL rather than an opaque UDF to customize relational operators. The DSL can force the programmer to expose intent. This is the approach taken by SparkSQL~\cite{armbrust2015spark}, for example. Using the DSL, Spark can analyze the AST of the code and understand the semantics of an expression such as \texttt{employee1.startYear === employee2.endYear + 1}. However, after such an AST tree has been optimized, it is directly converted into Java byte code; the semantics are lost and a code snippet like \texttt{employee2.endYear + 1} cannot be extracted separately and re-invoked in the future. This means that it cannot be directly used to perform workload-based optimization. %Also SparkSQL DSL can not work with highly nested objects and arbitrary UDFs as to our knowledge. 
Weld IR~\cite{palkar2017weld} and Lara IR~\cite{kunft2019intermediate} map python code to a high-level functional language. However, %similar with SparkSQL DSL, the IR and the code generation are fully decoupled. This means 
if you want to reuse the code for some part of the IR, e.g. a function that maps an object to a key, you need first extract the IR fragment, modify it into a complete Weld IR program, and generate code from it using a compiler. This process is difficult to automate. There are other provenance graph IRs proposed in recent, like Juneau for Jupyter Notebook~\cite{zhang2020finding}. While we mainly use the PC's lambda calculus IR~\cite{zou2018plinycompute} to demonstrate the key ideas proposed in this paper, these ideas can also be extended to these existing DSL/IRs to bridge the gap. 
}}

\eat{

%\section{System Implementation}
%\label{sec:dsl}
\eat{
\subsection{DSL/IR Subsystem}
We provide a set of subgraph-reuse-friendly DSL/IR functionalities to serve as the function $h_{\mathcal{W}\leftarrow\mathcal{A}}$ that extracts from an opaque workload a representation of the more detailed data flows and control flows to describe and unobscure the workload. The main idea is to embed a DSL in a high-level programming language, and the compilation of the DSL generates an IR. The design of DSL/IR targets at three goals:

\noindent
(1) The DSL must be fully declarative and easy to use, so that the programmer only specifies what to compute, and the system optimizer automatically decides how to compute (including the selection of join ordering, selection of the partitioner), based on the hardware environments, input data characteristics, etc.. This ensures the data independence of our proposed automatic partitioning process.

\noindent
(2) The IR must facilitate the reuse of a connected subgraph of the IR so that it can run separately like a standalone executable. This ensures that $\forall F_i \in \mathcal{F}$, $F_i$ can be reused independently.

\noindent
(3) The IR must facilitate the analysis of the desired partitioning logic for each partition-required operator, so that the implementation of the partitioner candidate enumeration function $\hat{h}_{\mathcal{W}\leftarrow\mathcal{F}}$ is feasible.

To fulfill above goals, the corresponding design decisions made are as follows:

\noindent
(1) In the DSL, we provide a set of high-level declarative operators such as \texttt{join}, \texttt{aggregate}, \texttt{select} (i.e. map), \texttt{multiselect} (i.e. flatten), \texttt{sort}, \texttt{partition} and so on, which can be further customized by UDFs composed by lower-level declarative lambda calculus constructs~\cite{barendregt1984lambda}. Although it look similar to dataflow languages such as Spark~\cite{zaharia2012resilient} with Weld~\cite{palkar2017weld} embedded (so that our DSL is as easy to use as these languages), there is a distinction in the declarativeness. For example, the \texttt{join} can take any number of collections of arbitrary objects as inputs with lambda calculus specifying how to join. This is different with the \texttt{join} in other dataflow languages, which can only take two collections of key-value pairs, and leads to an undesirable fact that the join ordering must be controlled by the programmers. 

\noindent
(2) To facilitate the reuse of any subcomputation embedded in UDFs, the IR is designed into a DAG where each node represents an atomic computation (i.e. a lambda term) that can be isolated and executable. Once the DSL is compiled, the binary codes for these atomic computations \eat{as well as the logic that executes a flexible combination of them} are generated. Therefore, it is easy to reuse and execute a subgraph that combines arbitrary atomic computations. As far as to our knowledge, we are the first to propose such IR design to facilitate the decomposition of UDFs and the reuse of their subcomputations. To do the same in existing dataflow languages, you need analyze and glue assembly code together at LLVM level, or glue high-level DSL fragments together and then compile these to executable code~\cite{crotty2015tupleware, palkar2017weld}.

\noindent
(3) To facilitate the analysis of subcomputations, we allow programmers to indirectly control the number of nodes in an IR graph to a reasonable level by encapsulating a bulk of user defined logic not relevant with partitioning (e.g. a complex UDF that specifies \texttt{join} projection logic) into an opaque lambda term.

}

\eat{
Today's analytics platforms such as Spark, Flink, Google Dataflow provide dataflow APIs like following~\cite{misale2017pico}: map$\langle$f$\rangle$(x) = $\{f(v) | v \in x \}$, filter$\langle$f$\rangle$(x) = $\{v | v \in x, f(v) == true \}$, flatmap$\langle$f$\rangle$(x) = $\{v' | v \in x, v' \in f(v)\}$, groupByKey(x) = $\{(k, \{v | (k, v)\in x\})\}$, join(x, y) = $\{(k, \{v_x, v_y\}) | (k, v_x)\in x, (k, v_y)\in y\}$, etc..

The core UDF-centric data abstractions in these platforms, such as RDD in Spark, DataSet in Flink, PCollection in Google dataflow, and so on, compared to relational dataframes, can flexibly represent a distributed collection of nested objects that share the same type. However for partitioning-sensitive operators like \texttt{join}, it only accepts two inputs, which must have the $\langle$key, value$\rangle$ pair type. Such abstraction exposes following problems:

\noindent 
(1) It is difficult to extract sub computations as a standalone executable from the UDFs that are written in high-level language, such as Scala, or in embedded DSLs like Weld~\cite{palkar2017weld}. You need analyze and glue assembly code together at LLVM level, or glue high-level DSL fragments together and then compile it to executable code.

\noindent
(2) The UDFs related with one computation are not encapsulated well. For example, the logic about how a \texttt{join} key or \texttt{sort} key is extracted, is defined outside of the \texttt{join} or \texttt{sort} operator. In order to identify a candidate partition function, it requires additional efforts to break down and glue related functions. 

\noindent
(3) Their binary join operator based on $\langle$key, value$\rangle$ pairs is not declarative, because while a user specifies the key mapping in multiple joins, the join ordering is also implicitly determined by the user.  As a result, in these systems, tuning application implementations such as join ordering for good performance becomes a heavy burden for programmers.

Therefore, we choose to design a DSL based on relational algebra operators. Each operator will be further customized by a few user defined functions that can be described using lambda calculus expressions. We choose the design not only for the ease to create persistent partitionings across applications, but also to ensure the declarativity of user-programming. }
\eat{The adoption of relational operators greatly  facilitates the identification of  partition candidates, which are always embedded in the join filtering functions, aggregation key projection functions, and so on. }

\eat{Under the hood, each lambda calculus expression will return a tree of lambda term abstractions and higher order lambda composition functions, which can be traversed by a system optimizer for extracting partitioning predicates as lambda terms from the UDFs.}

\eat{
\subsection{Formalization}
\label{sec:operations}

}

\eat{We can have a lower level lambda calculus to support recursion and loop inside of one lambda term. Although capturing such complex structure can enable optimizations such as loop fusion and tiling, it has hardly any benefits for data partitioning, which is the focus of this paper. So we omit the discussions for such structures here.}

\eat{
\noindent
\textbf{About manual and intra-application partitioning.} Besides the automatic persistent partitioning that we advocate in our work, based on our DSL, programmers can also choose to manually create persistent partitionings. If a \texttt{Set} is defined with a partition scheme identifier (enum type, which can specify one of range partitioning, hash partitioning, random partitioning, and round-robin partitioning); and optionally a partition computation (a lambda term object) $f_{keyProj}$ that specifies how to extract the partition key from each object of type in the \texttt{Set}. Also the \texttt{partition} operator can be used to perform an intra-application partitioning. 

\eat{
Our investigation with $6$ graduate students and $3$ professionals also show that it usually takes about $1$-week to $4$-weeks' time for someone new to this language to successfully write an application using the DSL proposed in this work. This learning curve is similar to other dataflow language such as Spark APIs and Flink APIs. }}

%\subsection{IR Functionalities}

%\label{sec:examples}
\eat{
At execution time, UDF-centric applications coded in our declarative DSL will be translated into an IR, which is a directed acyclic graph (DAG). Each node is an executable atomic computation, which is annotated with the information regarding this atmoic computation (e.g. computation type, invoked method name, projected attribute name, etc.). Each edge represents a dataflow, for which the destination node processes the data output from the source node; or a control flow, where the execution of the destination node depends on the output from the source node. For example, the IR created for the code in Listing.~\ref{code4} is illustrated in Fig.~\ref{fig:partition-candidate}. 

The supported atomic computations includes unary physical operators such as to apply a lambda abstraction function to or create hash on a vector of objects; and binary physical operators such as to compare two vectors of objects, to join two vectors of hashed objects, and to filter a vector of objects based on an associated vector of boolean values. These lower-level atomic computations are fully transparent to the programmers. We implement each atomic computations using C++ meta templated programming for high-performance.

As mentioned, the IR is generated while compiling the DSL. So each subgraph can be executed separately without the need for re-compliation.
For example, the two partitioner candidates highlighted in Fig.~\ref{fig:partition-candidate} can be executed directly from the IR, requiring no compilation and code generation. Thus, any partitioner candidate can be reused if the corresponding historical IR exists. Easy reuse of historical IR subgraphs is a critical enabler for automatically creating partitionings at storage time.

We propose several unique IR functionalities for creating persistent partitioning. The first functionality is to extract and index a subgraph of the IR as a partitioner candidate using the source scanner node and the output node that are unique for each partitioner candidate. The second functionality is to store and cache historical IRs and subgraph indexes, which facilitates efficient reuse of partitioner candidates. The third one is an IR matching functionality that is often used in query optimization, for determining whether the partitioner of the input datasets matches the desired partitioner of an partition-required operator, so that it can be utilized to avoid a shuffling operation. The IR matching is similar to a DAG isomorphism problem~\cite{babai2016graph}, except that in our problem each partitioner DAG derived from Alg.~\ref{alg:search-partitioner-candidate} and Alg.~\ref{alg:merge-partitioner-candidate} must have one and only one source scanner node and one and only one destination output node. Therefore each partitioner DAG can be seen as a set of all possible paths enumerated from the scanner node to the output node. Then a globally unique signature can be derived from each partitioner DAG by sorting and concatenating sub-signatures of each path, where each sub-signature is simply the concatenation of all labels in the path. In the same time, a shorter and higher-level signature that simply encodes the identifier of the source scanner node, the destnation output node, and the number of nodes in the DAG is used to serve as a bloom filter for fast filtering of non-identical partitioner DAGs.}

\eat{Due to space limitation we will not elaborate more details on these functionalities.}

\eat{
\subsubsection{Extraction and Index of Partitioner Candidates}
\label{sec:extract}
As mentioned, in the graph IR, each node represents an executable lambda term, and each edge represents a dataflow or control flow. A subgraph of the IR is also executable, and can be extracted as a function. Therefore, we can analyze IR to extract partitioner candidates as subgraphs in the IR.

Each IR graph has a set of source nodes that directly connect to the input datasets and have no parents.  \eat{Souce nodes are highlighted in red in the examples of Sec.~\ref{sec:examples}.} Given an IR graph, the partitioner candidate analyzer will first check whether the Then it will iterate the set of source nodes, and in each iteration, it will start from an empty outcome subgraph, add one source node as the root of the subgraph, and traverse and include all of the connected edges into the outcome subgraph, until it meets the ending lambda term or a binary higher order lambda function such as boolean comparison operation (i.e. ==, >, !=) of a partition-sensitive UDF like join filter function ($f_{fil}@join$), aggregation key projection function ($f_{keyProj}@aggregate$), and sort key extraction function ($f_{keyProj}@sort$). At the end of each iteration, the analyzer will return the final outcome subgraph.

Suppose there are a large number of application IR graphs, and each application may yield multiple partitioner candidates for all input datasets. Then if we describe a partitioner candidate as an independent subgraph (i.e. a set of edges), the storage overhead associated will linearly increase with the number of IRs. 

To address the problem, we further identify that actually the subgraph for a partition candidate always has only one source node that has no parent nodes and only one destination node that has no children nodes. This is because a partitioner candidate is to extract a key object from a source object, so it must preserves such 1-1 mapping. Based on this observation, we can index a partitioner candidate by the ID of its source node, ID of its destination node, and the ID of the IR graph, denoted as <\texttt{IR{ref}}, \texttt{src}, \texttt{dest}>. Introducing such indexing can significantly reduce the required storage space for partitioner candidates.

\subsubsection{Storage of IRs and Partitioner Candidates} In Lachesis, all of the IRs are stored as a graph in the storage formats of PC object model~\cite{zou2018plinycompute}, in which the object has exactly same representation in memory and on disk by replacing c-style pointer into offset-based handle, which eliminates (de)serialization overhead. In our implementation, each IR is indexed by its application ID, and stored as <\texttt{appId}, \texttt{IR}>. Again, each node or subgraph in the IR can be executed independently on corresponding inputs.

Based on the compact index scheme we introduced for describing a partitioner candidate, we further store all partitioner candidates as key-value pairs. The key is the user defined type (UDT) of the input object, and we obtain the UDT at runtime via the dynamic dispatching approach as described in PlinyCompute~\cite{zou2018plinycompute}. The value part is a <\texttt{appId}, \texttt{src}, \texttt{dest}> triple that can be utilized to locate the subgraph in the IR indexed by \texttt{appId}, as well as a few bits to specify information for fast matching at query optimization time which we will describe in Sec.~\ref{sec:match}.  We cache all pairs in a hashmap so that given a \texttt{Set} of objects with a certain user defined type, we can quickly find all possible partitioner candidates that accepts input data of the same UDT type.

\subsubsection{IR Matching}
\label{sec:match}
Once a \texttt{Set} of objects is partitioned by a selected partitioner candidate (See Sec.~\ref{sec:rl} for the selection strategy) at storage time, the \texttt{Set} will be associated with a partition key projection function ($f_{partition}$) represented in the compact index format. If one or more partitioned \texttt{Set}s are input to a running application, the query optimizer needs to determine at runtime  whether the persistent partitionings associated with the input datasets are matching the expected co-partition pattern and can be utilized to avoid one or more shuffling operations.

To address the problem, if the IR of the current application has one or more partition-sensitive operations (e.g. join filter function ($f_{fil}@join$), aggregation key projection function ($f_{keyProj}@aggr-egate$), and sort key extraction function ($f_{keyProj}@sort$)), for each of these operations, we will extract the partitioner candidate for each input dataset using the approach described in Sec.~\ref{sec:extract} as expected partitioner, and match each expected partitioner to the persistent partitioning provided by the corresponding input \texttt{Set}. If all input \texttt{Set}s match, the optimizer will decide to avoid a shuffling stage for executing this partition-sensitive operation.

In the partitioner matching process, if the two partitioners have the same index (<\texttt{appId}, \texttt{src}, \texttt{dest}>), they must match. Otherwise, the partitioner associated with the \texttt{Set} is extracted from a different application, and we need further retrieve the node and edge information of the indexed subgraph from that application's IR. Then the matching problem is transformed into a graph isomorphism problem, to 
answer the question whether the subgraphs associated with the two partitioners are identical or not.

Different with general graph matching problem, a node that is a lambda term abstracted in an opaque UDF  (i.e. \texttt{func(x, f)}, such as the json\_parse function in Fig.~\ref{fig:example1-ir}) has opaque semantic, and thus such node cannot be exactly matched. Therefore when extracting partitioner candidates, we add a bit to the partitioner representation to specify whether any opaque lambda term is included in its subgraph. If the index doesn't match and the bit is set, then there is no need to run the graph matching and it returns a negative matching result directly. We use several other bits for similar fast filtering purposes. For example, four bits are used to specify the hash of the sequence of nodes in breadth-first traversal ordering (nodes at the same layer are also radix-sorted by name).

}

%\subsection{Historical Workflow Analysis}

\eat{
Then the system will enumerate partitioner candidates from these groups as described in Sec.~\ref{sec:dsl}. To select a partitioner candidate using the DRL model as described in Sec.~\ref{sec:rl}, the system will extract features for each partitioner candidate based on historical workflow analysis. These features include:}

\eat{
\subsection{Cost-based Optimizer}
\label{sec:rb}
We first propose a cost-based approach to automatically select a candidate lambda term for partitioning a dataset to be generated.

\subsubsection{Cost Model}
The overall benefit of each candidate is computed as the sum of the benefit estimated for each \texttt{historical consumer stage}, which once used the lambda term to extract join keys or aggregation keys from a dataset created by a \texttt{historical job stage equivalent}. The  benefit of the $i$-th \texttt{historical consumer stage} is further estimated as the probability that the \texttt{historical consumer stage} will be executed again in the next future time period $T$, denoted as $p_{i}$, multiplied by the estimated output size of this future execution of the $i$-th \texttt{historical consumer stage}, denoted as $s_i$. So the benefit model is as illustrated in Eq.~\ref{eq:cost}.
\begin{equation}
\label{eq:cost}
B = \sum{p_{i} \times s_{i}}
\end{equation}

We need take $s_i$ into consideration, because the partition-preserving operations like transformation, filtering and so on can significantly change the data size in the pipeline, which will significantly affect the running time of the consumer job stage. For example, for a certain consumer job stage, if there is a filtering operation that removed a lot of the data to be generated, the actual input to the join/aggregation operation in this consumer job stage could be very small in size, then there is not much performance gain can be brought to this consumer by pre-partitioning the data.

\subsubsection{Estimation of Model Parameters}
$p_{i}$ for a job stage is computed from the job stage's $\lambda$ value, where $\lambda$
is the rate (per time period) at which the job stage is used.
If we model the arrival time of the next execution of each job stage as a Poisson point process \cite{kleinrock1976queueing}, then the probability that the job stage is used in the next t time 
periods is $1 - e^{-\lambda t}$ (this follows from the cumulative density of the exponential distribution, which models the time-until-next arrival for a Poisson point process).

There are a number of ways that $\lambda$ for a job stage can be estimated.
We can collect the number of executions $n$ of the job stage in the 
last $t'$ time periods, and estimate the rate of usage per time period as $\lambda \approx n/t'$.
The historical workflow analysis component keeps track of all executions for all job stages, and estimate $\lambda$ by using a sliding time window.  
\vspace{3pt}
To estimate $s_i$ for a particular job stage $i$, we first collect all the historical workflows executions, each of which leads from a \texttt{historical job stage equivalent}, which we call as the source stage, to the $i$-th \texttt{historical consumer job stage}, which we call as a destination stage. For all such workflow executions, we compute the  selection ratio, which is defined as the output size of the destination stage to the input size of the source stage. We group these selection ratios into a histogram, and select the middle point of the most frequent range as the estimation for the future selection ratio of this $i$-th \texttt{historical consumer job stage}. We further estimate $s_i$ as the multiplication of the input size of current job stage that is going to generate the data, and the estimated selection ratio.

\vspace{3pt}
User can specify a heterogeneous replication factor $K$, and the system will select the $K$ lambda terms that have the top estimated values for performance benefit to partition the data in parallel. As a result the data generated will have $K$ replicas, and each replica is partitioned by a different lambda term.

The query optimizer that we design will select a replica to work on by matching the replica's associated lambda term with the IR of the query to optimize, using the IR matching approach as described in Sec.~\ref{sec:match-history}.
}

}

\section{Conclusion}
In this paper, we argue that automatically creating persistent data partitionings for Big Data applications is an important and challenging task for UDF-centric workloads.
We propose \textit{Lachesis} to address the problem, which includes a unique set of functionalities for extracting, reusing, and matching of sub-computations in UDFs. \textit{Lachesis} also provides a data placement optimizer based on a deep reinforcement learning approach and historical workflow analysis. The evaluation results demonstrate that \textit{Lachesis} can bring up significant performance speedup for various Big Data integration and analytics applications such as Big Data integration, deep learning model serving, web analytics, and analytics queries. The proposed approach is effective with different data sizes and different environments. Most importantly, \textit{Lachesis} significantly reduces the efforts required on the part of enterprise IT professionals and data scientists who may not have sufficient systems tuning skills for creating partitionings for UDF-centric analytics applications.

\eat{\section{Acknowledgements} 
Thanks to two ASU graduate students: Pradeep Jampani for helping with the implementation and benchmark of the sparse matrix multiplication, and Srihari Jayakumar for helping with the benchmark of the dense matrix multiplication. Thanks to Donpaul Stephens for
his insightful feedbacks on an earlier version of this
paper. The work presented in this paper has been supported by the ASU faculty start-up funding, AWS Cloud Credits for Research program, Google GCP research credits program, and DARPA MUSE award No. FA8750-14-2-0270.}

\eat{
\begin{appendix}
\section{Lambda Calculus DSL formalization}
Our DSL is mainly based on two core data abstractions: typed \texttt{Set} and typed \texttt{Map}. All data involved in a Big Data analytics workflow can be represented using these two data structures.

A \texttt{Set} represents a collection of immutable
objects that share the same type, such as the input data, and output data. It can be associated with a partition
scheme (i.e. range partition, hash partition, random partition, round-robin partition, customized partition) and a key projection function that extracts the partition key from each object, which is required for range partition and hash partition.

\begin{DSL}
Set$\langle$T, partitionKeyT$\rangle$ {
  isPartitioned: bool; 
  partitionScheme: enum{Hash, Range, ...};
  $f_{keyProj}: T \rightarrow partitionKeyT$;
  D: Vector$\langle$T$\rangle$;
}
\end{DSL}

In addition, a \texttt{Map}  is used to represent a hash map of $\langle$key, value$\rangle$ pairs generated intermediately for aggregation and join probing, which can be mutated, described as following. The $\langle$key, value$\rangle$ pairs are always hash partitioned on its key.

\begin{DSL}
Map$\langle$KeyT, ValueT$\rangle$ {
  D: HashMap$\langle$KeyT, ValueT$\rangle$;
}
\end{DSL}

Some examples of the formalized relational operators are described as following:

\noindent
\textbf{Partition-Required Operators}

\noindent
(1) A $join$ operator that takes arbitrary number of inputs, a filter function $f_{fil}$ (i.e. the \texttt{WHERE} caluse) and a transformation function $f_{proj}$ (i.e. the \texttt{SELECT} clause), maps $n$ input \texttt{Set}s of objects having type \texttt{$T_1,...,T_n$} to an output \texttt{Set} of objects having type \texttt{$O$}:

\begin{DSL}
join$\langle$$T_1$,...,$T_n$,O$\rangle$$\{$
$f_{fil}$:$\{T_1$,...,$T_n\}\rightarrow$ bool, $f_{proj}$:$\{T_1,...,T_n\}\rightarrow$ O$\}(S_1,...,S_n)$
$=\{f_{proj}(v_1,...,v_n) | f_{fil}(v_1,...,v_n)==true, v_i\in S_i, i\in{1...n}\}$
\end{DSL}

\noindent
(2) An $aggregate$ operator takes a key projection function $f_{keyProj}$, a value projection function $f_{valueProj}$, an aggregation function $f_{+}$ that aggregates two values that have the same key, and an output function $f_O$ that constructs output object having type $O$ from an aggregated \texttt{$\langle$k, v$\rangle$} pair. In the end, it transforms an input \texttt{Set} having type \texttt{T} to an output \texttt{Set} of objects having type \texttt{$O$}: 

\begin{DSL}
aggregate$\langle$T,K,V,O$\rangle$$\{$
$f_{keyProj}$:T$\rightarrow$ K, $f_{valueProj}$:T$\rightarrow$ V,
$f_{+}$:(V, V)$\rightarrow$ V, 
$f_O$:(K, V)$\rightarrow$ O$\}(x)$
$= \{f_O(k,f_+(\{v|k=f_{keyProj}(a), v=f_{valueProj}(a), a\in x\}))\}$
\end{DSL}

\noindent
In practice, two functions are implemented based on $f_{+}$: a local aggregation function that aggregates a local \texttt{$\langle$k, v$\rangle$} pair to an intermediate \texttt{Map$\langle$K, V$\rangle$} structure, and a global aggregation function that aggregates a shuffled \texttt{Map$\langle$K, V$\rangle$} structure to a same \texttt{Map$\langle$K, V$\rangle$} structure that conducts final aggregation for a data partition.

\noindent
(3) A $sort$ operator takes a  comparator function $f_{keyProj}$ and sort the input \texttt{Set} of objects according to the the key (of type \texttt{K}, on which a comparator function must be defined) projected from each object :

\begin{DSL}
sort$\langle$T, K$\rangle$$\{$$f_{keyProj}$:T$\rightarrow K\}(S)$$=\{v|v\in S\}$
\end{DSL}

\noindent
(4) A $partition$ operator takes a key projection function $f_{keyProj}$ and partitions the input \texttt{Set} of objects according to the key (of type \texttt{K}, on which a hash function must be defined) projected from each object:

\begin{DSL}
partition$\langle$T, K$\rangle$$\{$$f_{keyProj}$:T$\rightarrow K\}(S)$$=\{v|v\in S\}$
\end{DSL}

\noindent
\textbf{Non-Partition-Required Operators}

\noindent
(1) A $select$ operator that takes a filter function $f_{fil}$ (i.e. the \texttt{WHERE} clause) and a transformation function $f_{proj}$ (i.e. the \texttt{SELECT} clause), maps an input \texttt{Set} of objects $S$, having type \texttt{T} to an output \texttt{Set} of objects having type \texttt{O}:

\begin{DSL}
select$\langle$T, O$\rangle$$\{$
$f_{fil}$: T $\rightarrow$ bool, $f_{proj}$: T$\rightarrow$ O$\} (S)$
$= \{f_{proj}(v) | v \in S, f_{fil}(v) == true\}$
\end{DSL}

\noindent
(2) A $multiselect$ operator flattens an input \texttt{Set} of objects having type \texttt{T} to an output \texttt{Set} of objects having type \texttt{O}:

\begin{DSL}
multiselect$\langle$T,O$\rangle$$\{$
$f_{fil}$:T$\rightarrow$ bool, $f_{proj}$:T$\rightarrow$ Vector<O>$\}(S)$
$= \{v' | v \in S, f_{fil}(v) == true, v' \in f_{proj}(v)\}$
\end{DSL}

To make functions  (e.g. $f_{partition}$, $f_{fil}$, $f_{compare}$, and etc.) decomposable to executable partitioner candidates, we further propose a set
of lambda calculus expressions to define these functions.

The expressions consist of a set of built-in \emph{lambda abstraction} 
families \cite{miller1991logic} to create lambda terms for input objects, as 
well as a set of \emph{higher-order functions} \cite{chen1993hilog}
that take as input one or more lambda terms, and return a new lambda term.  These built-in lambda abstraction families 
include:

\noindent
(1) \texttt{member($x$, attribute)}, takes an object $x$ as input, and returns a function returning one of the object's member variables (i.e. $\lambda x.(x$->$attribute)$ in lambda calculus formalization~\cite{barendregt1984lambda}. Here, the first $x$ is the variable, and the part after $.$ is the function to be applied to the variable.);

\noindent
(2) \texttt{method($x$, methodName)}, which returns a function calling a method on the object (i.e. $\lambda x.(x$->$methodName()))$;

\noindent
(3) \texttt{literal($l$)}, returns a function returning $l$ (i.e. $\lambda x.l$);

\noindent
(4) \texttt{self($x$)}, returns an identity function (i.e. $\lambda x.x$);

\noindent
(5) \texttt{func($x$, f)}, returning a function calling
opaque user defined function $f(x)$ to express complex but not performance-critical logic, for balance of ease of user programming and ease of system query optimization (i.e. $\lambda x.f(x)$).

\noindent
Then a set of higher-order functions are provided to compose above
basic lambda terms into a new lambda term that can be regarded as a
tree of lambda terms, which include:

\noindent
(1) The boolean comparison operations: \texttt{==}, \texttt{>}, \texttt{!=}, etc.;

\noindent
(2) The boolean
operations: \texttt{\&\&}, \texttt{||}, \texttt{!}, etc.;

\noindent
(3) The arithmetic operations: \texttt{+}, \texttt{-}, \texttt{*}, etc.;

\noindent
(4) The construction operation: \texttt{construct(O, $l_1$,...,$l_n$)}, which returns a lambda term that constructs a new object of type \texttt{O} using values returned from lambda terms $l_1$,...,$l_n$;

\noindent
(5) Parentheses that define the explicit ordering for lambda term composing: \texttt{()};

\noindent
(6) index operator: \texttt{[]}.

\noindent
(7) conditional branch: \texttt{$l_1$?$l_2$:$l_3$}, or \texttt{if($l_1$) $l_2$ else $l_3$}.

\section{DSL/IR Examples}
\label{sec:more}

\subsection{Three-way Join}
In this example, we join three Reddit datasets: (1) Reddit submissions, which is a $6$ terabytes' JSON dataset including all reddit posts submitted in Aug $2019$; (2) Reddit accounts, which is $1.2$ terabytes' CSV dataset containing the information about 78 millions of reddit users; and (3) the subreddits dataset, which is a $1.6$ terabytes' JSON dataset including the information about all reddit forums. The goal is to form a training dataset for automatically recommending posts to reddit users. We show the join\_filter function {$f_{fil}$} expressed in lambda calculus in Listing.~\ref{reddit_filter}. We omit the project function $f_{proj}$ of this join operator here.

\begin{lstlisting}[language=C,frame=single,caption=Reddit join: $f_{fil}$ expressed in lambda calculus DSL, label=reddit_filter, breaklines=true, basicstyle=\small, columns=fullflexible,keywordstyle=\bfseries\color{black}, otherkeywords = {lambda, join_filter, func,=,[,],==,[],&,&,&&}]
lambda<bool> join_filter(string submission, string account, string subreddit) {
    return (func(submission, json_parse)["author"] == func(account, csv_parse)[1]) && (func(submission, json_parse)["subreddit_id"] ==(func(subreddit, json_parse)["id"]);
}
\end{lstlisting}

The above expression will return a tree of lambda terms as illustrated in Fig.~\ref{fig:example1-ir}.
The logical query optimizer will traverse the tree of lambda terms, extract partitioner candidates as a subgraph in the IR, and store the IR as a persistent PC object~\cite{zou2018plinycompute}, as well as the index of each partitioner candidate into the \textit{Lachesis}, while applying other optimization techniques, such as removing the redundant lambda term, fuse operators and so on. 

Listing~\ref{reddit_dataflow} shows the code of using the binary join operator in dataflow-based DSL
~\cite{palkar2017weld, zaharia2010spark} to implement the logic of Listing~\ref{reddit_filter}. You will see it depends on the programmers to determine the join ordering, whether to first join submissions with accounts, or first join submissions with subreddits, thus it is not declarative. In contrary, if our proposed DSL is used as in Listing~\ref{reddit_filter}, the join ordering can be automatically determined by traversing the IR via a query optimizer. In addition, the dataflow-based DSL requires to trace how is a join key extracted from a series of transformations, which is more complex than our proposed DSL where 

\begin{figure}[H]
\centering{%
   \includegraphics[width=3.4in]{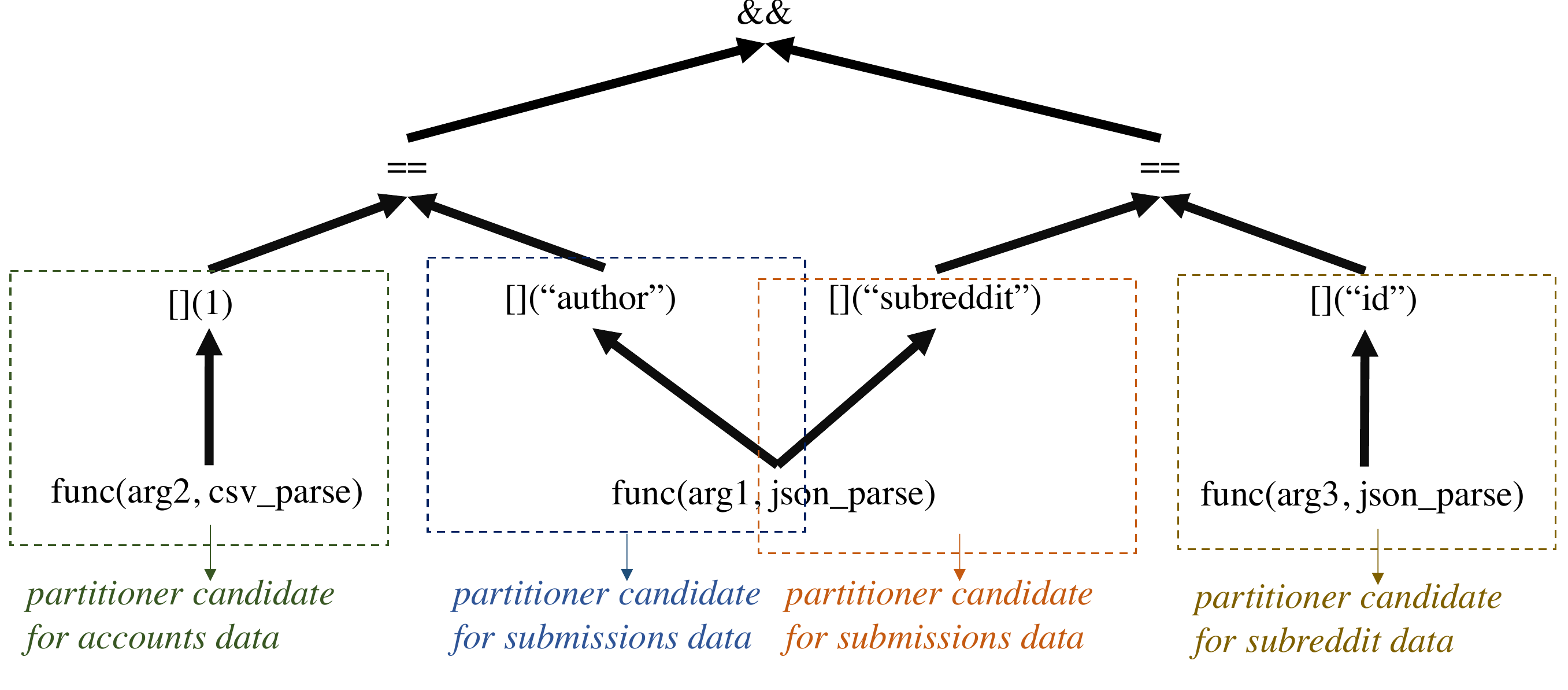}  
}
\caption{\label{fig:example1-ir}
IR fragments for the 3-way reddit datasets join.
}
\end{figure}

\begin{lstlisting}[language=C,frame=single,caption=Reddit join: expressed in dataflow-based DSL, label=reddit_dataflow, breaklines=true, basicstyle=\small, columns=fullflexible,keywordstyle=\bfseries\color{black}, otherkeywords = {map, join, =>}]
submission_jsons = submissions.map(submission => json_parse(submission))
submission_pairs = submission_jsons.map(submission_json => (submission_json["author"], (submission_json["subreddit"], submission_json))
account_pairs = accounts.map(account => (csv_parse(account)[1], account))
//join submissions with accounts first
intermediate_pairs = submission_pairs.join(account_pairs).map((author, ((subreddit, submission), account)) => (subreddit, (submission, account)))
//then join with subreddits
subreddit_pairs = subreddits.map(subreddit =>
(json_parse(subreddit)["id"]
result = subreddit_pairs.join(intermediate_pairs)
\end{lstlisting}

\subsection{Log Merge}
In this example, we want to develop a universal tool to merge-sort arbitrary set of large log files into one dataset that is ordered by timestamp~\cite{logmerge}. Each log file may have hundreds of gigabytes to terabytes in size, and may use different timestamp formats, such as ISO-8610 standard, cloud-init standard, Unix time, and so on. So the sort key extraction function ($f_{keyProj}$) needs to match multiple regex expression. Its implementation in our proposed DSL is illustrated in Listing.~\ref{sort_compare}. 

\begin{lstlisting}[language=C,frame=single,caption=Merge-Sort: $f_{keyProj}$ expressed in lambda calculus DSL, label=sort_compare, breaklines=true, basicstyle=\small, columns=fullflexible,keywordstyle=\bfseries\color{black}, otherkeywords = {lambda, key_proj, func,member, if, else, =,[,],==,[],&,&,&&}]
lambda<DateTime> key_proj(string logrec) {
    lambda<Match> match = func(logrec, iso8601_patternmatch);
    if (member(match, res))
        return member(match, datetime);
    lambda<Match> match = func(logrec, cloudinit_patternmatch);
    if (member(match, res))
        return member(match, datetime);
    return member(func(logrec, unix_patternmatch), datetime);
}
\end{lstlisting}

The corresponding IR is illustrated in Fig.~\ref{fig:sort-ir}. It is obvious that the whole \texttt{key\_proj()} function ($f_{keyProj}$) of the sort operator can serve as the partitioner candidate to extract partition keys from log records for range partitioning. This indicates that the user can simply put all of the projection logic into one opaque lambda term as illustrated in Listing.~\ref{sort_opaque}, because there is no need to decompose the sort key extraction logic in this case.

\begin{figure}[H]
\centering{%
   \includegraphics[width=2in]{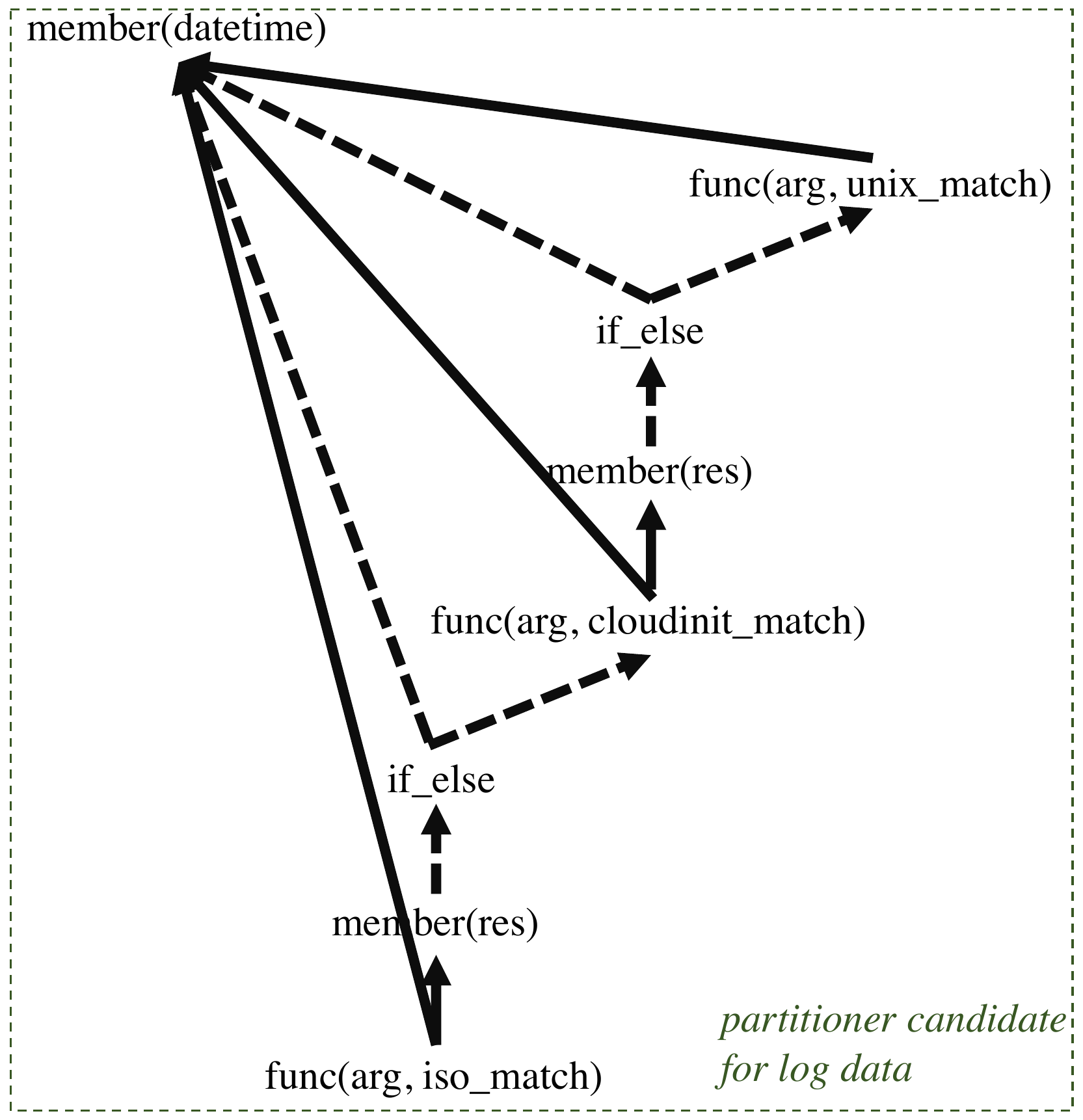}  
}
\caption{\label{fig:sort-ir}
IR example for the sort key projection function, which also serves as a partitioner candidate for the log files. (Solid line represents the dataflow and dashed line represents the control flow.)
}
\end{figure}

\begin{lstlisting}[language=C,frame=single,caption=Merge-Sort: $f_{keyProj}$ expressed in a single lambda term, label=sort_opaque, breaklines=true, basicstyle=\small, columns=fullflexible,keywordstyle=\bfseries\color{black}, otherkeywords = {lambda, key_proj, func,member, if, else, =,[,],[],&,&,&&}]
lambda<DateTime> key_proj(string logrec) {
    return func(logrec)[](string arg){
        Match match = iso8601_patternmatch(logrec);
        if(match.res)
            return match.datetime;
        match = cloudinit_patternmatch(logrec);
        if(match.res)
            return match.datetime;
        return unix_patternmatch(logrec).datetime;
    }
}
\end{lstlisting}

\subsection{Similarity Join}
It is possible that the input data may be transformed through a series of operators before it is being joined. For example, assume we need detect Reddit posts submitted in July $2019$ that are similar to Reddit posts submitted in Aug $2019$, a similarity join~\cite{vernica2010efficient, chen2019customizable} between the July $2019$ dateset and the Aug $2019$ dataset can solve the problem. The application consists of three operators: two flatten operators (\texttt{multiselect}) that derives multiple locality sensitive hash (LSH) signatures from each post for both datasets, followed by a \texttt{join} operator that returns all pairs of Aug post and July post that share at least one LSH. The projection UDF $f_{proj}$ of the \texttt{multiselect} operator and the filter UDF $f_{fil}$ of the \texttt{join} operator are illustrated in Listing.~\ref{multiselect} and Listing.~\ref{similarity-join} respectively.

\begin{lstlisting}[language=C,frame=single,caption=$f_{proj}$ of $multiselect$ in lambda calculus DSL, label=multiselect, breaklines=true, basicstyle=\small, columns=fullflexible,keywordstyle=\bfseries\color{black}, otherkeywords = {lambda, proj, func,member, vector, pair, if, else, =,[,],==,[],&,&,&&}]
lambda<vector<pair<string, string>>> proj(string submission) {
    //the multi_probe is a UDF invoked from a LSH library, it returns a list of strings as LSH signatures.
    return func(submission, multi_probe);
}
\end{lstlisting}

\begin{lstlisting}[language=C,frame=single,caption=$f_{fil}$ of $join$ expressed in lambda calculus DSL, label=similarity-join, breaklines=true, basicstyle=\small, columns=fullflexible,keywordstyle=\bfseries\color{black}, otherkeywords = {lambda, key_proj, func,member, vector, if, else, =,[,],==,[],&,&,&&}]
lambda<bool> join_filter(pair<string, string> lsh_aug_sub_pair, pair<string, string> lsh_jul_sub_pair) {
    return lsh_aug_sub_pair[0] == lsh_jul_sub_pair[0];
}
\end{lstlisting}

The end-to-end IR of the application is illustrated in Fig.~\ref{fig:similarity-join-ir}. Different with other two examples, the partitioner candidates extracted from this application span the lambda terms of multiple operators.

\begin{figure}[H]
\centering{%
   \includegraphics[width=3.5in]{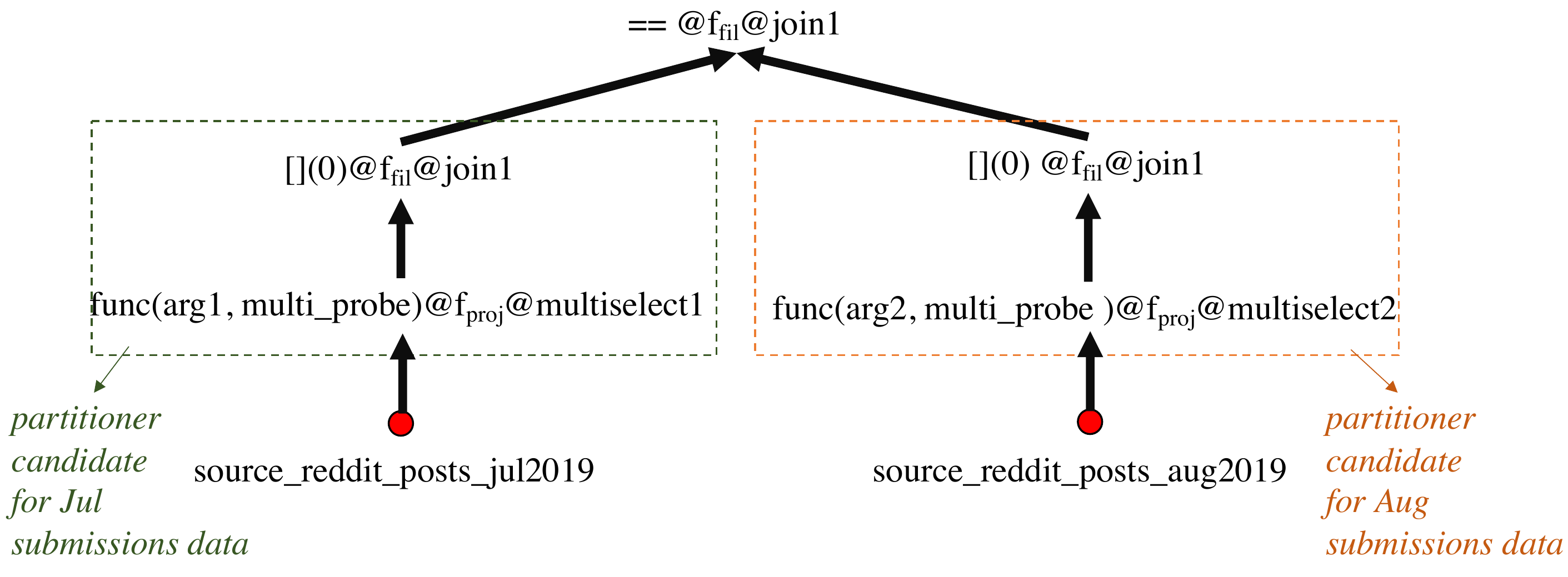}  
}
\caption{\label{fig:similarity-join-ir}
IR for the Similarity Join example.
}
\end{figure}

\end{appendix}}

%\balance

\bibliographystyle{abbrv}
\bibliography{refs}  % vldb_sample.bib is the name of the Bibliography in this case

\end{document}